\documentclass[a4paper,onecolumn,notitlepage,nofootinbib,superscriptaddress,prd]{revtex4-1}
 \usepackage{amsmath}
\usepackage{graphicx}
\usepackage{color}
\usepackage{bm}
\newcommand{\intq}{\int_{\mathbf{q}}}
\newcommand{\deltadir}{\delta^{(\mathrm{D})}}

\renewcommand{\mathbf}{\vec}
\newcommand{\bk}{\mathbf{k}}
\newcommand{\bkappa}{\mathbf{\kappa}}
\newcommand{\bq}{\mathbf{q}}

\newcommand{\bx}{\mathbf{x}}

\newcommand{\be}{\begin{equation}}
\newcommand{\ee}{\end{equation}}

\newcommand{\hMpcsq}{h^{-2}\ \text{Mpc}^2}
\newcommand{\ihMpc}{h\ \text{Mpc}^{-1}}

\renewcommand{\vec}{\bm}
\newcommand{\derd}{\text{d}}
\newcommand{\deltakron}{\delta^\text{(K)}}

\newcommand{\cssq}{c_\text{s}^2}

\newcommand{\lcdm}{$\Lambda$CDM}
\newcommand{\Eprint}{arXiv: }
\newcommand{\bkone}{\mathbf{k}_{1}}
\newcommand{\bktwo}{\mathbf{k}_{2}}
\newcommand{\bkthree}{\mathbf{k}_{3}}
\newcommand{\bkfour}{\mathbf{k}_{4}}
\newcommand{\el}{\@fleqntrue\@mathmargin0pt}
\newcommand{\cl}{\@fleqnfalse}
\renewcommand{\vec}{\bm}
\newcommand{\ii}{\text{i}}

\allowdisplaybreaks
\begin{document}
\title{Precise Calibration of the One-Loop Trispectrum in the Effective Field Theory of Large Scale Structure}

\author{Theodore Steele}
\email{ts715@cam.ac.uk}
\affiliation{Department of Applied Mathematics and Theoretical Physics, University of Cambridge, Wilberforce Road, CB3 0WA}
\author{Tobias Baldauf}
\affiliation{Department of Applied Mathematics and Theoretical Physics, University of Cambridge, Wilberforce Road, CB3 0WA}

\date{\today}

\begin{abstract}
The Large-Scale Structure (LSS) of the universe has the potential to provide decisive answers to the remaining open questions in cosmology.  Early attempts at modelling it analytically focused on using perturbation theory. However, small-scale effects introduced by gravitational collapse cannot be described perturbatively and this failure of perturbation theory is reflected even on the largest scales.  The Effective Field Theory of Large Scale Structure (EFTofLSS) has emerged as a consistent method for describing LSS on large scales by introducing counterterms that account for the effects of small-scale dynamics.  So far studies of the EFT have mostly focused on the two and three point functions with little attention devoted to the four point function or trispectrum.
The trispectrum probes cubic interactions arising from non-linear clustering, biasing, and primordial non-Gaussianities, and constitutes a key element of the covariance matrix of the power spectrum.  In this paper, we present explicit calibrations of the EFT counterterms for the one-loop trispectrum. Specifically, we find clear evidence for non-zero EFT corrections.  We define two one-parameter ans\"{a}tze for the counterterm of the one-loop propagator and show that they provide a good correction to the residual at scales below $k\sim 0.07 \ihMpc$.  We then take the amplitudes of the linear and quadratic counterkernels calculated in our previous paper on the bispectrum and use them in the remaining counterterms, establishing consistency of the counterterms in the two, three and four point function.  We also show that the commonly used EdS approximation for the growth of the density fields leads to errors that are of the same magnitude as loop corrections to the trispectrum on large scales.
\end{abstract}
\maketitle

\tableofcontents


\section{Introduction}
The Cosmic Microwave Background (CMB) has provided us with detailed information on the history and composition of the Universe \cite{Aghanim:2018eyx,Akrami:2018vks}. While the analysis of the mostly linear primary CMB fluctuations is fairly straightforward, its information content has now been almost exhausted due to its two-dimensional nature.
The Large-Scale Structure (LSS) of the Universe, the distribution of dark and baryonic matter on scales larger than $\sim 10$ Mpc, is going to be the next major probe in cosmology, with its three-dimensional nature promising orders of magnitude more information than the CMB. The analysis of LSS is complicated by its non-linear nature.
$N$-body simulations can be used to calculate the non-linear clustering statistics for a given cosmological model, but only provide a sparse sampling of parameter space due to their computational cost. The simulations are also affected by numerical errors and usually do not include baryonic effects. 
A more time efficient manner of evaluating theory is to establish an analytical description of non-linear LSS.  Traditionally, the most common such technique was Standard Perturbation Theory (SPT) \cite{Bernardeau:2001qr}.  However, SPT assumes that LSS can be described perturbatively even down to the smallest scales, which is unrealistic due to the highly non-linear effects of gravitational collapse.  Indeed, these effects not only make perturbative descriptions of small scales impossible, but resonate up to larger scales and make any perturbative model that does not explicitly take them into account inherently inaccurate.  This has been addressed with the development of a modified form of SPT known as the Effective Field Theory of Large Scale Structure (EFTofLSS) \cite{Baumann:2010tm,Carrasco:2012cv,Carrasco:2013mua,Pajer:2013jj,Mercolli:2013bsa}.

As the seeds of cosmological structures arise from stochastic processes, the main objects of study in LSS are statistics, specifically the Fourier transforms of the two, three, and four point correlation functions of the overdensity fields $\delta$: the power spectrum, 
\begin{equation}
\langle \delta_\text{A}(\mathbf{k}_{1})\delta_\text{B}(\mathbf{k}_{2})\rangle = (2\pi)^{3}\delta^{(\mathrm{D})}(\mathbf{k}_{1}+\mathbf{k}_{2})P_\text{AB}(k_{1})~,
\label{1}
\end{equation}
bispectrum, 
\begin{equation}
\langle \delta_\text{A}(\mathbf{k}_{1})\delta_\text{B}(\mathbf{k}_{2})\delta_\text{C}(\mathbf{k}_{3})\rangle = (2\pi)^{3}\delta^{(\mathrm{D})}(\mathbf{k}_{1}+\mathbf{k}_{2}+\mathbf{k}_{3})B_\text{ABC}(k_{1},k_{2},k_{3})~,
\end{equation}
and trispectrum, 
\begin{equation}
\langle \delta_\text{A}(\mathbf{k}_{1})\delta_\text{B}(\mathbf{k}_{2})\delta_\text{C}(\mathbf{k}_{3})\delta_\text{D}(\mathbf{k}_{4})\rangle = (2\pi)^{3}\delta^{(\mathrm{D})}(\mathbf{k}_{1}+\mathbf{k}_{2}+\mathbf{k}_{3}+\mathbf{k}_{4})T_\text{ABCD}(k_{1},k_{2},k_{3},k_{4},k_{5},k_{6})~,
\label{2}
\end{equation}
where $k_5=|\vec k_1+\vec k_2|$ and $k_6=|\vec k_2+\vec k_3|$.

The power spectrum is the principle object of study in LSS perturbation theory and has been widely investigated \cite{Carrasco:2012cv,Carrasco:2013mua,Baldauf:2015aha}, while less attention has been paid to higher order correlators such as the bispectrum \cite{Angulo:2014tfa,Baldauf:2014qfa,Lazanu:2018yae,Steele:2020tak} and trispectrum \cite{PhysRevD.71.063001,Bertolini:2015fya,Bertolini:2016bmt,Gualdi:2020eag}.  However, these higher order correlators provide the opportunity to break parameter degeneracies that may arise when using only the power spectrum to place constraints on fundamental parameters. 

With four density fields one may also construct the covariance matrix of the power spectra between two momentum shells centred at $\mathbf{k}_{i}$ and $\mathbf{k}_{j}$, 
\begin{equation}
C_{\text{ABCD},ij}=\langle \delta_\text{A}(\bk_{i})\delta_\text{B}(\bk_{i})\delta_\text{C}(\bk_{j})\delta_\text{D}(\bk_{j})\rangle-\langle \delta_\text{A}(\bk_{i})\delta_\text{B}(\bk_{i})\rangle\langle \delta_\text{C}(\bk_{j})\delta_\text{D}(\bk_{j})\rangle.
\end{equation} 
This can be decomposed into Gaussian and non-Gaussian components: $C_{\text{ABCD},ij}=C_{\text{ABCD},ij}^{\mathrm{G}}+C_{\text{ABCD},ij}^{\mathrm{NG}}$, which can be defined in terms of the power spectrum
\begin{equation}
C_{\text{ABCD},ij}^{\mathrm{G}}=\frac{2}{V}\frac{(2\pi)^{3}}{V_{k}}\delta^{(\mathrm{K})}_{ij}P_\text{AB}(k_{i})P_\text{CD}(k_{j})
\end{equation}
 and connected trispectrum 
\begin{equation}
C_{\text{ABCD},ij}^{\mathrm{NG}}=\frac{1}{V}\bar{T}_\text{ABCD}({\mathbf{k}}_{i},-{\mathbf{k}}_{i},{\mathbf{k}}_{j},-{\mathbf{k}}_{j})~,
\end{equation}
where $V$ is the overall volume of the space being surveyed.  We have divided Fourier space into measurement shells of volume $V_{k}$ and $\bar{T}$ is the trispectrum averaged over the volume of a chosen such shell \cite{Bertolini:2016bmt}.

From a practical point of view, modelling the trispectrum is essential for a complete calculation of the covariance matrix describing the correlations between power spectra.  Besides its direct relevance for understanding the covariance matrix, the trispectrum also provides interesting insights into cubic counterterms in the EFT itself as well as cubic bias terms in the EFT for biased tracers \cite{PhysRevD.71.063001}. Like the bispectrum, the trispectrum can thus be used to break parameter the degeneracies present in the power spectrum. This information can also be extracted with simplified ``cubic field'' estimators \cite{Abidi:2018eyd}, but to obtain insights on the full scale dependence, we decide to study the full trispectrum.  Trispectrum estimators have been presented in \cite{PhysRevD.71.063001,Tomlinson:2019bjx,Gualdi:2020eag} but have not been used to constrain the EFT counterterms. The trispectrum is also the natural statistic to study cubic primordial non-Gaussianities \cite{Regan:2010cn,Fergusson:2010gn,Chen:2009zp}
such as for instance the cubic local $g_\text{NL}$ model
\cite{Desjacques:2009jb,Roth:2012yy}.  

\begin{figure}[t]
\centering
\includegraphics[width=0.49\textwidth]{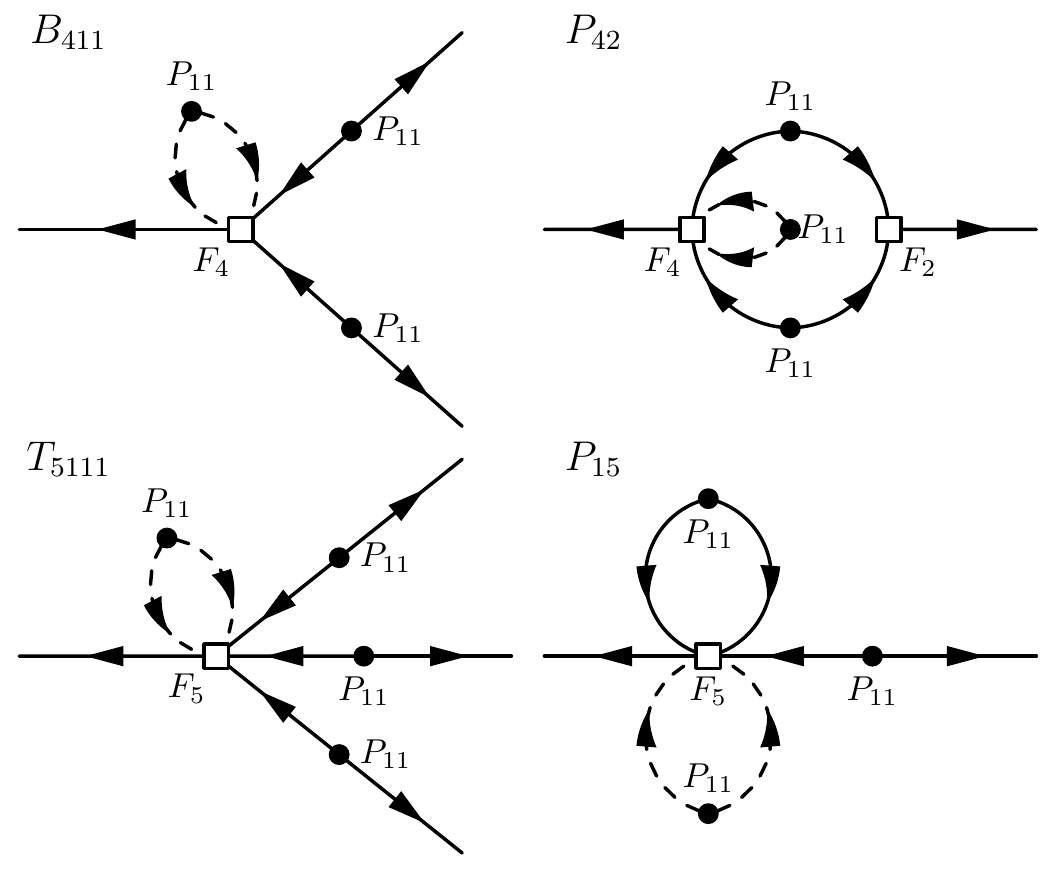}
\caption{Feynman diagram representations of one-loop bispectrum and trispectrum, showing their connection to the two-loop power spectrum. Hard loops for which $q/k\to \infty$ are shown as dashed lines.  We refer to these loops connected to only one kernel as daisy diagrams or free loops.  The bispectrum propagator term $B_{411}$ is connected to the $P_{42}$ contribution in the two-loop power spectrum. When two of the external legs in $T_{5111}$ are connected to each other to form a soft loop, it can be connected to $P_{15}$.}
\label{fig:onetwoloops}
\end{figure}

The cubic gravitational interactions probed by the trispectrum also provide connections with higher loops in the bispectrum and power spectrum.
Fig.~\ref{fig:onetwoloops} shows the connection between the leading UV-sensitive diagrams in the one-loop bispectrum and trispectrum and how they are connected to UV-sensitive contributions in the two-loop power spectrum. To highlight this connection we will consider the loop momentum in $B_{411}$ and $T_{5111}$ to be hard, i.e. $q_{1}\gg k$.  This hard loop which only contains one single $F$ kernel is highlighted by a dashed line and referred to as a daisy diagram or free loop.
When the two external legs in $B_{411}$ are then connected by an $F_2$ kernel to form a soft loop (loop momentum $q_{2}\sim k$), they contribute to the single-hard limit of $P_{42}$. Likewise, if two of the external legs in $T_{5111}$ are connected with each other to form a soft loop, we obtain the single-hard limit of $P_{51}$. In \cite{Baldauf:2015aha} these single-hard diagrams were identified as the relevant templates for the two-loop power spectrum counterterms.

A previous study of the EFTofLSS trispectrum which focused on the covariance of the power spectrum found inconsistent results when comparing analytic theory to two different simulation suites \cite{Bertolini:2015fya}. While one of the simulation suites showed evidence for EFT corrections the other one did not. This apparently contradictory result is probably caused by incomplete convergence of the estimated covariance matrices. Measurements of the power spectrum covariance require large suites of simulations and even then convergence is slow.  Here we aim to improve upon these results and resolve these inconsistencies such that future covariance matrix calculations can be used for data analysis and putting accurate constraints on fundamental physical parameters.  We hypothesise that these systematics may be avoided if one were to explicitly study the trispectrum rather than the covariance of the power spectrum, particularly when taking into account the various lessons learned relating to the study of higher order correlators in \cite{Steele:2020tak}.  

The recent publication \cite{Gualdi:2020eag} made predictions regarding the trispectrum's potential to improve constraints on primordial non-Gaussianity. Averaging over 5000 simulation realisations and comparing their results to perturbation theory, they showed that they had obtained good measurements of the trispectra on the grid.  Constructing a covariance matrix from power spectra, bispectra, and their newly measured trispectra, they quantified the corrections made by the inclusion of trispectra and the extent to which these can be used to provide improved constraints on primordial non-Gaussianity. They found a significant improvement in their constraining potential and that the trispectrum was the only correlator able to offer meaningful constraints at $k>0.1\ihMpc$ at $z=0.5$. However, they restricted the perturbative modelling to tree-level, placing no constraints on EFT parameters.

In our recent paper \cite{Steele:2020tak} we studied the one-loop bispectrum using the technique of realisation perturbation theory (referred to as gridPT by \cite{Taruya:2018jtk,Taruya:2020qoy}, a name we shall adopt hereafter). This technique had previously been used to assess SPT \cite{Roth:2011test,Taruya:2018jtk,Taruya:2020qoy} but until recently has not been used to constrain the EFT counterterms.  Using this technique, we are able to provide precise calibrations of the one-loop counterterms with a minimal number of numerical realisations. We also tested two commonly used approximations: the use of simpler EdS growth factors instead of the correct $\Lambda$CDM growth factors and the notion that the UV-limits of one-loop terms can be taken as approximations for the corresponding counterterms.  We showed that neither of these approximations was sufficiently accurate for studies of loop corrections to the bispectrum.  Here we expand upon that research, showing that the EdS approximation is insufficiently accurate for studies of loop corrections to the trispectrum and applying gridPT to the study of the trispectrum and obtaining one of the first constraints of the one-loop trispectrum counterterms.   

This paper is structured as follows: we begin by introducing SPT and the EFTofLSS before moving on to more detailed discussions of the power spectrum, bispectrum, and trispectrum, and their respective counterterms.  We discuss how they can be measured and how their counterterms can be connected.  We then move on to discuss how we have regularised the one-loop trispectrum, showing that the results from our bispectrum paper can be applied to the regularisation of three of the four relevant counterterms while the fourth requires a new parametrisation to regularise.  Finally, we summarise and discuss our results and note on possible directions for future research.

\subsection{Standard Perturbation Theory}
\label{SPTSec}
Cosmological perturbation theory is based on treating dark matter as a collection of collisionless point particles which interact only gravitationally \cite{Bernardeau:2001qr}, the one-particle phase space density of which is defined as 
\begin{equation}
f_{n}(\mathbf{x},\mathbf{p})=\delta^{(\mathrm{D})}(\mathbf{x}-\mathbf{x}_{n})\delta^{(\mathrm{D})}(\mathbf{p}-ma\mathbf{v}_{n})~,
\end{equation}
such that $f_{n}(\mathbf{x},\mathbf{p})\mathrm{d}^{3}\vec x\, \mathrm{d}^{3}\vec p$ is the probability of finding the particle $n$ in the infinitesimal phase space volume defined by $\mathrm{d}^{3}\vec x\, \mathrm{d}^{3}\vec p$.  The distribution of such a particle obeys the Boltzmann equation
\begin{equation}
\frac{Df_{n}}{Dt}=\frac{\partial f_{n}}{\partial t}+\frac{\mathbf{p}}{ma^{2}}\cdot\frac{\partial f_{n}}{\partial \mathbf{x}}-m\sum_{\bar{n}\neq n}\frac{\partial \phi_{\bar{n}}}{\partial \mathbf{x}}\cdot\frac{\partial f_{n}}{\partial \mathbf{p}}=0~,
\label{BE1}
\end{equation}
where $\phi_{n}$ is the Newtonian gravitational potential for a single particle.  This Newtonian approximation is valid on large scales as relativistic effects only become noticeable on scale below that of gravitational collapse, which are out of range for perturbation theory.

For a given number of particles, the total phase space is defined as
\begin{equation}
f(\mathbf{x},\mathbf{p})=\sum_{n} f_{n}(\mathbf{x},\mathbf{p})~,
\end{equation}
such that $f(\mathbf{x},\mathbf{p})\mathrm{d}^{3}\vec x\, \mathrm{d}^{3}\vec p$ is the probability of finding any particle in the phase space volume $\mathrm{d}^{3}\vec x\, \mathrm{d}^{3}\vec p$.  Eq.~\eqref{BE1} can be summed over to give the Boltzmann equation for such an ensemble of particles
\begin{equation}
\frac{Df}{Dt}=\frac{\partial f}{\partial t}+\frac{\mathbf{p}}{ma^{2}}\cdot\frac{\partial f}{\partial \mathbf{x}}-m\sum_{n,\bar{n},\bar{n}\neq n}\frac{\partial \phi_{\bar{n}}}{\partial \mathbf{x}}\cdot\frac{\partial f_{n}}{\partial \mathbf{p}}=0~,
\label{BE}
\end{equation}

We can define the first three moments of the phase space distribution function as
\begin{equation}
\rho(\bx,a)=\frac{m}{a^{3}}\int \mathrm{d}^{3}\vec p \, f(\bx,\mathbf{p},a)~,
\end{equation}
\begin{equation}
\pi^{i}(\bx,a)=\frac{1}{a^{4}}\int \mathrm{d}^{3}\vec p \,f(\bx,\mathbf{p},a)p^{i}~,
\end{equation}
and
\begin{equation}
\sigma^{ij}(\bx,a)=\frac{1}{m}\int \mathrm{d}^{3}\vec p\, f(\bx,\mathbf{p},a)p^{i}p^{j}-\frac{\pi^{i}\pi^{j}}{\rho}~,
\end{equation}
where $\rho(\bx,a)$ is the comoving mass density, $\pi^{i}(\bx,a)$ is the comoving momentum density, and $\sigma^{ij}(\bx,a)$ is the comoving velocity dispersion tensor which vanishes in our universe due to homogeneity on large scales.  Setting $\sigma^{ij}$ to zero and taking the first three moments of Eq.~\eqref{BE} allows one to find the Eulerian equations of motion for a perfect fluid:
\begin{equation}
\partial_{\eta}\delta+\nabla \cdot \left[\left(1+\delta\right)v^{i}\right]=0~,
\label{eq1}
\end{equation}
\begin{equation}
\partial_{\eta} v^{i}+\mathcal{H} v^{i}_{l}+\partial^{i}\phi+v^{i}_{l}\partial_{j}v^{i}=-\frac{1}{a\rho}\partial_{j}\tau^{ij}~,
\label{eq3}
\end{equation}
\begin{equation}
\triangle \phi=\frac{3}{2}\mathcal{H}^{2}\Omega_{\mathrm{m}}\delta~,
\end{equation}
where $\eta$ is the conformal time defined by $a\derd \eta=\derd t$, $\phi$ is the Newtonian gravitational potential, $\mathcal{H}=aH$ is the conformal Hubble parameter, $\tau^{ij}$ is the stress-energy tensor, which contains terms which account for the effects of small scale gravitational collapse and in SPT is taken to vanish, $\delta=\rho / \bar{\rho}-1$ is a perturbation in the density field with background $\bar{\rho}$, and $\mathbf{v}$ is the velocity field of the fluid defined such that $v^{i}=\pi^{i}/\rho$ \cite{Mercolli:2013bsa}.  One may define the velocity divergence field as $\theta=\partial_{i}v^{i}$ and the vorticity of the velocity field as $\omega^{i}=\epsilon^{ijk}\partial_{j}v_{k}$ such that
\begin{equation}
v^{i}=\frac{\partial^{i}}{\partial^{2}}\theta-\epsilon^{ijk}\frac{\partial_{j}}{\partial^{2}}\omega_{k}~.
\end{equation}
Observations indicate that our universe has negligible vorticity on large scales, so we can set $\omega^{i}=0$ and write the Euler equation in terms of the velocity divergence field as
\begin{equation}
\partial_{\eta}\theta+\mathcal{H}\theta+v^{j}\partial_{j}\theta+\partial_{i}v^{j}\partial_{j}v^{i}+\triangle\phi=\tau_{\theta}~,
\label{eq2}
\end{equation}
where
\begin{equation}
\tau_{\theta}=-\partial_{i}\left[\frac{1}{\rho}\partial_{j}\tau^{ij}\right]
\end{equation}
vanishes in SPT.  On large scales Fourier modes evolve mostly independently of one another, preserving primordial statistics.  This is what makes large scale structure such an effective way of probing the non-Gaussianities of inflation and also makes it convenient to Fourier transform Eqs.~(\ref{eq1}$-$\ref{eq2}) as
\begin{equation}
\partial_{\eta}\delta(\bk,a)+\theta(\bk,a)=\mathcal{S}_{\alpha}(\bk,a),
\label{eom1}
\end{equation}
and
\begin{equation}
\partial_{\eta}\theta(\bk,a)+\mathcal{H}\theta(\bk,a)+\frac{3}{2}\Omega_{\mathrm{m}}\mathcal{H}^{2}\delta(\bk,a)=\mathcal{S}_{\beta}(\bk,a)~,
\label{eom2}
\end{equation}
where the two non-linear source terms $S_{\alpha}$ and $S_{\beta}$ may be defined as
\begin{equation}
S_{\alpha}(\bk,a)=-\int \frac{\mathrm{d}^{3}\mathbf{q}}{(2\pi)^{3}}\alpha(\mathbf{q},\mathbf{k}-\mathbf{q})\theta(\mathbf{q},\tau)\delta(\mathbf{k}-\mathbf{q},\tau)~,
\end{equation}
\begin{equation}
S_{\beta}(\bk,a)=-\int \frac{\mathrm{d}^{3}\mathbf{q}}{(2\pi)^{3}}\beta(\mathbf{q},\mathbf{k}-\mathbf{q})\theta(\mathbf{q},\tau)\theta(\mathbf{k}-\mathbf{q},\tau)+\tau_{\theta}(\bk,a)~.
\label{eq:Sbeta}
\end{equation}
The kernels $\alpha$ and $\beta$, defined as
\begin{equation}
\alpha(\mathbf{k}_{1},\mathbf{k}_{2})=\frac{\mathbf{k}_{1}\cdot (\mathbf{k}_{1}+\mathbf{k}_{2})}{k_{1}^{2}}
\end{equation}
and
\begin{equation}
\beta(\mathbf{k}_{1},\mathbf{k}_{2})=\frac{1}{2}(\mathbf{k}_{1}+\mathbf{k}_{2})^{2}\frac{\mathbf{k}_{1}\cdot\mathbf{k}_{2}}{k_{1}^{2}k_{2}^{2}}~,
\end{equation}
encapsulate the coupling between modes.
Assuming an Einstein-de-Sitter universe, Eqs.~\eqref{eom1} and \eqref{eom2} can be solved by a power series ansatz in the density and velocity fields:
\begin{equation}
\delta(\bk,a)=\sum^{\infty}_{n=1}a^{n}(t)\delta_{n}(\mathbf{k})~,
\label{pl1}
\end{equation}
\begin{equation}
\theta(\bk,a)=-\mathcal{H}\sum^{\infty}_{n=1}a^{n}(t)\theta_{n}(\mathbf{k})~.
\label{pl2}
\end{equation}
as in an EdS universe the linear density field evolves linearly with the cosmological scale factor $a$ and the $n$th-order perturbations evolve as $a^{n}$.  These can be generalised to other cosmological models such as $\Lambda$CDM by replacing the scale factor with an appropriate growth factor $D^{n}$ as we described in detail in \cite{Steele:2020tak}.

Inserting Eqs.~\eqref{pl1} and \eqref{pl2} into Eqs.~\eqref{eom1} and \eqref{eom2} shows that at a given time, the momentum dependence of the two fields can be given by convolutions in terms of the linear density field:
\begin{equation}
\delta_{n}(\mathbf{k})=\int_{\mathbf{q}_{1}}...\int_{\mathbf{q}_{n}}(2\pi)^{3}\delta^{(\mathrm{D})}(\mathbf{k}-\mathbf{q}_{1}...-\mathbf{q}_{n})F_{n}(\mathbf{q}_{1},...,\mathbf{q}_{n})\delta_{1}(\mathbf{q}_{1})...\delta_{1}(\mathbf{q}_{n})~,
\label{eq:deltan}
\end{equation}
\begin{equation}
\theta_{n}(\mathbf{k})=\int_{\mathbf{q}_{1}}...\int_{\mathbf{q}_{n}}(2\pi)^{3}\delta^{(\mathrm{D})}(\mathbf{k}-\mathbf{q}_{1}...-\mathbf{q}_{n})G_{n}(\mathbf{q}_{1},...,\mathbf{q}_{n})\delta_{1}(\mathbf{q}_{1})...\delta_{1}(\mathbf{q}_{n})~,
\label{rec2}
\end{equation}
where $\int_{\mathbf{q}_{n}}=\int_{0}^{\infty} \mathrm{d}^{3}\mathbf{q}_{n}/(2\pi)^{3}$, for kernels $F$ and $G$ given by
\begin{align}
F_{\mathrm{n}}(\bk_{1},...,\bk_{\mathrm{n}})=&\sum^{n-1}_{m=1}\frac{G_{m}(\bk_{1},...,\bk_{m})}{(2n+3)(n-1)}[(2n+1)\alpha(\bkappa_{1}^m,\bkappa_{m+1}^n)F_{\mathrm{n}-m}(\bk_{m+1},...,\bk_{\mathrm{n}})\nonumber\\&+2\beta(\bkappa_{1}^m,\bkappa_{m+1}^n)G_{\mathrm{n}-m}(\bk_{m+1},...,\bk_{\mathrm{n}})]~,\label{eq:recursf}\\
G_{\mathrm{n}}(\bk_{1},...,\bk_{\mathrm{n}})=&\sum^{n-1}_{m=1}\frac{G_{m}(\bk_{1},...,\bk_{m})}{(2n+3)(n-1)}[3\alpha(\bkappa_{1}^m,\bkappa_{m+1}^n)F_{\mathrm{n}-m}(\bk_{m+1},...,\bk_{\mathrm{n}})\nonumber\\&+2n\beta(\bkappa_{1}^m,\bkappa_{m+1}^n)G_{\mathrm{n}-m}(\bk_{m+1},...,\bk_{\mathrm{n}})]~\label{eq:recursg},
\end{align}
where $\kappa_{a}^b=\sum_{i=a}^b k_i$
and $F_{1}=G_{1}=1$.  These kernels are generally used in their symmetrised forms, $F^{(s)}$ and $G^{(s)}$, which can be obtained by summing the kernels over all possible permutations of their variables.  

\subsection{The Effective Fluid Approach}
\label{EFC}
As with SPT, the EFTofLSS \cite{Carrasco:2012cv,Baumann:2010tm,Pajer:2013jj,Mercolli:2013bsa} approach is based on treating dark matter as a fluid whose behaviour can be described by a Boltzmann equation
\begin{equation}
\frac{Df}{Dt}=\frac{\partial f}{\partial t}+\frac{\mathbf{p}}{ma^{2}}\cdot \frac{\partial f}{\partial \mathbf{x}}-m\sum_{n,\bar{n};\bar{n}\neq n}\frac{\partial \phi_{\bar{n}}}{\partial \mathbf{x}}\cdot \frac{\partial f_{n}}{\partial \mathbf{p}}=0~.
\end{equation}

Because the perturbative approach only works at low momenta, a cutoff must be introduced to turn the dark matter distribution into a low-energy effective fluid.  This cutoff takes the form of a smoothing function,
\begin{equation}
W_{\Lambda}(k)=e^{-\frac{1}{2}\frac{k^{2}}{\Lambda^{2}}}~,
\end{equation}
whose purpose is to smooth out quantities with a wavenumber $k\geq \Lambda$.  Incorporating this cutoff into the Boltzmann equation, we end up with the smoothed Boltzmann equation
\begin{equation}
\frac{Df}{Dt}=\frac{\partial f}{\partial t}+\frac{\mathbf{p}}{ma^{2}}\cdot \frac{\partial f}{\partial \mathbf{x}}-m\sum_{n,\bar{n};\bar{n}\neq n}\int \mathrm{d}^{3}x' W_{\Lambda}(\mathbf{x}-\mathbf{x}')\frac{\partial \phi_{\bar{n}}}{\partial \mathbf{x}}\cdot \frac{\partial f_{n}}{\partial \mathbf{p}}=0~,
\label{EFTBoltz}
\end{equation}
from which SPT is recovered when $\Lambda\rightarrow\infty$.  For practical reasons described in Sec.~\ref{sec:simgrid} we choose to use the cutoff $\Lambda\approx 0.3 \ihMpc$. Taking only the first two moments of Eq.~\eqref{EFTBoltz} gives us the equations of motion
\begin{equation}
\partial_{\eta}\rho+3\mathcal{H}\rho+\partial_{i}(\rho v^{i})=0
\end{equation}
and
\begin{equation}
\partial_{\eta}v^{i}+\mathcal{H}v^{i}+v^{j}\partial_{j}v^{i}+\partial_{i}\phi=-\frac{1}{\rho}\partial_{j}[\tau^{ij}]_{\Lambda},
\label{tau}
\end{equation}
which are the same as Eqs.~\eqref{eq1} and \eqref{eq3} but for the incorporation of the non-zero stress-energy tensor $\tau^{ij}$, which accounts for the non-perturbative effects beyond the chosen cutoff.  It is ultimately this stress tensor that we study in order to regularise the effective fluid and come up with a complete description of large scale structure beyond our chosen cutoff.  We do this by deriving analytic forms for the counterterms and fitting these to data from simulations.

All correlators featuring density fields of higher order than $\delta_{1}$ are cutoff dependent.  For the remainder of this paper, we do not explicitly include the cutoffs in the correlators' arguments unless it is required to illustrate a specific point.

\subsubsection{The Stress Tensor}
The explicit form of $\tau^{ij}$ can be found by summing all terms which are compatible with the symmetries of the model and finding appropriate parameters from simulations by which to multiply them \cite{Baumann:2010tm,Carrasco:2012cv}.  In this case, we must allow for Galilean invariance, homogeneity, isotropy, conservation of momentum, and conservation of mass.  The second derivatives of the gravitational potential,  $\partial_{i}\partial_{j}\phi$ are compatible with all of these symmetries and are therefore our natural building blocks.  Constructing all possible combinations gives
\begin{equation}
\begin{split}
\tau_{ij}=&c_{1}^{1}\delta_{ij}^{(\text{K})}\partial^{2}\phi+c_{2}^{1}\partial_{i}\partial_{j}\phi\\&+c_{1}^{2}\delta_{ij}^{(\text{K})}(\partial^{2}\phi)^{2}+c_{2}^{2}\partial^{2}\phi\partial_{i}\partial_{j}\phi+c_{3}^{2}(\partial_{i}\partial_{j}\phi)^{2}+c_{4}^{2}\partial_{i}\partial_{j}\phi\partial_{k}\partial_{l}\phi\\&+c_{1}^{3}\delta_{ij}^{(\text{K})}(\partial^{2}\phi)^{3}+c_{2}^{3}(\partial^{2}\phi)^{2}\partial_{i}\partial_{j}\phi+c_{3}^{3}\partial^{2}\phi(\partial_{i}\partial_{j}\phi)^{2}+c_{4}^{3}\partial^{2}\phi\partial_{i}\partial_{j}\phi\partial_{j}\partial_{k}\phi\\&+c_{5}^{3}(\partial_{i}\partial_{j}\phi)^{2}\partial_{k}\partial_{l}\phi+c_{6}^{3}\partial_{i}\partial_{j}\phi\partial_{j}\partial_{k}\phi\partial_{k}\partial_{l}\phi+c_{7}^{3}\partial_{i}\partial_{j}\phi\partial_{j}\partial_{k}\phi\partial_{k}\partial_{i}\phi+...~,
\end{split}
\end{equation}
for constants $c_{n}^{m}$, where the upper index specifies the number of building blocks attached to the constant and by extension its order in perturbation theory, due to the power series basis of PT.  

The various terms in the stress-energy tensor can be grouped into two categories: viscosity terms, which provide the correlated corrections to the perturbative terms, and subleading noise terms, which account for the self-coupling of non-perturbative (strongly coupled) small scale modes analogous to the one-halo term in the halo model.  In this paper we focus on the viscosity terms and leave the noise terms for future work.

At cubic order, where we regularise $\delta_{3}$, there are two viscosity terms, such that
\begin{equation}
    \tau^{i}_\theta\vert_{1}=-d_{\phi}^{2}\partial^{i}\Delta\phi+\frac{d_{u}^{2}}{\mathcal{H}f}\partial^{i}\Delta u~,
\end{equation}
for free parameters $d_{\phi}^{2}$ and $d_{u}^{2}$ which we combine in the new variable $d^{2}\equiv d_{\phi}^{2}+d_{u}^{2}$.  

At quartic order, where we regularise $\delta_{4}$,  this expands to become
\begin{equation}
\tau^{ij}_\theta\vert_{2}=-d^{2}\delta^{ij}_{(\text{K})}\delta_{2}+c_{1}\delta^{ij}_{(\text{K})}(\Delta\phi)^{2}+c_{2}\partial^{i}\partial^{j}\phi\Delta\phi+c_{3}\partial^{i}\partial_{k}\phi\partial^{j}\partial^{k}\phi~.
\end{equation}
Defining the tidal tensor
\begin{equation}
    s^{ij}\equiv \partial^{i}\partial^{j}\phi - \frac{1}{3}\delta^{ij}_{(\mathrm{K})}\Delta\phi~,
\end{equation}
and $s^2=s_{ij}s^{ij}$ one can obtain 
\begin{align}
    \tau_{\theta}\vert_{1}&=-d^{2}\Delta\delta_{1}~,\\
    \tau_{\theta}\vert_{2}&=-d^{2}\Delta\delta_{2}-e_{1}\Delta\delta^{2}_{1}-e_{2}\Delta s^{2}-e_{3}\partial_{i}\left[s^{ij}\partial_{j}\delta_{1}\right]~,
    \label{eq:tau12}
\end{align}
where $e_{i}$ are functions of $c_{i}$ and $d_{u}^{2}$, leaving us with four free parameters: $d^{2}$, $e_{1}$, $e_{2}$, and $e_{3}$, and the counterterm density fields which regularise our model are given by $\tilde{\delta}_{n}=\tau_{\theta}\vert_{n}$.  

We wish to relate these terms to perturbation theory as defined above.  As such, we can formulate a definition for the present time counterterm density field in terms of kernels and products of the linear field analogously to Eq.~\eqref{eq:deltan} by defining a set of counterkernels $\tilde{F}$, such that
\begin{equation}
        \tilde{\delta}_{\mathrm{n}}(\bk)\equiv \int_{\bq_{1}}...\int _{\bq_{\mathrm{n}}} \deltadir\left(\bk-\sum_{i=1}^{n}\bq_{i}\right)\tilde{F}_{\mathrm{n}}(\bq_{1},...,\bq_{\mathrm{n}})\prod_{i=1}^{n}\delta_{1}(\bq_{i})~.
\end{equation}

The parameters $d^{2}$ and $e_{1,2,3}$ in Eq.~\eqref{eq:tau12} are time dependent.  Splitting them into time dependent and independent components and integrating to the present time, we can generate the new parameters $\epsilon_{1,2,3}$ and $\gamma_{1,2}$, such that we can define the first two present time counter kernels as
\begin{align}
    \label{F1t}
    \tilde{F}_{1}(k)=&-\gamma_{1}k^{2}~,\\
\tilde{F}_{2}(\bk_{1},\bk_{2})=&-\left[\sum_{i=1}^{3}\epsilon_{i}E_{i}(\bk_{1},\bk_{2})+\gamma_{2}\Gamma(\bk_{1},\bk_{2})\right] ,
\label{eq:symmf2tilde}
\end{align}
for momentum dependent kernels $E_{1,2,3}$ and $\Gamma$ (see  \cite{Steele:2020tak} for explicit forms and derivation).  In much of the literature, $\gamma_{1}$ is referred to as the speed of sound, $\cssq$, when calculated from the power spectrum.

\subsubsection{The Power Spectrum}
\label{PSPT}
In the case of a Gaussian field, the ensemble average of an odd number of variables vanishes and that of an even number of variables can be rewritten as a product of ensemble averages of all possible pairs of variables in accordance with Wick's Theorem:
\begin{align}
&\langle \delta(\mathbf{k}_{1})\ldots\delta(\mathbf{k}_{n+1})\rangle=0~,\\
&\langle \delta(\mathbf{k}_{1})\ldots\delta(\mathbf{k}_{n})\rangle=\sum_{\mathrm{pairings}}\prod_{\mathrm{pairs}}\langle \delta(\mathbf{k}_{i})\delta(\mathbf{k}_{j})\rangle~,
\end{align}
for even $n$.  This makes the power spectrum the only spectrum of interest in a truly Gaussian field and the simplest and most common object of study in a field which is only weakly non-Gaussian. As our universe is primarily Gaussian on large scales, the power spectrum provides the majority of our understanding of LSS with the bispectrum and trispectrum providing further statistics to help us understand the non-Gaussianities in the fields and extract the full information content.

\begin{figure}[h]
\centering
\includegraphics[width=16cm]{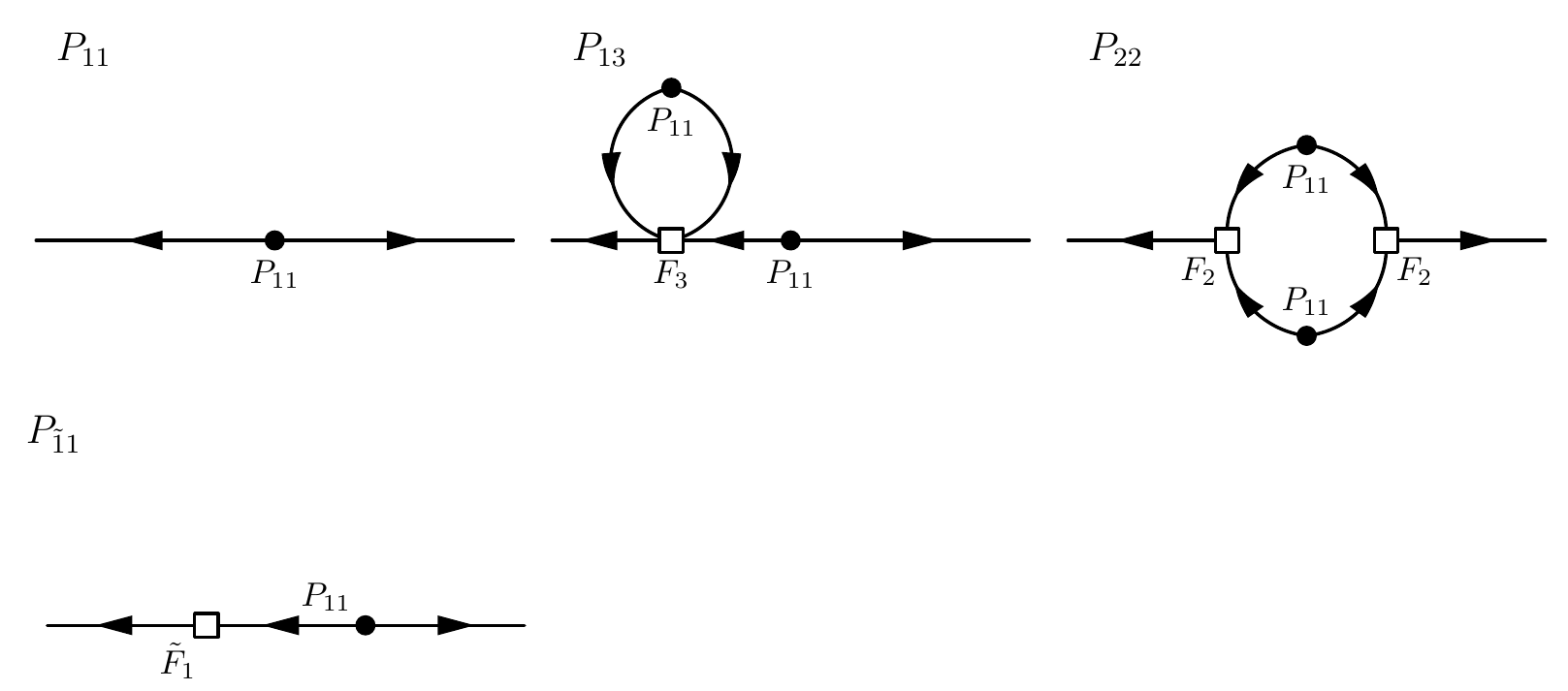}
\caption{Feynman diagram representations of the linear and one-loop contributions to the power spectra of LSS together with the $P_{\tilde{1}1}$ one-loop counterterm.}
\label{FPS}
\end{figure}

There are four contributions to the one-loop power spectrum: the linear power spectrum $P_{11}$, and the three one-loop contributions $P_{13}$, $P_{31}$, and $P_{22}$, where $P_{13}=P_{31}$ and
\begin{align}
P_{31}(k)=&3\int_{\mathbf{q}}~F_{3}^{(\text{s})}(\mathbf{k},\bq,-\bq)P_{11}(k)P_{11}(q)~,\\
P_{22}(k)=&2\int_{\mathbf{q}}~|F_{2}^{(\text{s})}(\mathbf{k}-\mathbf{q},\mathbf{q})|^{2}P_{11}(|\mathbf{k}-\mathbf{q}|)P_{11}(q)~,
\end{align}
all of which are represented diagrammatically in Fig.~\ref{FPS}.  

$P_{22}$ can be regularised by a subleading noise term which makes a contribution substantially smaller than the viscosity term that regularises $P_{31}$ and which will not be addressed here.  For a given cutoff $\Lambda$, $P_{31}$ can be regularised with the counterterm $P_{\tilde{1}1}=\tilde{F}_{1}(k)P_{11}$ where $\tilde{F}_{1}(k)=-c_\text{s}^{2}(\Lambda)k^{2}$, such that
\begin{equation}
P_{\textrm{nn}}(k)=P_{11}(k)+P_{22}(k)+2P_{31}(k)-2c_\text{s}^{2}(\Lambda)k^{2}P_{11}(k)~,
\label{eq:cs}
\end{equation}
for non-linear power spectrum $P_{\mathrm{nn}}$.  In theory $c_\text{s}^{2}$ should be a constant, but studies up to any given order in perturbation theory will inevitably result in estimators of this quantity to be scale dependent due to higher order terms which have not been accounted for.  In  \cite{Baldauf:2015aha}, it was shown that $c_\text{s}^{2}$ becomes roughly constant for all scales of interest to the EFTofLSS when two-loop terms are taken into account.

\subsubsection{The Bispectrum}
As non-Gaussianities appear in LSS, it becomes necessary to look beyond the power spectrum and study higher order correlators.  The tree-level bispectrum is given by
\begin{equation}
B_{211}(k_{1},k_{2},k_{3})=2P_{11}(k_{2})P_{11}(k_{3})F_{2}^{(s)}(\mathbf{k}_{2},\mathbf{k}_{3})+2~\mathrm{perms.}~.
\end{equation}
and the one-loop contributions are

\begin{align}
&B_{222}(k_{1},k_{2},k_{3})=8\int_{\mathbf{q}}P_{11}(q)P_{11}(|\mathbf{k}_{2}-\mathbf{q}|)P_{11}(|\mathbf{k}_{3}-\mathbf{q}|)F_{2}^{(s)}(-\mathbf{q},\mathbf{k}_{3}+\mathbf{q})\nonumber \\&~~~~~~~~~~\times F_{2}^{(s)}(\mathbf{k}_{3}+\mathbf{q},\mathbf{k}_{2}-\mathbf{q})F_{2}^{(s)}(\mathbf{k}_{2}-\mathbf{q},\mathbf{q})~,\\
&B_{321a}(k_{1},k_{2},k_{3})=6P_{11}(k_{3})\int_{\mathbf{q}}P_{11}(q)P_{11}(|\mathbf{k}_{2}+\mathbf{q}|)F_{3}^{(s)}(-\mathbf{q},-\mathbf{k}_{2}+\mathbf{q},-\mathbf{k}_{3})\nonumber \\&~~~~~~~~~~~\times F_{2}^{(s)}(\mathbf{q},\mathbf{k}_{2}-\mathbf{q})	+5~\mathrm{perms.}~,\\
&B_{321b}(k_{1},k_{2},k_{3})=6P_{11}(k_{2})P_{11}(k_{3})F_{2}^{(s)}(\mathbf{k}_{2},\mathbf{k}_{3})\int_{\mathbf{q}}P_{11}(q)F_{3}^{(s)}(\mathbf{k}_{3},\mathbf{q},-\mathbf{q})+5~\mathrm{perms.}~,\\
&B_{411}(k_{1},k_{2},k_{3})=12P_{11}(k_{2})P_{11}(k_{3})\int_{\mathbf{q}}P_{11}(q)F_{4}^{(s)}(\mathbf{q},-\mathbf{q},-\mathbf{k}_{2},-\mathbf{k}_{3})+2~\mathrm{perms.}~,
\end{align}
such that at one-loop order, the bispectrum is given by
\begin{equation}
B_{\mathrm{tree+1-loop}}=B_{211}+B_{222}+B_{321a}+B_{321b}+B_{411}~.
\end{equation}

\begin{figure}[h]
	\centering
	\includegraphics[width=16cm]{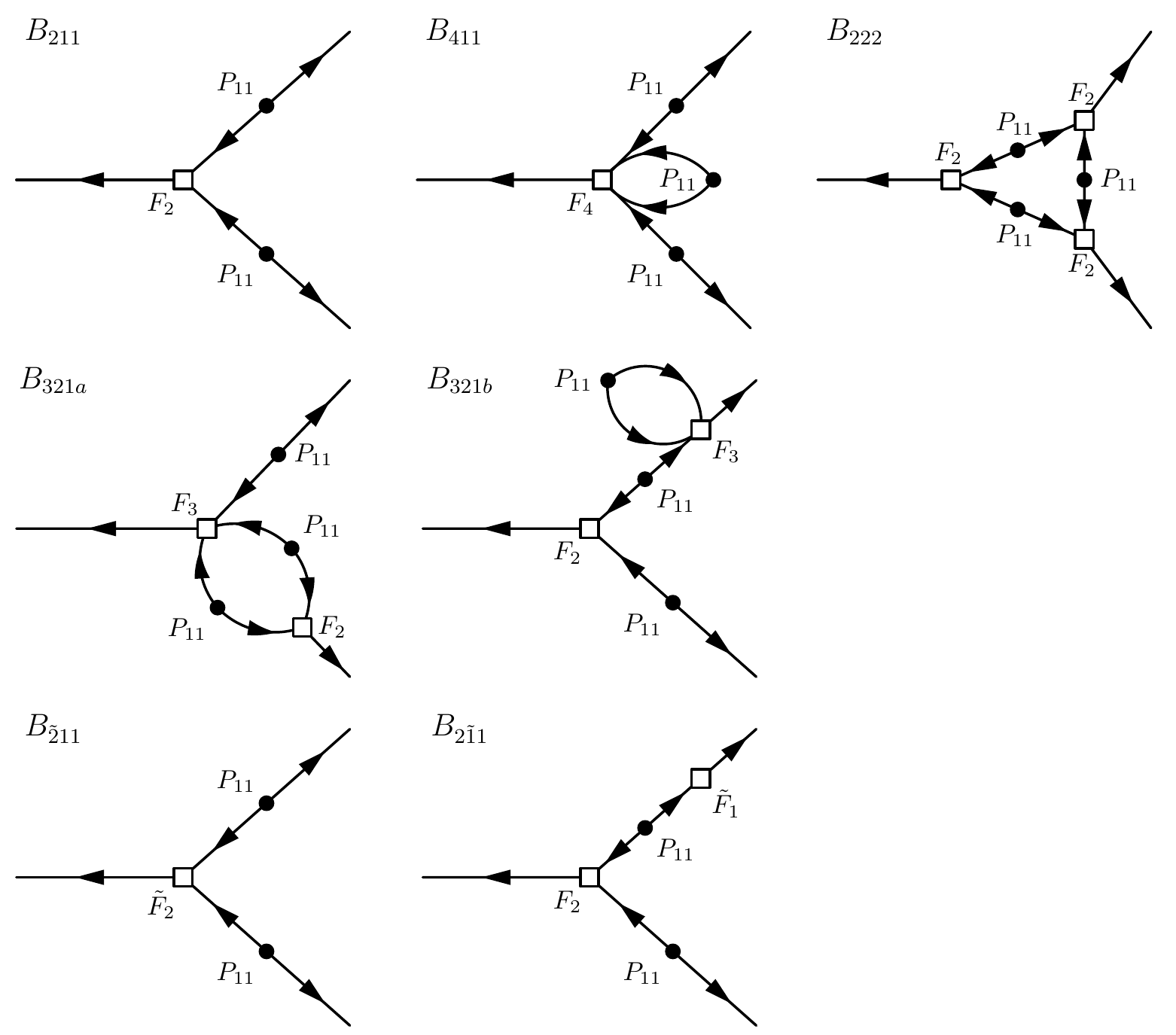}
	\caption{Feynman diagram representations of the tree-level and one-loop contributions to the bispectrum together with the one-loop counterterms.}
	\label{fig:BisFeyn}
\end{figure}

Each loop term needs regularisation but the counterterms for $B_{222}$ and $B_{321a}$ are subleading noise term such that for most purposes we can focus on the remaining two counterterms,
\begin{equation}
B_{\tilde{2}11}(k_{1},k_{2},k_{3})= 2! \tilde{F}_{2}^{(s)}(\bk_{2},\bk_{3})P_{11}(k_{2})P_{11}(k_{3})~\label{B411c},
\end{equation}
which regularises $B_{411}$, and
\begin{equation}
B_{\tilde{1}21}(k_{1},k_{2},k_{3})= 2! \tilde{F}_{1}^{(s)}(-\bk_{1})F_{2}^{(s)}(\bk_{1},\bk_{3})P_{11}(k_{1})P_{11}(k_{3})~,
\end{equation}
which regularises $B_{321b}$.  These counterterms are shown diagrammatically alongside the tree-level and one-loop terms in Fig.~\ref{fig:BisFeyn}. 

In \cite{Steele:2020tak} we employed a variety of methods to constrain $\tilde{F}_{1}$ and $\tilde{F}_{2}$ from simulations using gridPT; in Sec.~\ref{sec:expandedkernels} we will use these constraints for the  regularisation of the corresponding interactions in the trispectra.

\subsubsection{The Trispectrum}
\label{sec:tri}
The trispectrum contains the non-Gaussian information in the covariance matrix of the power spectrum.  As the covariance matrix of the power spectrum is a key component of cosmological parameter inference and the power spectrum is already well understood up to two-loop level, it is important to develop our understanding of the trispectrum in preparation for upcoming surveys.

\begin{figure}[h!]
	\centering
	\includegraphics[width=16cm]{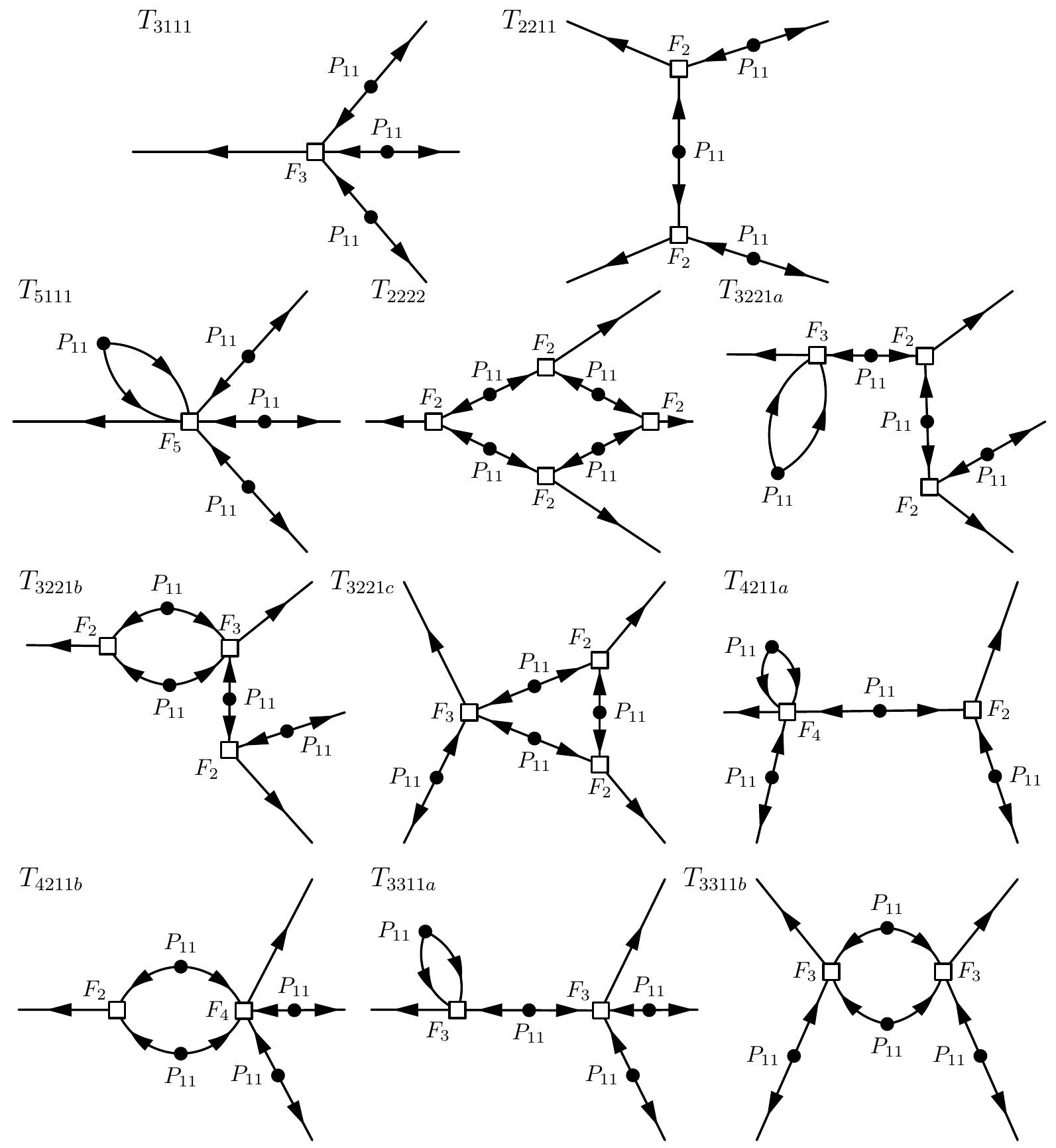}
	\caption{The tree-level (top row) and one-loop (other rows) contributions to the trispectrum.  Note that $T_{5111}$, $T_{3221a}$, $T_{4211a}$ and $T_{3311a}$ contain free loops.  This results in them constituting the leading UV-sensitive terms in the trispectrum. Most of them can be related to UV-sensitive interactions already encountered in the one-loop power spectrum and bispectrum, but $T_{5111}$ constitutes a new UV-sensitive interaction.}
	\label{FT}
\end{figure}

Up to one-loop order the trispectrum is given by
\begin{equation}
\begin{split}
T_{\mathrm{tree+1-loop}}=&~T_{3111}+T_{2211}+T_{5111}+T_{4211a}+T_{4211b}+T_{3221a}+T_{3221b}+T_{3221c}\\& +T_{3311a}+T_{3311b}+T_{2222}~,
\end{split}
\end{equation}
where $T_{3111}$ and $T_{2211}$ are the tree-level contributions.  The contribution are given by
\begin{align}
&T_{3111}(\bk_{1},\bk_{2},\bk_{3},\bk_{4})=3!F_{3}^{(s)}(\mathbf{k}_{2},\mathbf{k}_{3},\mathbf{k}_{4})P_{11}(k_{2})P_{11}(k_{3})P_{11}(k_{4})+3~\mathrm{perms.}~,\label{eq:T3111PT}\\
&T_{2211}(\bk_{1},\bk_{2},\bk_{3},\bk_{4})=(2!)^{2}F_{2}^{(s)}(-\mathbf{k}_{3}-\mathbf{k}_{4},\mathbf{k}_{4})F_{2}^{(s)}(\mathbf{k}_{3}+\mathbf{k}_{4},\mathbf{k}_{2})P_{11}(|\mathbf{k}_{3}+\mathbf{k}_{4}|)P_{11}(k_{2})P_{11}(k_{3})\nonumber\\&~~~~~~~~~~~~~~~~+11~\mathrm{perms.}~,\\
&T_{5111}(\bk_{1},\bk_{2},\bk_{3},\bk_{4})=\frac{5!}{2!}\int_{\mathbf{q}}F_{5}^{(s)}(\mathbf{q},-\mathbf{q},\mathbf{k}_{2},\mathbf{k}_{3},\mathbf{k}_{4})P_{11}(q)P_{11}(k_{2})P_{11}(k_{3})P_{11}(k_{4})+3~\mathrm{perms.}~,\\
&T_{4211a}(\bk_{1},\bk_{2},\bk_{3},\bk_{4})=~4!\int_{\mathbf{q}}	F_{4}^{(s)}(\mathbf{q},-\mathbf{q},\bk_{2}+\bk_{3},\bk_{4})F_{2}^{(s)}(-\bk_{2}-\bk_{3},\bk_{3})\nonumber\\&~~~~~~~~~~~~~~~~\times P_{11}(q)P_{11}(|\bk_{2}+\bk_{3}|)P_{11}(k_{3})P_{11}(k_{4})+23~\mathrm{perms.}~,\\
&T_{4211b}(\bk_{1},\bk_{2},\bk_{3},\bk_{4})=4!\intq F_{4}^{(s)}(\bk_{1}+\mathbf{q},-\mathbf{q},\bk_{3},\bk_{4})F_{2}^{(s)}(-\mathbf{q}-\bk_{1},\mathbf{q})P_{11}(|	\mathbf{q}+\bk_{1}|)\nonumber\\&~~~~~~~~~~~~~~~~\times P_{11}(q)P_{11}(k_{3})P_{11}(k_{4})+11~\mathrm{perms.}~,\\
&T_{3221a}(\bk_{1},\bk_{2},\bk_{3},\bk_{4})=3!2!\intq F_{3}^{(s)}(\mathbf{q},-\mathbf{q},-\bk_{1})F_{2}^{(s)}(\bk_{1},\bk_{3}+\bk_{4})F_{2}^{(s)}(-\bk_{3}-\bk_{4},\bk_{4})\nonumber\\&~~~~~~~~~~~~~~~~\times P_{11}(q)P_{11}(k_{1})P_{11}(|\bk_{3}+\bk_{4}|)P_{11}(k_{4})+23~\mathrm{perms.}~,\\
&T_{3221b}(\bk_{1},\bk_{2},\bk_{3},\bk_{4})=3!2!\intq F_{3}^{(s)}(\mathbf{q}+\bk_{1},-\mathbf{q},\bk_{3}+\bk_{4})F_{2}^{(s)}(-\bk_{1}-\mathbf{q},\mathbf{q})F_{2}^{(s)}(-\bk_{3}-\bk_{4},\bk_{4})\nonumber\\&~~~~~~~~~~~~~~~~\times P_{11}(q)P_{11}(|\bk_{1}+\mathbf{q}|)P_{11}(|\bk_{3}+\bk_{4}|)P_{11}(k_{4})+23~\mathrm{perms.}~,\\
&T_{3221c}(\bk_{1},\bk_{2},\bk_{3},\bk_{4})=3!(2!)^{2}\intq F_{3}^{(s)}(\mathbf{q},-\bk_{1}-\bk_{4}-\mathbf{q},\bk_{4})F_{2}^{(s)}(-\mathbf{q},\mathbf{q}-\bk_{2})F_{2}^{(s)}(\bk_{2}-\mathbf{q},\bk_{1}+\bk_{4}+\mathbf{q})\nonumber\\&~~~~~~~~~~~~~~~~\times P_{11}(q)P_{11}(|\bk_{1}+\bk_{4}+\mathbf{q}|)P_{11}(|\mathbf{q}-\bk_{2}|)P_{11}(k_{4})+11~\mathrm{perms.}~,\\
&T_{3311a}(\bk_{1},\bk_{2},\bk_{3},\bk_{4})=\frac{(3!)^{2}}{2!}\intq F_{3}^{(s)}(\mathbf{q},-\mathbf{q},-\bk_{1})F_{3}^{(s)}(\bk_{1},\bk_{3},\bk_{4})P_{11}(q)\nonumber\\&~~~~~~~~~~~~~~~~\times P_{11}(k_{2})P_{11}(k_{3})P_{11}(k_{4})+11~\mathrm{perms.}~,\\
&T_{3311b}(\bk_{1},\bk_{2},\bk_{3},\bk_{4})=\frac{(3!)^{2}}{2!}\intq F_{3}^{(s)}(\mathbf{q},-\mathbf{q}-\bk_{1}-\bk_{4},\bk_{4})F_{3}^{(s)}(-\mathbf{q},\mathbf{q}+\bk_{1}+\bk_{4},\bk_{3})\nonumber\\&~~~~~~~~~~~~~~~~\times P_{11}(q)P_{11}(|\mathbf{q}+\bk_{1}+\bk_{4}|)P_{11}(k_{3})P_{11}(k_{4})+11~\mathrm{perms.}~,\\
&T_{2222}(\bk_{1},\bk_{2},\bk_{3},\bk_{4})=(2!)^{4}\intq F_{2}^{(s)}(\mathbf{q},-\bk_{1}-\mathbf{q})F_{2}^{(s)}(-\mathbf{q},\mathbf{q}-\bk_{2})F_{2}^{(s)}(\bk_{2}-\mathbf{q},\mathbf{q}-\bk_{2}-\bk_{3})\nonumber\\&~~~~~~~~~~~~~~~~\times F_{2}^{(s)}(\bk_{2}+\bk_{3}-\mathbf{q},\mathbf{q}+\bk_{1}) P_{11}(q)P_{11}(|\mathbf{q}-\bk_{2}|)P_{11}(|\mathbf{q}-\bk_{2}-\bk_{3}|)\nonumber\\&~~~~~~~~~~~~~~~~\times P_{11}(|\mathbf{q}+\bk_{1}|)+2~\mathrm{perms.}~,\label{eq:TPT}
\end{align}
each of which is shown diagrammatically in Fig.~\ref{FT}.  

\begin{figure}[h!]
	\centering
	\includegraphics[width=14cm]{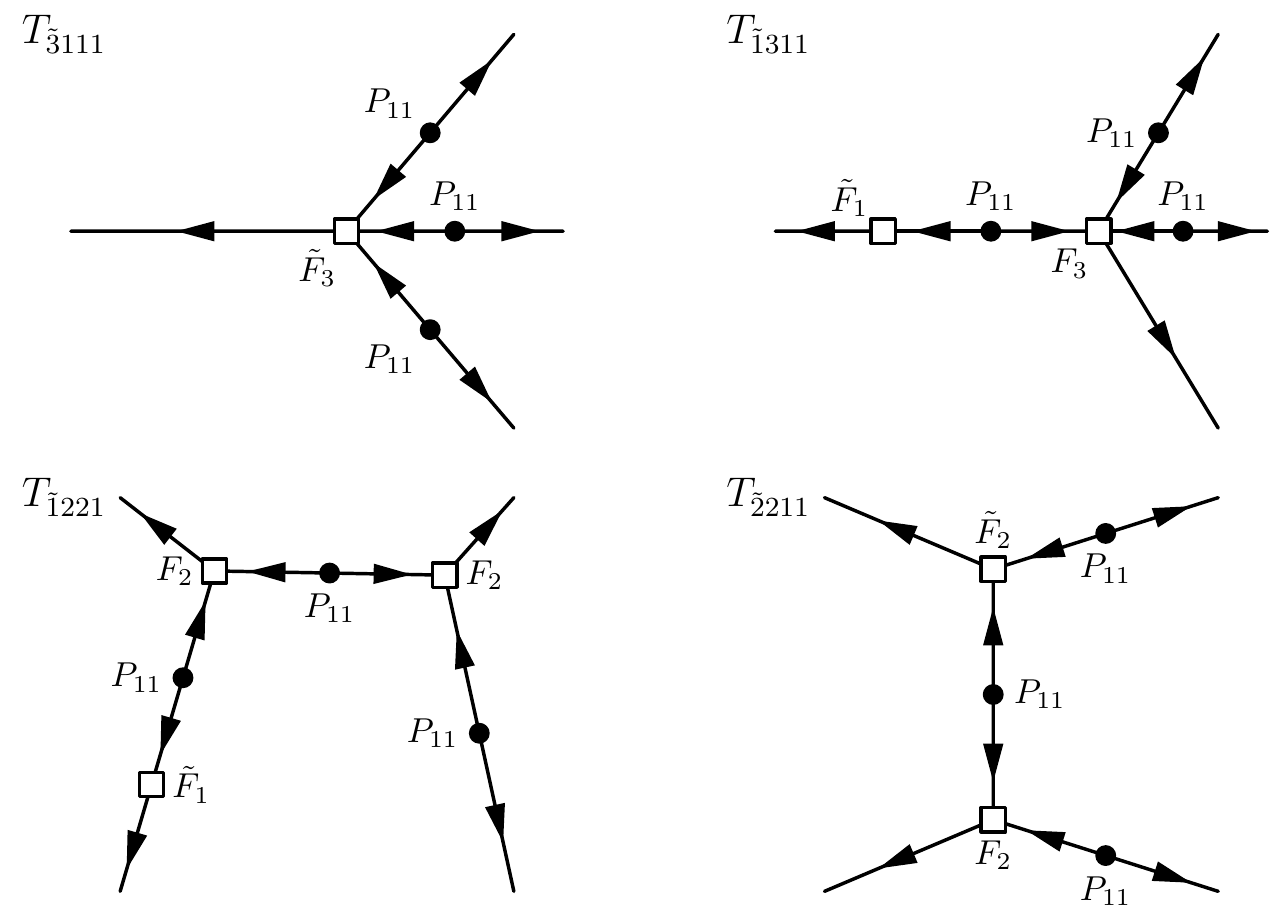}
	\caption{The leading counterterm diagrams for the one-loop trispectrum. Note that $T_{\tilde{2}211}$ contains the counterkernel $\tilde{F}_{2}$, which regularises the one-loop bispectrum and $T_{\tilde{1}221}$ and $T_{\tilde{1}311}$ contain the counterkernel $\tilde{F}_{2}$, which regularises the one-loop power spectrum (and some bispectrum terms). This leads to the opportunity of a consistency check by using counterterm amplitudes constrained in the power spectrum and bispectrum and assessing their ability to regularise the trispectrum.}
	\label{trireno}
\end{figure}

As with other correlators, each one-loop term requires regularisation.  However, the counterterms for $T_{2222}$, $T_{4211b}$, $T_{3221b}$, $T_{3221c}$, and $T_{3311b}$ are subleading noise terms and we focus on the remaining four, which are given by
\begin{align}
\label{eq:T3t111}
&T_{\tilde{3}111}(\bk_{1},\bk_{2},\bk_{3},\bk_{4})=3! \tilde{F}_{3}^{(s)}(\mathbf{k}_{2},\mathbf{k}_{3},\mathbf{k}_{4})P_{11}(k_{2})P_{11}(k_{3})P_{11}(k_{4})+3~\mathrm{perms.}~,\\
&T_{\tilde{2}211}(\bk_{1},\bk_{2},\bk_{3},\bk_{4})=(2!)^{2} \tilde{F}_{2}^{(s)}(\mathbf{k}_{2}+\mathbf{k}_{3},\mathbf{k}_{4})F_{2}^{(s)}(-\mathbf{k}_{2}-\mathbf{k}_{3},\mathbf{k}_{3})P_{11}(|\mathbf{k}_{2}+\mathbf{k}_{3}|)P_{11}(k_{3})P_{11}(k_{4})\nonumber\\&~~~~~~~~~~~~~~~~+11~\mathrm{perms.}~,\label{T4211c}\\
&T_{\tilde{1}311}(\bk_{1},\bk_{2},\bk_{3},\bk_{4})=3! \tilde{F}_{1}^{(s)}(-\bk_{1})F_{3}^{(s)}(\mathbf{k}_{2},\mathbf{k}_{3},\mathbf{k}_{4})P_{11}(k_{2})P_{11}(k_{3})P_{11}(k_{4})+3~\mathrm{perms.}~,\\
&T_{\tilde{1}221}(\bk_{1},\bk_{2},\bk_{3},\bk_{4})=(2!)^{2} \tilde{F}_{1}^{(s)}(-\bk_{1})F_{2}^{(s)}(\bk_{1},\bk_{3}+\bk_{4})F_{2}^{(s)}(-\bk_{3}-\bk_{4},\bk_{4})\nonumber
\\&~~~~~~~~~~~~~~~~\times P_{11}(k_{1})P_{11}(|\bk_{3}+\bk_{4}|)P_{11}(k_{4})+11~\mathrm{perms.}~,\label{eq:T1t221}
\end{align}
Feynman diagrams for each of which are shown in Fig.~\ref{trireno}. Notice that each counterterm depends upon one counterkernel and three of them depend upon $\tilde{F}_{1}$ and $\tilde{F}_{2}$, the same counterkernels found in the counterterms to the one-loop bispectrum which we studied in \cite{Steele:2020tak}.  This gives us the opportunity to model $T_{\tilde{2}211}$, $T_{\tilde{1}311}$, and $T_{\tilde{1}221}$ using the amplitudes of the counterparameters calibrated from the bispectrum.  This leaves only $T_{\tilde{3}111}$ left to be calibrated, which we shall discuss in Sec.~\ref{sec:alphabeta}.  It is also notable that the UV limit of $T_{5111}$ scales as $k_{1}^{2}P_{11}(k_{2})P_{11}(k_{3})P_{11}(k_{4})$ implying that, when fitted to an amplitude, it could be used as an estimator for $T_{\tilde{3}111}$.  We do not use explicitly use the UV limit of $T_{5111}$ as an estimator for $T_{\tilde{3}111}$ in this paper but do define two one-parameter estimators that scale as $k_{1}^{2}P_{11}(k_{2})P_{11}(k_{3})P_{11}(k_{4})$, which in theory should differ from $T_{5111,\mathrm{UV}}$ only by a constant that will be accounted for in the fitted parameter. Sec.~\ref{sec:alphabeta}.

\begin{figure}
    \centering
    \includegraphics{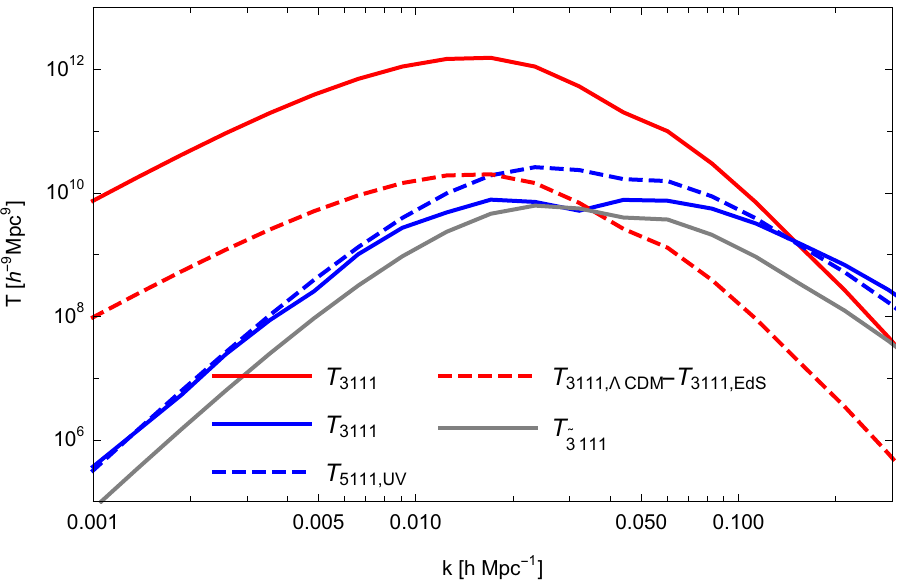}
    \caption{
The one-loop contributions to the equilateral trispectrum propagator.  On large scales the difference between the $\Lambda$CDM and EdS (red dashed) tree-level terms exceeds the one-loop trispectrum contribution from $T_{5111}$ (blue solid) and the counterterm $T_{\tilde{3}111}$ (gray) by up to three orders of magnitude. The blue dashed line shows the UV-limit of $T_{5111}$.}
    \label{fig:TriLCDM}
\end{figure}

Note that in previous studies of LSS trispectra, an approximate EdS cosmology has been assumed.  In Fig.~\ref{fig:TriLCDM} we show that the difference between the tree level propagator term for EdS and $\Lambda$CDM growth factors exceeds the one-loop terms and the one-loop counterterm on large scales, highlighting the importance of using the correct growth factors when studying loop corrections to the trispectrum (for details on the derivation of the \lcdm growth factors see \cite{Takahashi:2008yk,Fasiello:2016qpn,Steele:2020tak}).  
\subsection{Theoretical Errors}
Calculating higher and higher loops becomes more and more complicated, yet it is beneficial to estimate the size of sub-leading corrections. Such an estimation is possible in scale-free universes \cite{Pajer:2013jj}, where the non-linear scale is the only non-trivial reference scale and the the initial power spectrum thus can be expressed as $(k/k_{\mathrm{NL}})^{n}$.
Our universe is not a scaling universe, but its power spectrum can be approximated by a power-law with slope $n_\text{NL}\approx-3/2$ around the non-linear scale $k_\text{NL}$.  At $l$-loop order we can estimate the size of the loop corrections to the power spectrum as
\begin{equation}
    \Delta P_{l}(k)=P_{\mathrm{tree}}(k)\left[\frac{k}{k_{\mathrm{NL}}}\right]^{(3+n_\text{NL})l}~,
\end{equation}
where for our analysis, $k_{\mathrm{NL}}$ is taken to be $0.3 \ihMpc$.  For the bispectrum one can generalise this to  \cite{Baldauf:2016sjb,Steele:2020tak}
\begin{equation}
    \Delta B_{l}(k_{1},k_{2},k_{3})=B_{\mathrm{tree}}(k_{1},k_{2},k_{3})\left[\frac{k_{\mathrm{ext}}}{k_{\mathrm{NL}}}\right]^{(3+n_\text{NL})l}~,
    \label{eq:bierr}
\end{equation}
where $k_{\mathrm{ext}}$ is the mean external momentum magnitude of the kernel featuring the loop and is averaged as $k_{\mathrm{ext}}=(k_{1}+k_{2}+k_{3})/3$ when we are studying the entire configuration space of the bispectrum.

Extrapolating from Eq.~\eqref{eq:bierr}, we can generate an estimator for the theoretical error of the trispectrum at $l$-loops in
\begin{equation}
    \Delta T_{l}(\bk_{1},\bk_{2},\bk_{3},\bk_{4})=T_{\mathrm{tree}}(\bk_{1},\bk_{2},\bk_{3},\bk_{4})\left[\frac{k_{\mathrm{ext}}}{k_{\mathrm{NL}}}\right]^{(3+n_\text{NL})l}~.
\end{equation}

\subsection{Simulations and PT on the Grid}\label{sec:simgrid}
Initially, a random realisation $\delta_1$ of a Gaussian density field is generated on a grid.  We then simulate the evolution of this field until the present day, allowing for the generation of non-linearities; this gives us a representation of the real universe to which we can compare our theory.  We employ GADGET II \cite{Springel:2005mi}, a tree-PM hybrid $N$-body simulation code.  Once the simulations have been run, we sample the resulting density fields on a grid generated using the Cloud-In-Cell (CIC) mass assignment scheme.  On this grid, we then measure correlators by averaging the complex products of density fields in spherical shells $i=1,\ldots,N_\text{bin}$ with momentum magnitudes ranging from a chosen $k_{i,\mathrm{min}}$ to a chosen $k_{i,\mathrm{max}}$.  We refer to our choices of minimum and maximum momenta for these shells as the momentum binning.

From the realization of the Gaussian field $\delta_{1}$, one can then calculate the higher order fields using the recursion relations \cite{Roth:2011test,Taruya:2018jtk,Taruya:2020qoy,Steele:2020tak}
\begin{equation}
\begin{split}
\delta_{n}=&\sum^{n-1}_{m=1}\frac{1}{(2n+3)(n-1)}[(2n+1)(\theta_{m}\delta_{n-m}-\vec \Psi_{m}\cdot \vec \nabla \delta_{n-m})+\\&2(-\vec \Psi_{m}\cdot \vec \nabla \theta_{n-m}/2 - \vec \Psi_{n-m}\cdot \vec \nabla \theta_{m}/2 +K_{m,ij}K_{n-m,ij}+\theta_{m}\theta_{n-m}/3)]~,
\end{split}
\label{eq:rr1}
\end{equation}
\begin{equation}
\begin{split}
\theta_{n}=&\sum^{n-1}_{m=1}\frac{1}{(2n+3)(n-1)}[3(\theta_{m}\delta_{n-m}-\vec \Psi_{m}\cdot \vec \nabla \delta_{n-m})+\\&2n(-\vec \Psi_{m}\cdot \vec \nabla \theta_{n-m}/2 - \vec \Psi_{n-m}\cdot \vec \nabla \theta_{m}/2 +K_{m,ij}K_{n-m,ij}+\theta_{m}\theta_{n-m}/3)]~.
\end{split}
\label{eq:rr2}
\end{equation}
The displacement fields are given by 
\be
\vec \Psi_{\theta_m}(\vec k)=\ii\frac{\vec k}{k^2}\theta_m(\vec k)\, 
\ee
and equivalently the tidal tensor is given by
\be
K_{\theta_m,ij}(\vec k)=\left(\frac{k_i k_j}{k^2}-\frac13 \deltakron_{ij}\right)\theta_m(\vec k)\, .
\ee
This allows us to perform perturbation theory on the grid (referred to as gridPT in \cite{Taruya:2018jtk,Taruya:2020qoy} a name we shall adopt hereafter), making it possible to compare the individual contributions to their theoretical predictions, rather than simply calculating the fully non-linear field and comparing it to the full predicted spectrum.

In most studies to date, analyses of LSS correlators have relied upon measuring the fully non-linear correlators and comparing them to SPT. Thus, the residual of the power spectrum up to one-loop would be defined as $P_{\mathrm{nn}}-P_{11,\mathrm{SPT}}-2P_{31,\mathrm{SPT}}-P_{22,\mathrm{SPT}}$ and this is what is used to calculate the counterterms.  We make use of gridPT, in which we calculate individual contributions to the non-linear correlator up to a given order explicitly on the grid using Eqs.~\eqref{eq:rr1} and \eqref{eq:rr2}.  We can then subtract these correlators from the non-linear correlator.  This gives significantly smaller error bars for a given number of numerical realisations for a number of reasons.  Namely, SPT calculations would ordinarily be done at the centre of a momentum shell rather than integrating over its entire volume as is done on the grid and the fact that terms which should not be contributing to the non-linear correlator are not being correctly removed, as will be discussed in Sec.~\ref{sec:unphysical}.  Furthermore, gridPT uses the same phases of the modes and accounts for the discreteness of the largest modes present in the box.  GridPT also offers the opportunity for components of counterterms to be explicitly calculated on the grid, such as the $\Gamma$ and $\mathrm{E}_{i}$ components of $\tilde{F}_{2}$ as shown in Eq.~\eqref{eq:symmf2tilde}, which we will use for some of our counterterm calibrations in Sec.~\ref{sec:expandedkernels}.


\section{Regularising the One-Loop Trispectrum}
In order to perform an analysis of the one-loop trispectrum we must create both a perturbative calculation and a non-perturbative measurement of the trispectrum.  Employing gridPT, we measure the individual contributions to the trispectrum such as $T_{3111}$ and $T_{5111}$ on the same grid used to analyse the simulation results; this allows much greater precision than could have been achieved by comparing the measured fully non-linear trispectrum to analytically calculated one-loop PT, both because it ensures that there is identical momentum shell averaging between the perturbative and non-linear density fields and because it allows us to reduce the noise by comparing perturbative and non-linear modes that arise from the same seeds.  This leads to a number of interesting results which affect the counterterm calculations, as detailed in Secs.~\ref{sec:unphysical} and \ref{sec:alphabeta}.  We also evaluate regular numerical PT using a routine that relies upon the CUBA Vegas numerical integrator \cite{Hahn:2004fe} and use this for comparative purposes as well as for some of our counterterm calculations, as will be discussed in Sec.~\ref{sec:expandedkernels}.

All of the trispectra listed in Sec.~\ref{sec:tri} contribute to the auto trispectrum $T_{\mathrm{nnnn}}$.  However, we can also isolate individual counterterms by studying the partially non-linear trispectra:
\begin{align}
    T_{\mathrm{n}111}&=T_{3111}+T_{5111}+T_{\tilde{3}111}~,\\
    T_{\mathrm{n}211}&=T_{2211}+T_{4211}+T_{\tilde{2}211}~,\\
    T_{\mathrm{n}221}&=T_{1221}+T_{3221}+T_{\tilde{1}221}~,\\
    T_{\mathrm{n}311}&=T_{1311}+T_{3311}+T_{\tilde{1}311}~,
\end{align}
where the subscript n represents a non-linear density field.  This allows us to constrain each counterterm separately, potentially increasing the precision of our calculations compared to calibrating all parameters of the four counterterms simultaneously from the non-linear trispectrum $T_\text{nnnn}$.  

By convention, $T_{\mathrm{n}111}$ is referred to as the trispectrum propagator.  In this paper, we expand upon this convention and refer to all correlators which feature only one non-linear field as propagators.  It is these four propagator terms that we shall study in the analysis described below.

\subsection{Estimating the Trispectrum}
\begin{figure}[h!]
	\centering
	\includegraphics[width=0.3\textwidth]{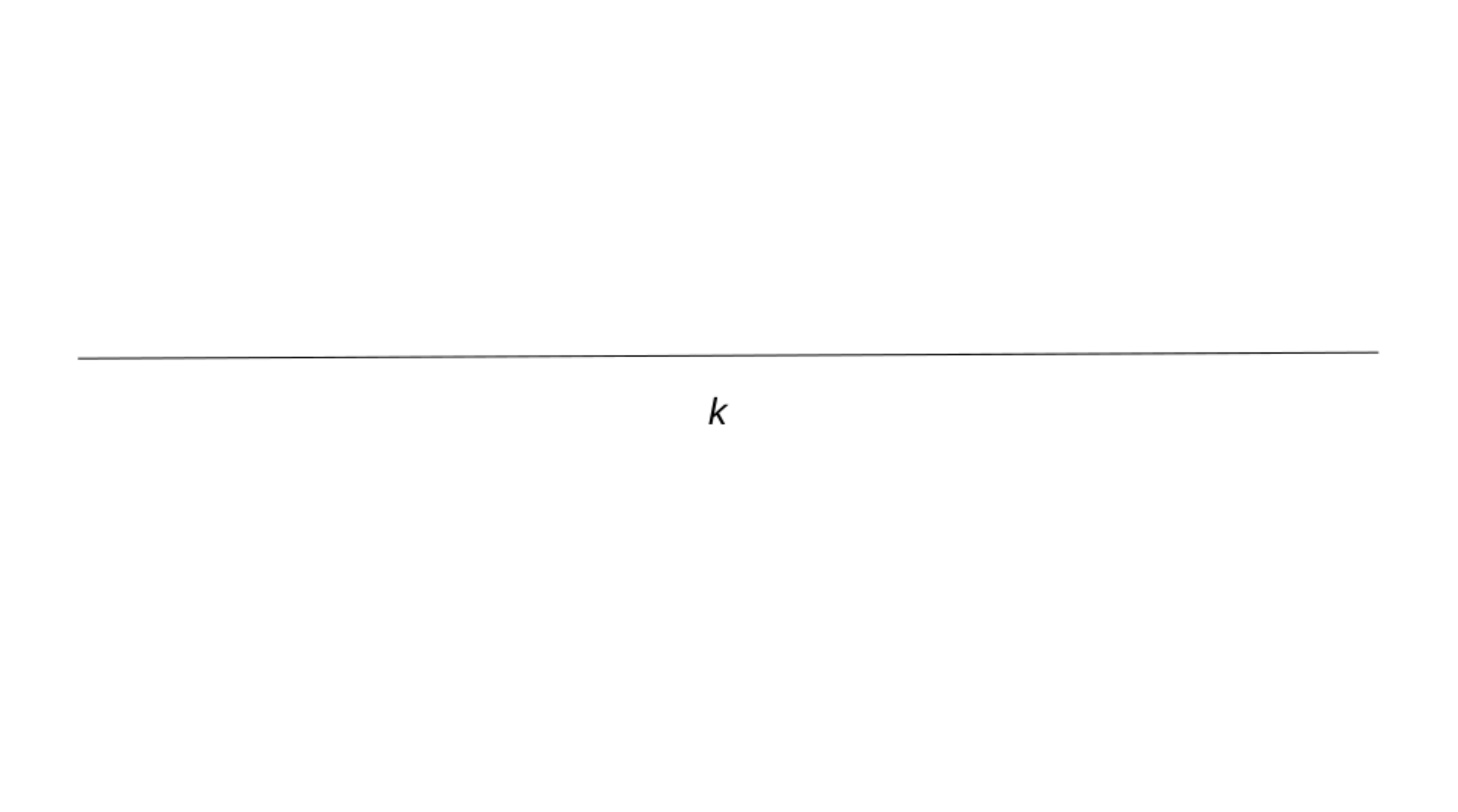}
	\hspace{3cm}
	\includegraphics[width=0.3\textwidth]{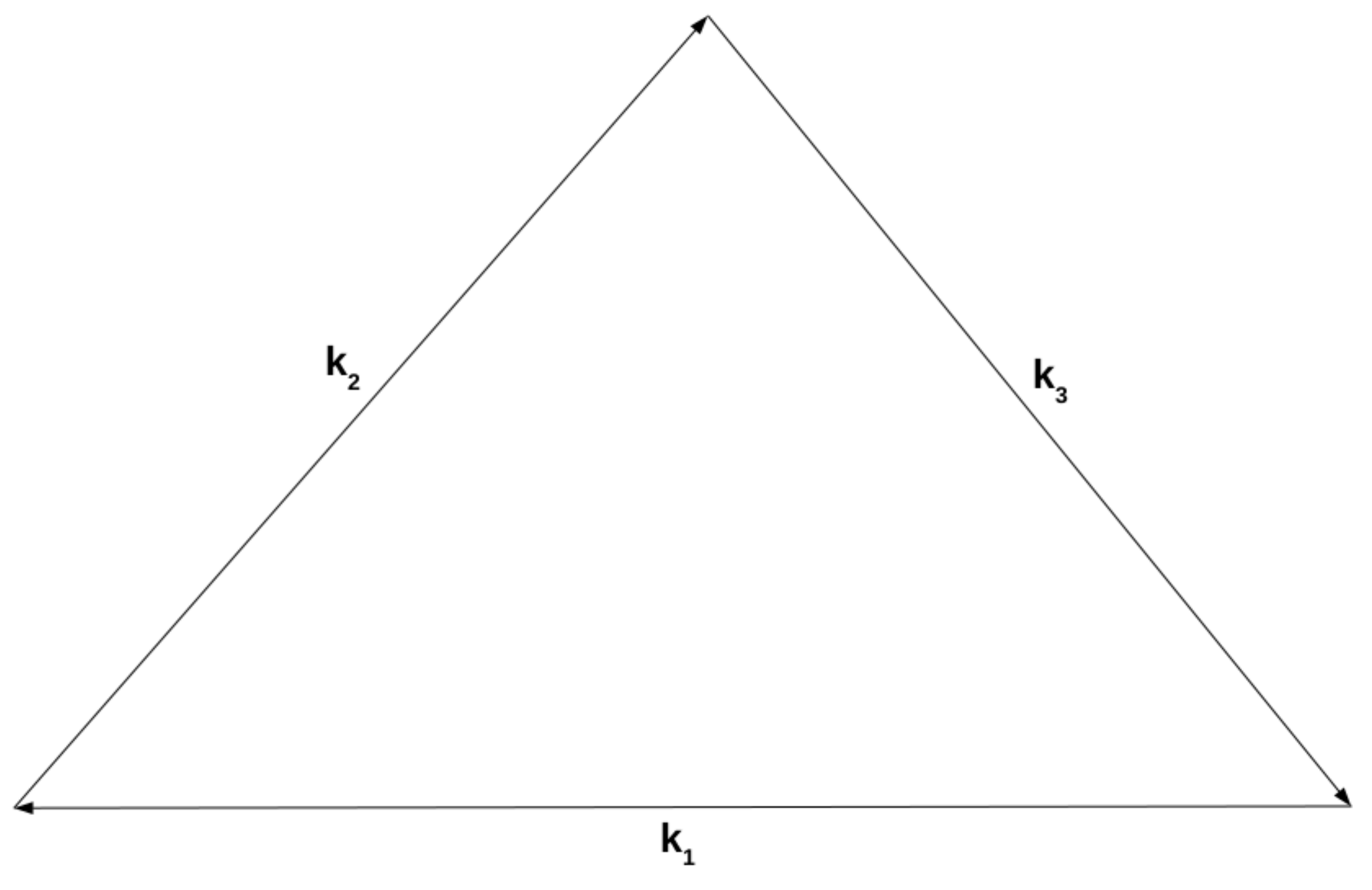}
	\caption{\emph{Left hand panel:} The shape of the power spectrum in momentum space:  As we are only connecting two points in a homogeneous and isotropic space, we can entirely define the power spectrum with one parameter: the momentum, $k$.  \emph{Right hand panel:} The shape of the bispectrum in momentum space.  It can be entirely defined by three parameters; either the magnitudes of two vectors and the angle that separates them, or the magnitudes of all three vectors. The requirement that the vectors connect to form a closed triangle allows us to deduce the remaining parameters defining all three vectors from basic trigonometry.}
	\label{fig:PSBSshapes}
\end{figure}

The space of possible shapes a correlator can take in real or momentum space is called its configuration space.  Ordinarily, we study correlators in momentum space with the magnitudes of their vectors chosen in bins ranging from a chosen $k_{\mathrm{min}}$ to a chosen $k_{\mathrm{max}}$, thus defining the shells in which we perform measurements. The power spectrum is the correlator of two fields and so takes the form of a single line which can be entirely defined by its magnitude, $k$.  Thus, it has a one dimensional configuration space which is numerically simple to study in full.  The bispectrum is the correlator of three fields and so takes the form of a triangle, with each density field represented by a vector. By convention, we refer to the vector defining the first density field as $\bk_{1}$, that defining the second density field as $\bk_{2}$, and that defining the third density field as $\bk_{3}$. We can entirely define the configuration of a bispectrum using three parameters; if we take the magnitudes of $\bk_{1}$ and $\bk_{2}$ together with the enclosed angle $\phi$, we can calculate $\bk_{3}$ from the requirement that these two vectors are connected into a closed triangle.  Alternatively, we could take the magnitudes of all three vectors and calculate $\phi$ using basic trigonometry.  Thus, the bispectrum has a three dimensional configuration space.  Sampling the entire space is more computationally intensive than working with fixed configurations but is not unfeasible; in \cite{Steele:2020tak} we studied the full configuration space of the bispectrum.  The momentum space shapes of the power spectrum and bispectrum are shown in Fig~\ref{fig:PSBSshapes}.

\begin{figure}[h!]
	\centering
	\includegraphics[width=10cm]{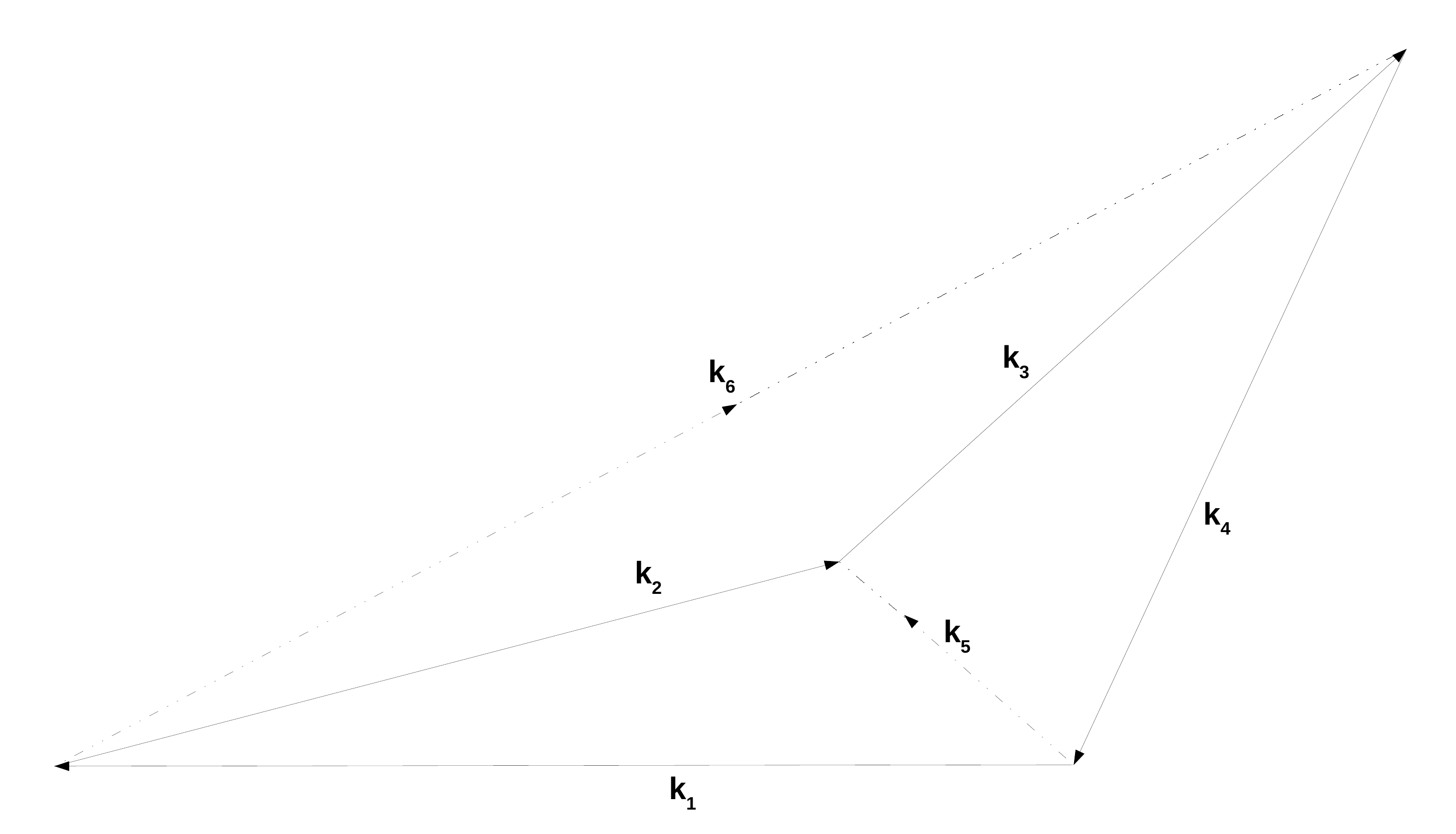}
	\caption{The shape of the trispectrum in momentum space.  The four density perturbations are connected by six legs to form a tetrahedron; by convention, we refer to four of those legs as the external legs, shown by solid lines, and two as diagonal legs, shown by dotted lines.}
	\label{fig:trishape}
\end{figure}

The trispectrum is the correlator of four fields and so takes the shape of a tetrahedron, as shown in Fig.~\ref{fig:trishape}.  This shape has six momentum vectors defining the four fields; four of them we label the external legs and the remaining two we call diagonal legs.  When we come to studying individual configurations, we will see that this distinction allows us to define trispectra using only their external legs and averaging over their diagonals.  

We can parametrise the trispectrum as
\begin{align}
    \bk_{1}&=(k_{1},0,0)~,\\
    \bk_{2}&=k_{2}\left(\sqrt{1-\mu_{1}^{2}},0,\mu_{1}\right)~,\\
    \bk_{3}&=k_{3}\left(\sqrt{1-\mu_{2}^{2}}~\mathrm{cos}(\phi_{2}),\sqrt{1-\mu_{2}^{2}} ~\mathrm{sin}(\phi_{2}), \mu_{2}\right)~,\\
    \label{eq:k4}
    \bk_{4}&=-\bk_{1}-\bk_{2}-\bk_{3}~,\\
    \bk_{5}&=\bk_{1}+\bk_{2}~,\\
    \bk_{6}&=\bk_{2}+\bk_{3}~.
\end{align}
with $\bk_{5}$ and $\bk_{6}$ being referred to hereafter as the diagonal legs.  The choice of these two as the diagonal legs is arbitrary, with the only requirement being that the external legs form a closed quadrilateral.  With three legs defined, the remaining three can be inferred from the requirement that every point be connected to every other in a closed tetrahedron; thus, we can define a configuration fully with six scalar parameters; those are usually chosen to be either the magnitude of three vectors and the three angles that separate them from one another in 3D space, $(k_1,k_2,k_3,\mu_1,\mu_2,\phi_2)$, or the magnitudes of all six legs, $(k_1,k_2,k_3,k_4,k_5,k_6)$.  The shape of the trispectrum in momentum space is shown in Fig.~\ref{fig:trishape}.

For the trispectrum, studying the full configuration space would be numerically non-trivial and we leave such a study to a future paper.  We choose to focus on configurations in which the external four legs are specified and the diagonals are allowed to vary.  In our CUBA calculations of the SPT contributions to the EFT, this takes the form of inputting the $k$-binning of three momenta, integrating over all possible values of the three angles, and calculating the resultant vectors $\bk_{1}$, $\bk_{2}$, and $\bk_{3}$ within the integrand. We use these and Eq.~\eqref{eq:k4} to calculate $\bk_{4}$ and its magnitude, $k_{4}$. An if-statement then determines whether or not $k_{4}$ is within the desired bin; if it is, we calculate the trispectrum with those external leg parameters and add it to the integral; if not, we simply do not.  In our analysis on the grid, we rely upon Dirac delta functions and their integral representations to ensure the validity of the configurations.  In both cases, the diagonal legs are implicitly integrated over as they are not required to fit into any particular $k$-bin. 

We use a trispectrum estimator, rather than relying upon calculations of the covariance matrix of measured power spectra.  Naively, one might try to measure the trispectrum by directly summing density fields in momentum space \cite{PhysRevD.71.063001}:
\begin{equation}
\hat{T}=\frac{1}{\mathcal{N}}
\int \frac{\derd^3 \bk_{1}}{(2\pi)^3}
\int \frac{\derd^3 \bk_{2}}{(2\pi)^3}
\int \frac{\derd^3 \bk_{3}}{(2\pi)^3} 
\int \frac{\derd^3 \bk_{4}}{(2\pi)^3}  ~(2\pi)^{3}\delta^{(\mathrm{D})}(\bk_{1}+\bk_{2}+\bk_{3}+\bk_{4})\delta(\bk_{1})\delta(\bk_{2})\delta(\bk_{3})\delta(\bk_{4})
\end{equation}
where the normalisation factor $\mathcal{N}$ is the volume of the space being integrated over, where the integrals run through the appropriate momentum bins for our chosen configurations $|\vec k|\in [k_i,k_{i+1}]$.  Here we are binning the magnitudes of $k_1, k_2, k_3, k_4$ and leave the diagonals unconstrained. We hereafter refer to this trispectrum estimator as the direct summation method.

However, this requires studying the entire 3D volume of the sampled space for each of three legs, giving a total of $N_{\mathrm{c}}^{9}$ measurements, where $N_{\mathrm{c}}$ is the number of cells per dimension.  The Dirac function would then give us $\bk_{4}$.
This is of course numerically intensive to the point of being unfeasible.  Instead, we use the integral representation of the Dirac function to give us
\begin{equation}
\begin{split}
\hat{T}&=\frac{1}{\mathcal{N}}\int \derd^{3}\bx\ \int \frac{{\derd}^{3}\bk_{1}}{(2\pi)^{3}}\int \frac{{\derd}^{3}\bk_{2}}{(2\pi)^{3}} \int \frac{{\derd}^{3}\bk_{3}}{(2\pi)^{3}} \int \frac{{\derd}^{3}\bk_{4}}{(2\pi)^{3}} ~e^{\ii\bx\cdot(\bk_{1}+\bk_{2}+\bk_{3}+\bk_{4})}\delta(\bk_{1})\delta(\bk_{2})\delta(\bk_{3})\delta(\bk_{4})\\&=\frac{1}{\mathcal{N}}\int \derd^{3}\bx\ \int \frac{{\derd}^{3}\bk_{1}}{(2\pi)^{3}}~e^{\ii\bx\cdot\bk_{1}}\delta(\bk_{1})\int \frac{{\derd}^{3}\bk_{2}}{(2\pi)^{3}} ~e^{\ii\bx\cdot\bk_{2}}\delta(\bk_{2})\int \frac{{\derd}^{3}\bk_{3}}{(2\pi)^{3}}~e^{\ii\bx\cdot\bk_{3}}\delta(\bk_{3}) \int \frac{{\derd}^{3}\bk_{4}}{(2\pi)^{3}} ~e^{\ii\bx\cdot\bk_{4}}\delta(\bk_{4})\\
&\equiv\frac{1}{\mathcal{N}}\int \derd^{3}\bx ~f_{1}(\bx)f_{2}(\bx)f_{3}(\bx)f_{4}(\bx),
\end{split}
\label{eq:FFT1}
\end{equation}
where $f$ is the Fourier transform of $\delta$ restricted to a momentum bin.  In practice this means setting the field to zero everywhere but in a shell in Fourier space and then Fourier transforming the resulting field.
 Our routine based upon this method works significantly faster than the direct summation routine and allows us to average over the diagonal legs of our configurations with minimal computational cost. In Appendix~\ref{App:FFT} we show how this method can be extended to account for a fixed length of the diagonals.  

Four momentum-space configurations were studied for this initial analysis of the trispectrum:
\begin{description}
\setlength{\itemsep}{0cm}
    \item[PPM]$\langle \delta(k_{i})\delta(k_{i+1})\delta(k_{i+1})\delta(k_{i-1}) \rangle$
    \item[PMM]$\langle \delta(k_{i})\delta(k_{i+1})\delta(k_{i-1})\delta(k_{i-1}) \rangle$
    \item[MMM]$\langle \delta(k_{i})\delta(k_{i-1})\delta(k_{i-1})\delta(k_{i-1}) \rangle$
    \item[PPP]$\langle \delta(k_{i})\delta(k_{i+1})\delta(k_{i+1})\delta(k_{i+1}) \rangle$
\end{description}
where $k_{i}$ represents the $i$th momentum bin, such that in all configurations one of the density fields was used as a reference while the others were positioned in the $k$-bins immediately above or below the reference.  Specifically, it is the non-linear density field in the propagator terms that is chosen to be in the reference bin.  We did not use configurations in which all four of the fields had the same momenta as allowing the diagonal legs to vary in this configuration leads to the intrusive inclusion of disconnected trispectra, i.e. pairs of anti-parallel power spectra in the measurement of the trispectra.

\begin{figure}[t]
	\centering
	\includegraphics[width=0.49\textwidth]{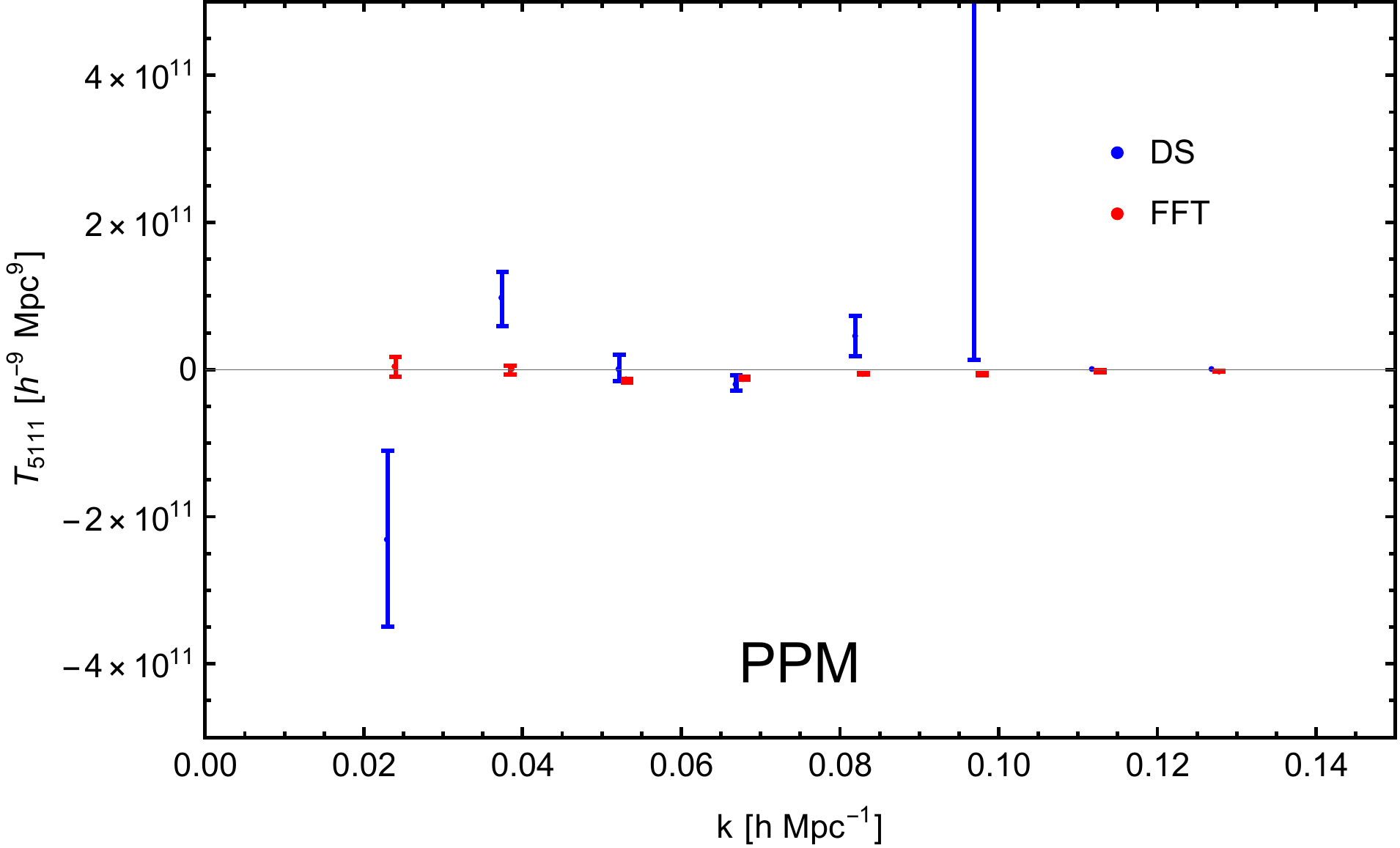}
	\includegraphics[width=0.49\textwidth]{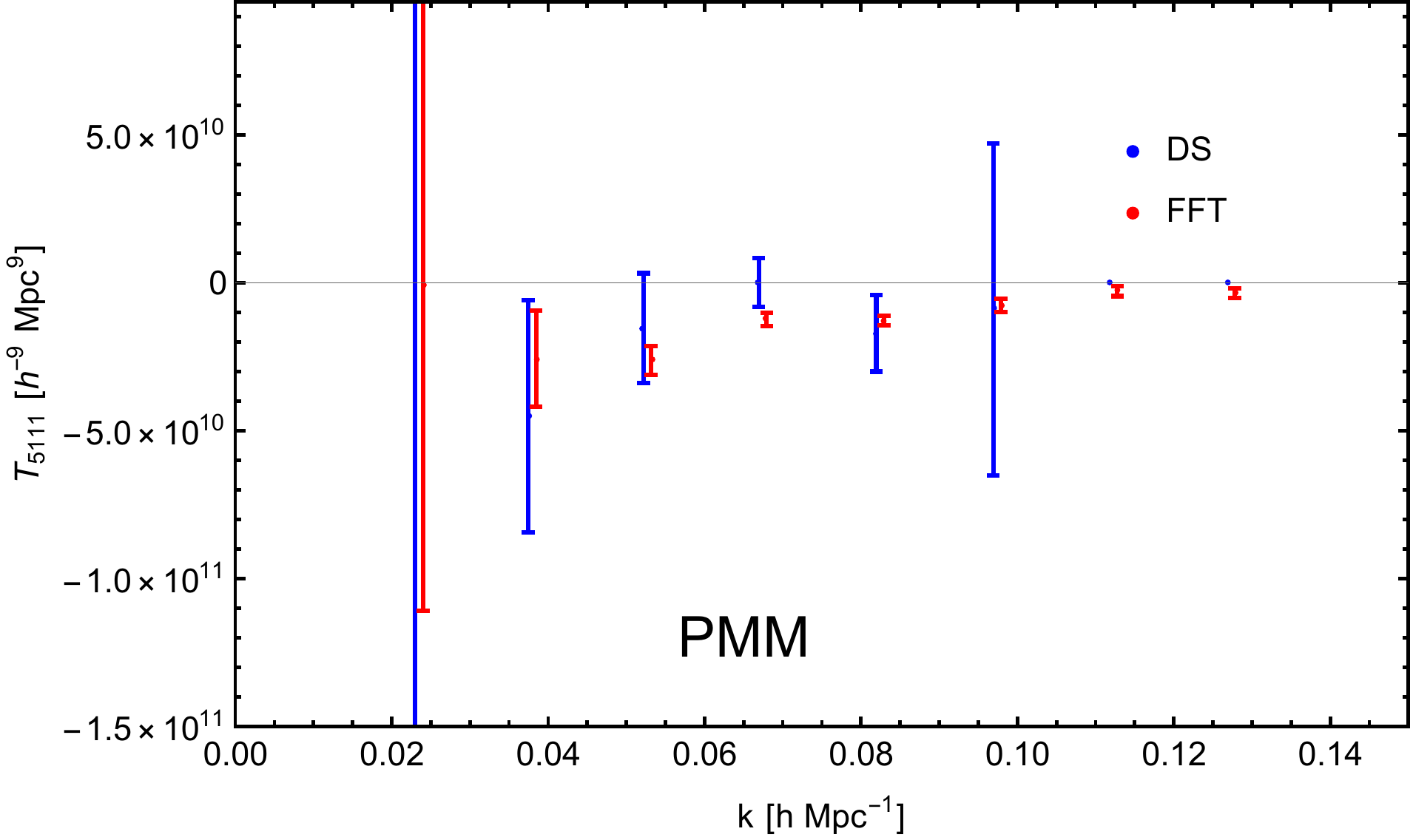}
	\includegraphics[width=0.49\textwidth]{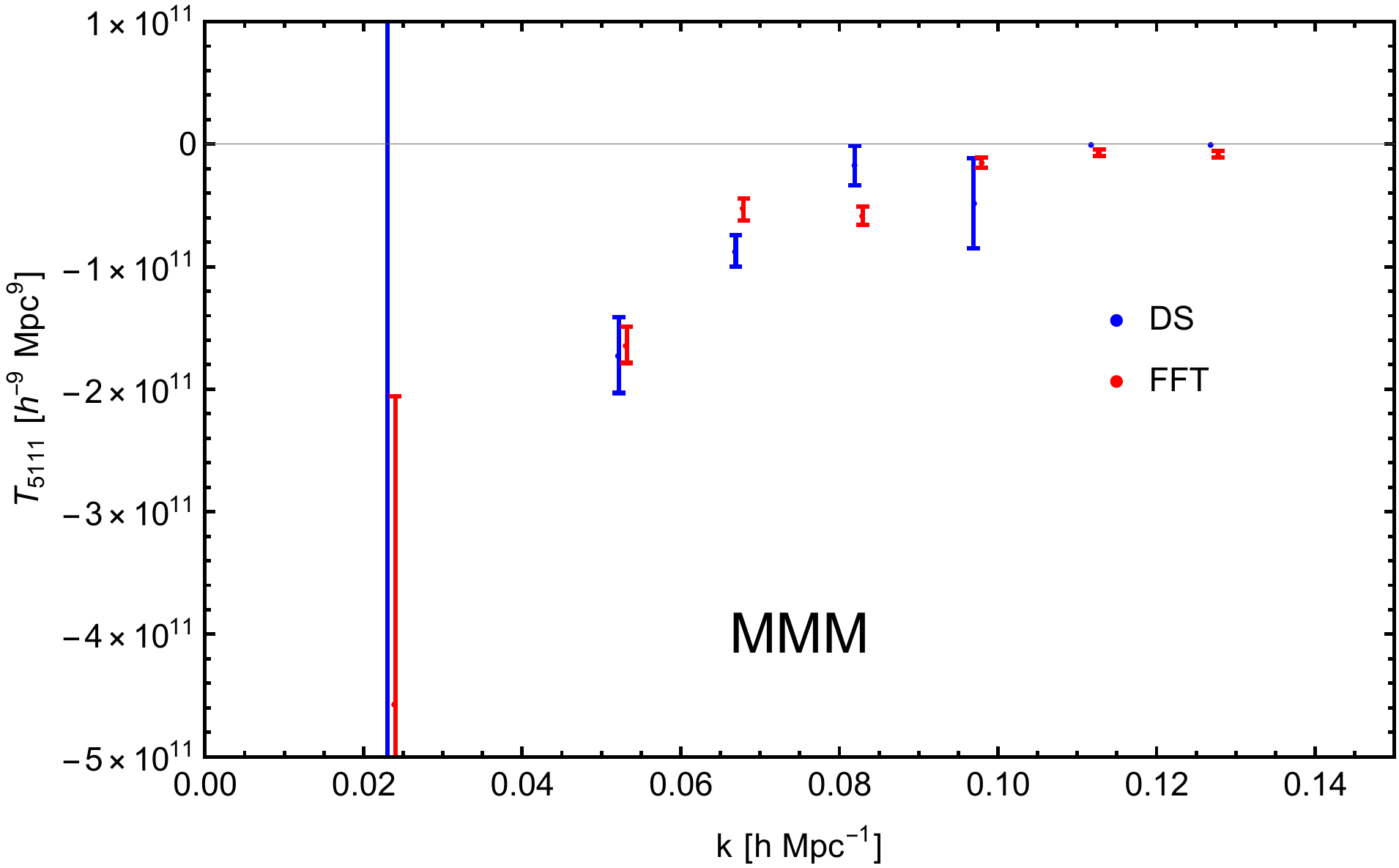}
	\includegraphics[width=0.49\textwidth]{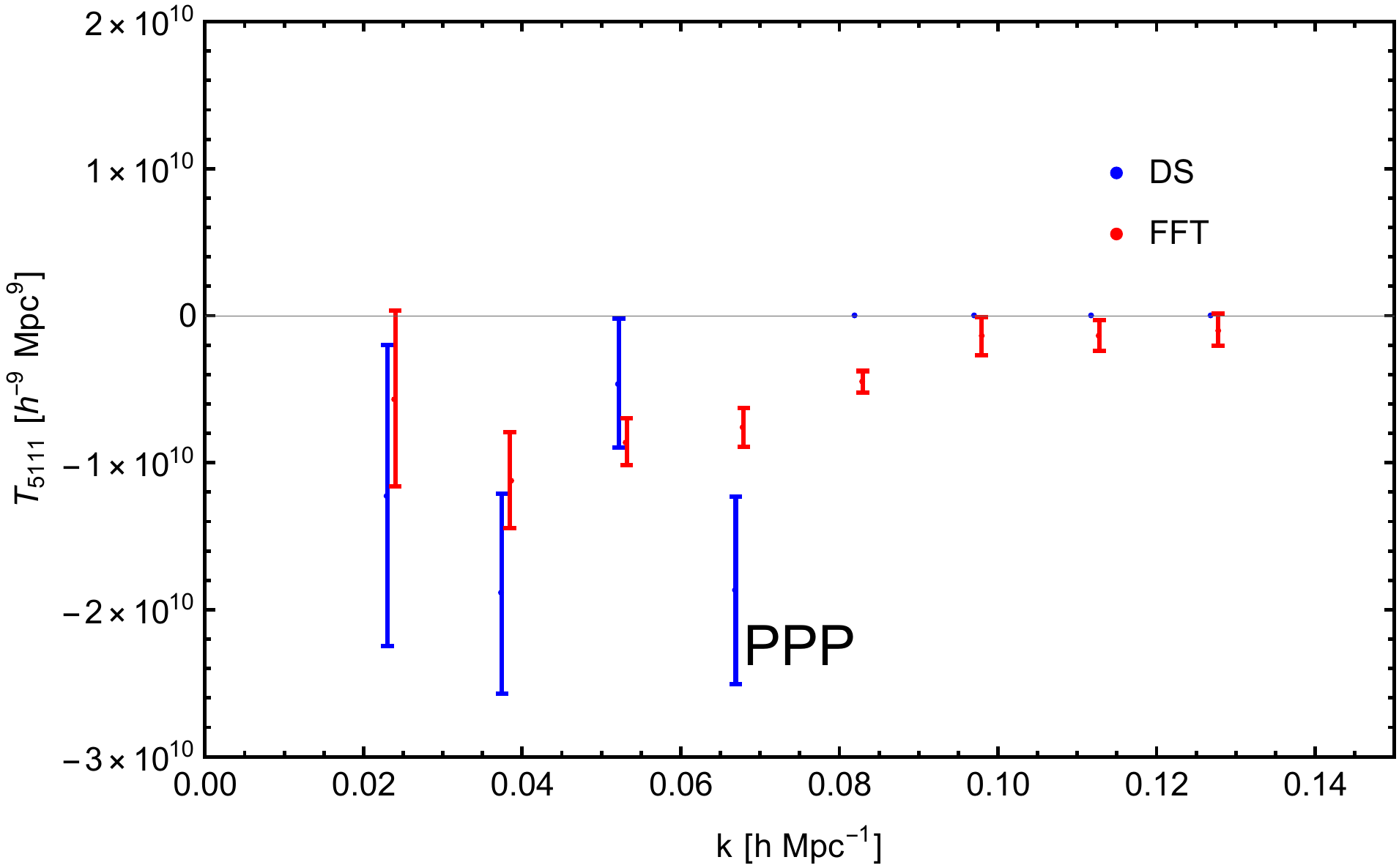}
	\caption{The measured $T_{5111}$ in the four sampled configurations with both the direct summation (DS) routine and the Fourier transform routine (FFT).  Notice that the DS routine only sampled a small fraction of the overall momentum space, resulting in it only measuring a few points for each configuration and each of those points being less precise than those measured by the FFT routine, which studied the entire grid.}
	\label{fig:DSFFT}
\end{figure}

In Fig.~\ref{fig:DSFFT} we plot the measured $T_{5111}$ from both the Direct Summation and Fast Fourier Transform routines.  Due to the computational intensity of the direct summation method, we were only able to sample a small fraction of the overall space, resulting in fewer data points with larger errors.  This clearly demonstrates the benefits of using the Fourier transform method, which we use throughout the remainder of the paper to the complete exclusion of the direct summation method.

\subsection{Cosmic Variance Cancellation}
\label{sec:unphysical}
In the infinite-volume limit, terms which describe disconnected pairs of power spectra, such as $T_{1111}$, should not contribute to the measured off-diagonal connected trispectrum.  Likewise, terms which  contain an odd number of linear density fields such as $T_{2111}$ and $T_{4111}$, should vanish in accordance with Wick's Theorem.  However, although their mean values will be zero, in a finite sample size they will have a non-zero variance.  Using gridPT, we were able to generate the correlators for these terms and found that they made a significant contribution to the measured non-linear trispectra.  With them explicitly measured, we were able to subtract them from the residuals, giving a significant reduction in cosmic variance.  We hereafter refer to terms with a vanishing mean and non-zero variance as mean zero terms.  
\begin{figure}[t]
	\centering
	\includegraphics[width=0.49\textwidth]{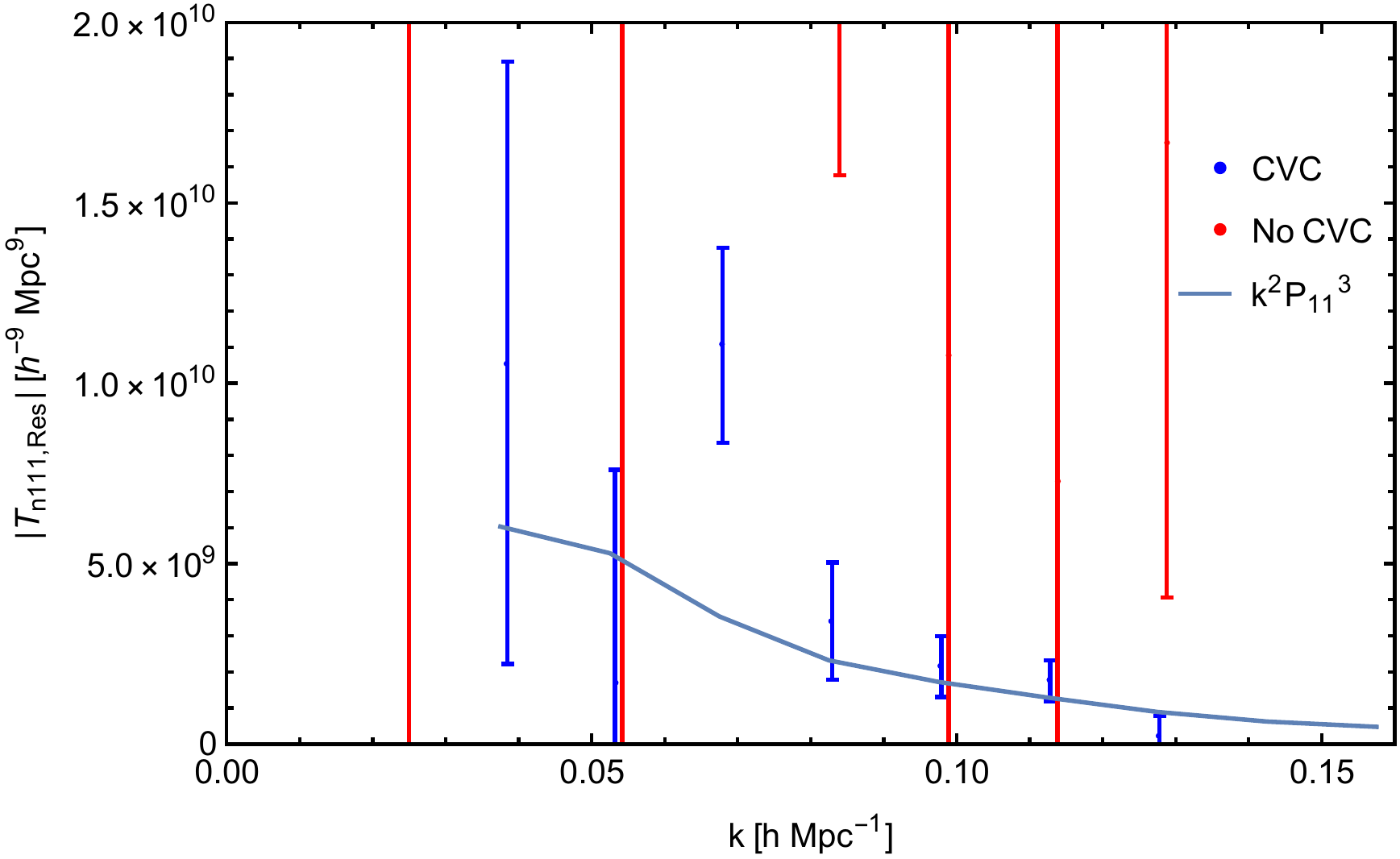}
	\caption{The measured residual of the trispectrum propagator $T_{n111}$ in configuration PPM with and without subtracting off the mean zero terms, together with a curve which scales as the counterterm would be expected to, that being $k_{1}^{2}P_{11}(k_{2})P_{11}(k_{3})P_{11}(k_{4})$. }
	\label{fig:CVC}
\end{figure}

When removing mean zero terms, we only remove those that have an highest order density field at most as high in order as the highest order density field in the corresponding one-loop contribution, such that we have 
\begin{align}
    T_{\mathrm{n}111,\mathrm{grid}}&\rightarrow T_{\mathrm{n}111,\mathrm{grid}}-T_{4111,\mathrm{grid}}-T_{2111,\mathrm{grid}}-T_{1111,\mathrm{grid}}~,\\
    T_{\mathrm{n}211,\mathrm{grid}}&\rightarrow T_{\mathrm{n}211,\mathrm{grid}}-T_{3211,\mathrm{grid}}-T_{1211,\mathrm{grid}}~,\\
    T_{\mathrm{n}221,\mathrm{grid}}&\rightarrow T_{\mathrm{n}221,\mathrm{grid}}-T_{2221,\mathrm{grid}}~,\\
    T_{\mathrm{n}311,\mathrm{grid}}&\rightarrow T_{\mathrm{n}311,\mathrm{grid}}-T_{2311,\mathrm{grid}}~.
\end{align}
In the left hand panel of Fig.~\ref{fig:CVC} we plot the residual of $T_{\mathrm{n}111}$ with and without subtraction of the mean zero terms, which we refer to as a form of cosmic variance cancellation (CVC).  

In studying the trispectrum we are interested in the kernels $F_{\mathrm{n}}$. Any measurement of the trispectrum will include measurements of the included power spectra and these power spectra are the source of the majority of any measurement's variance.  As such, we can significantly reduce the variance of our measurements by enacting cosmic variance cancellation on each of our trispectrum measurements by redefining 
\begin{equation}
Q= \frac{\hat{T}}{\hat{P}_{11}^{{n}}}
\label{fig:Tcvc}
\end{equation}
for some product of linear power spectra $P_{11}^{n}$ measured on the grid and corresponding to the power spectra found in the definitions of the trispectra.  We perform this on a realisation by realisation basis, dividing by power spectra from the same realisation and grid as each measured trispectrum, before averaging our results.  This isolates the kernels, effectively giving us a simulation measurement of the kernel itself, and in doing so removes all of the variance that came from the linear density fields. 

In theory the most precise possible measurements of the counterterms would come from using ${n}=3$ in Eq.~\eqref{fig:Tcvc} in order to fully remove the variance of the power spectrum from the measurements.  However, many contributions to the trispectrum feature power spectra that are not simple functions of one of the external density fields' momenta but are instead an averaged sum of the momenta between different fields.  This complicates the procedure and for the sake of convenience only power spectra of single density fields are removed from our measurements.  As such, bearing in mind the momentum arguments of the power spectra in Eqs.~\eqref{eq:T3111PT} to \eqref{eq:TPT}, we have that
\begin{align}
Q_{3111}(k_{1},k_{2},k_{3},k_{4},k_{5},k_{6})&=\frac{T_{3111}(k_{1},k_{2},k_{3},k_{4},k_{5},k_{6})}{P_{11}(k_{2})P_{11}(k_{3})P_{11}(k_{4})}~,\\
Q_{5111}(k_{1},k_{2},k_{3},k_{4},k_{5},k_{6})&=\frac{T_{5111}(k_{1},k_{2},k_{3},k_{4},k_{5},k_{6})}{P_{11}(k_{2})P_{11}(k_{3})P_{11}(k_{4})}~,\\
Q_{2211}(k_{1},k_{2},k_{3},k_{4},k_{5},k_{6})&=\frac{T_{2211}(k_{1},k_{2},k_{3},k_{4},k_{5},k_{6})}{P_{11}(k_{3})P_{11}(k_{4})}~,\\
Q_{4211}(k_{1},k_{2},k_{3},k_{4},k_{5},k_{6})&=\frac{T_{4211}(k_{1},k_{2},k_{3},k_{4},k_{5},k_{6})}{P_{11}(k_{3})P_{11}(k_{4})}~,\\
Q_{3221}(k_{1},k_{2},k_{3},k_{4},k_{5},k_{6})&=\frac{T_{3221}(k_{1},k_{2},k_{3},k_{4},k_{5},k_{6})}{P_{11}(k_{4})}~,\\
Q_{3311}(k_{1},k_{2},k_{3},k_{4},k_{5},k_{6})&=\frac{T_{3311}(k_{1},k_{2},k_{3},k_{4},k_{5},k_{6})}{P_{11}(k_{3})P_{11}(k_{4})}~,
\end{align}
and $T_{2222}$ is left as it is.  Note that the power spectra in the denominators occupy the momentum bin of the linear density fields in the trispectra.

\begin{figure}[p]
	\centering
	\includegraphics[width=0.49\textwidth]{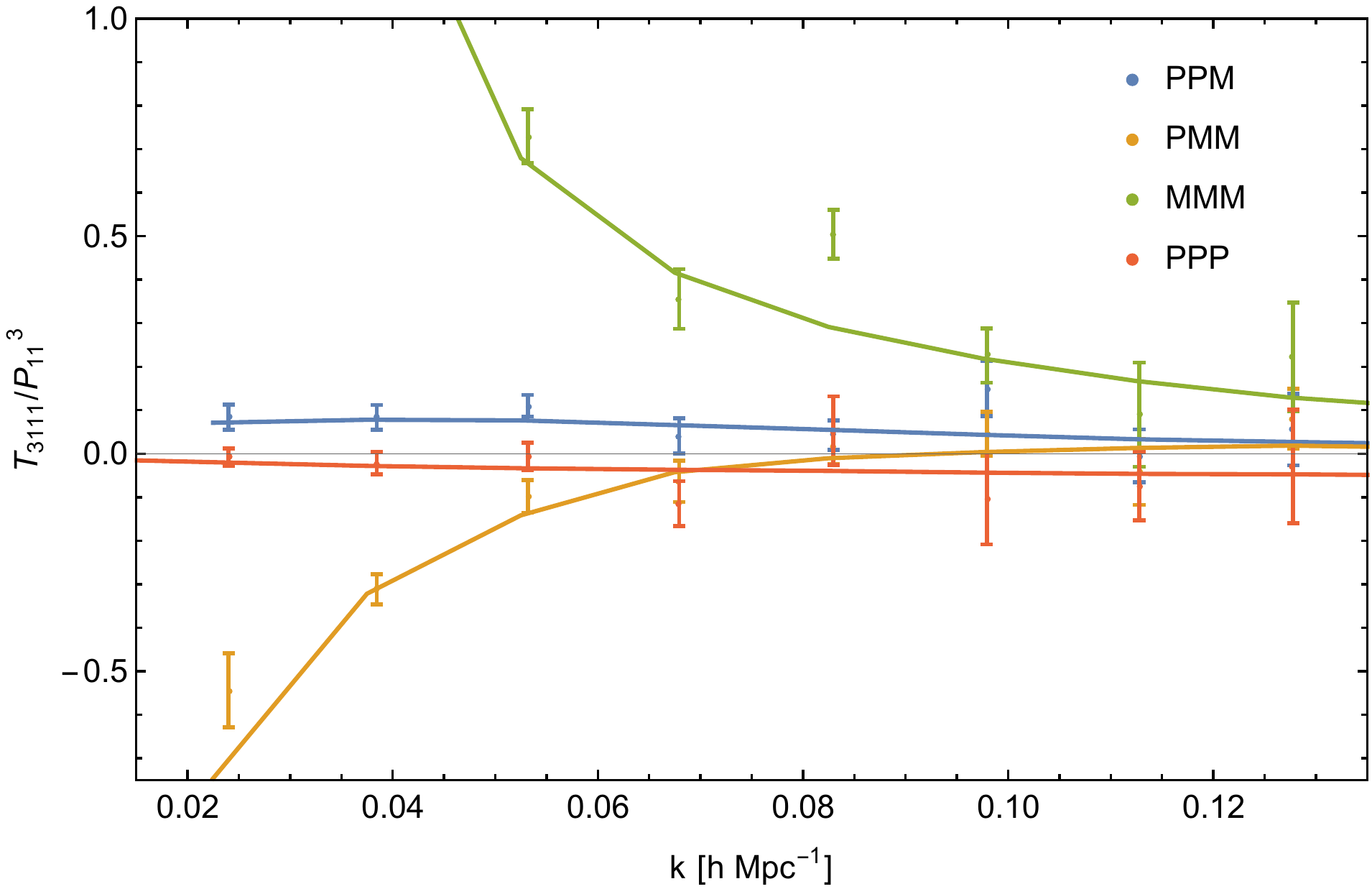}
	\includegraphics[width=0.49\textwidth]{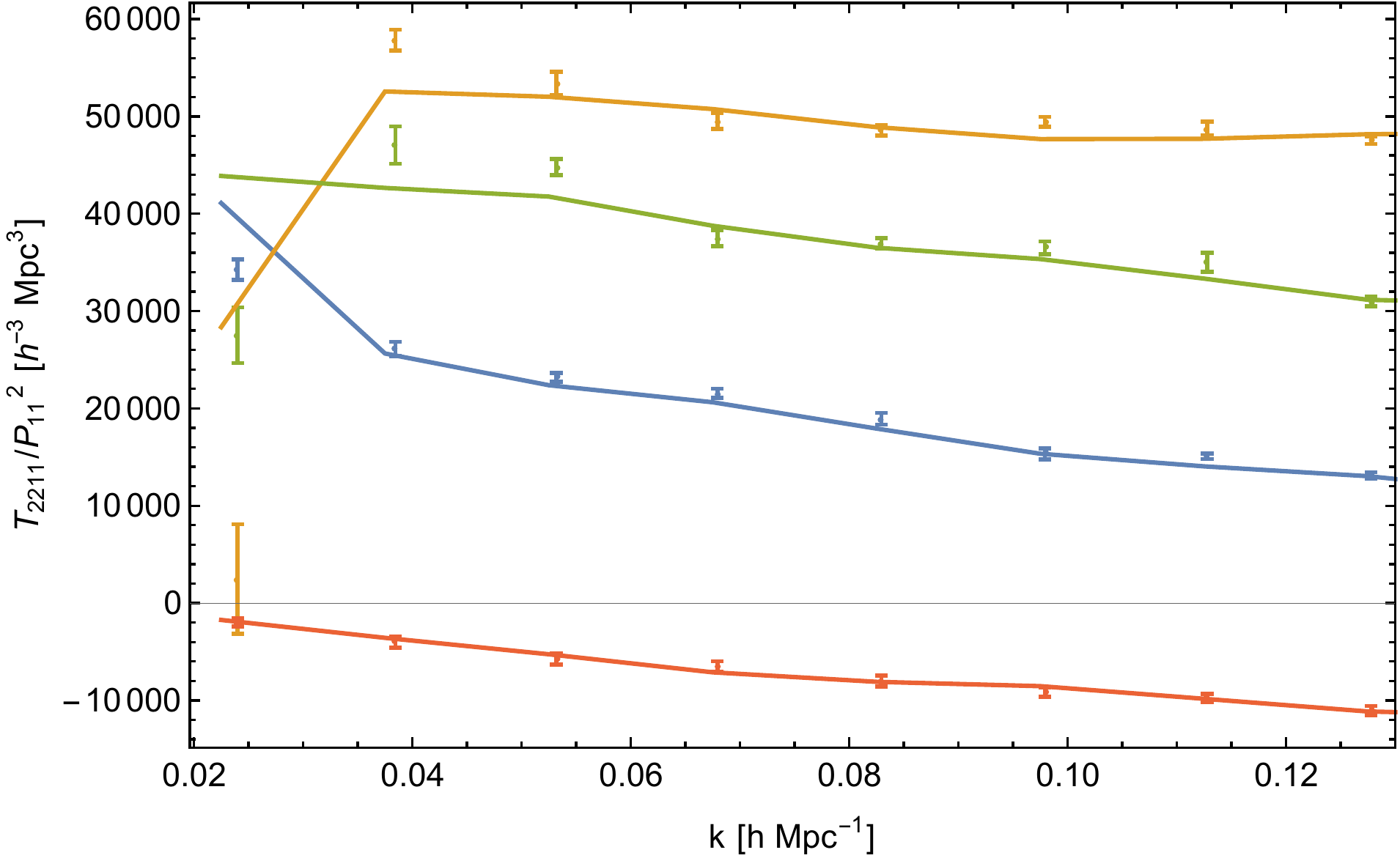}
	\includegraphics[width=0.49\textwidth]{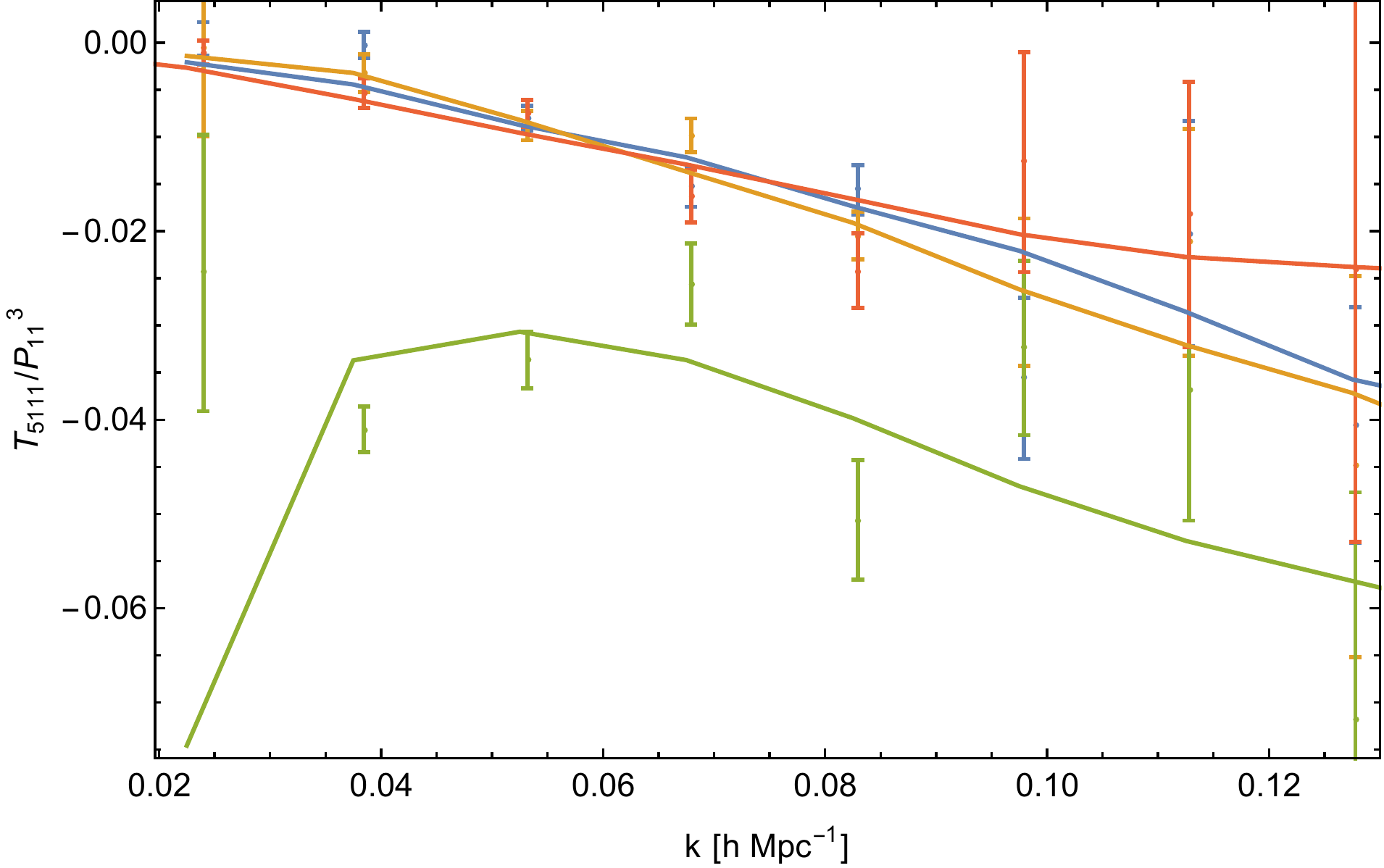}
	\includegraphics[width=0.49\textwidth]{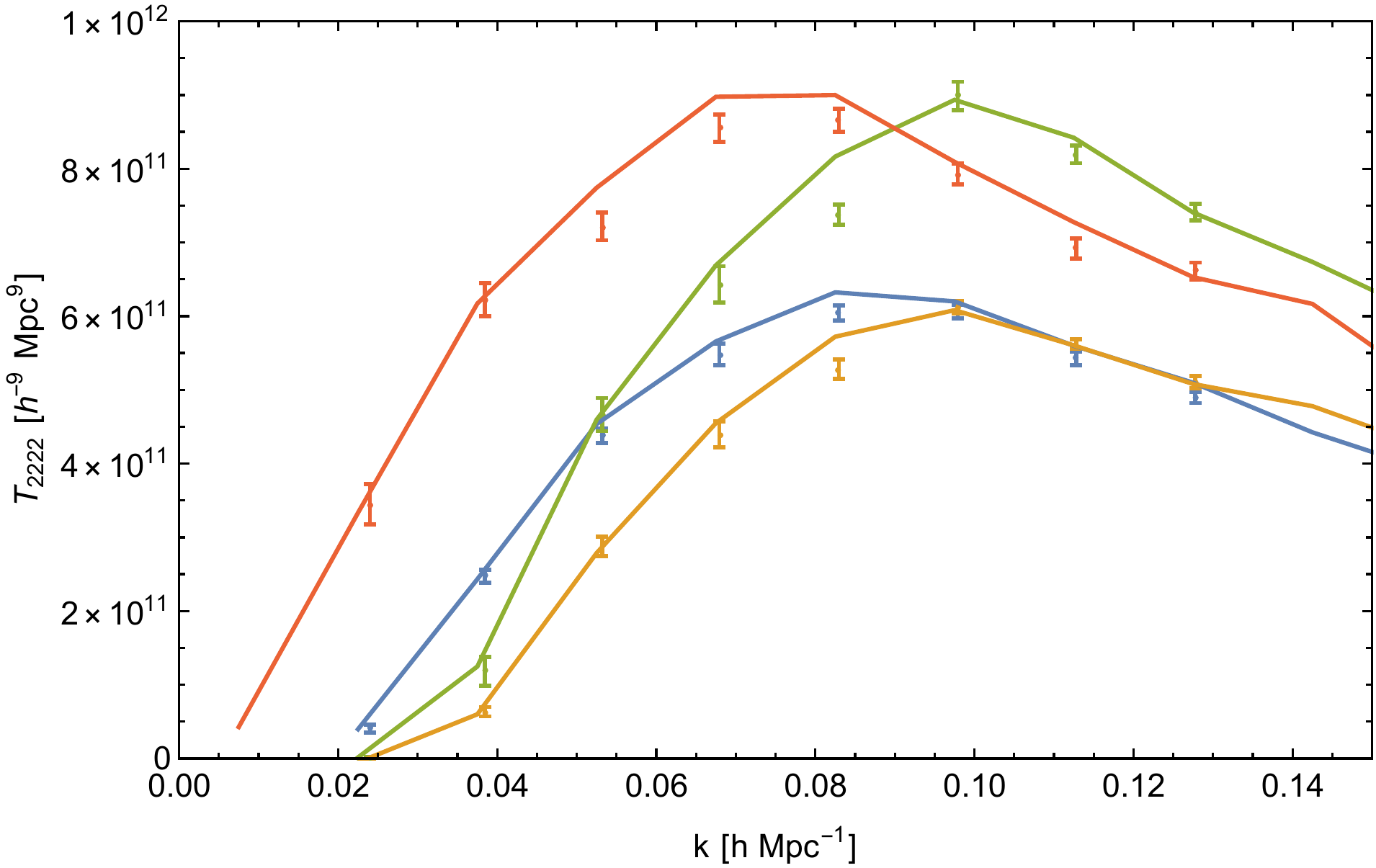}
	\includegraphics[width=0.49\textwidth]{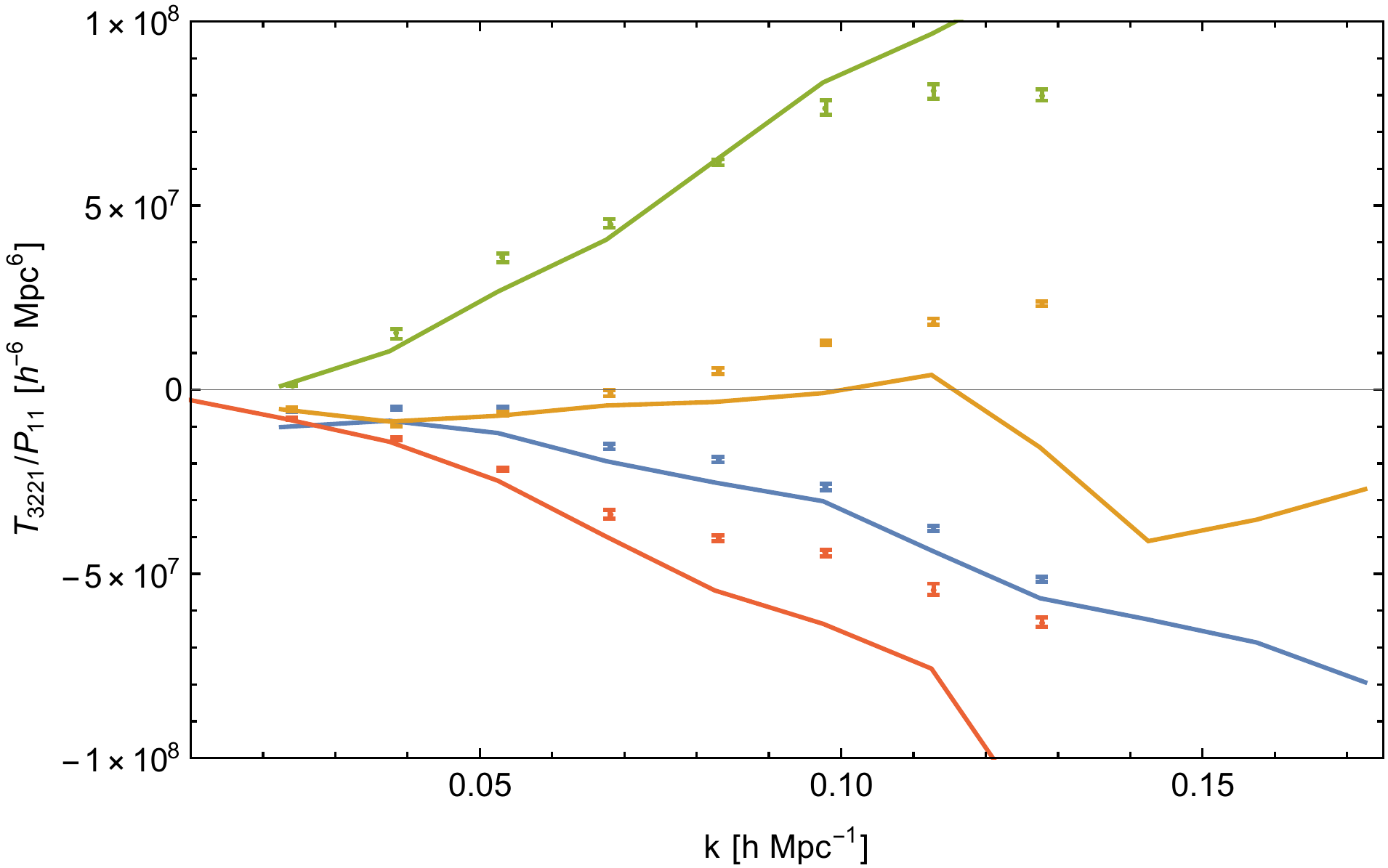}
	\includegraphics[width=0.49\textwidth]{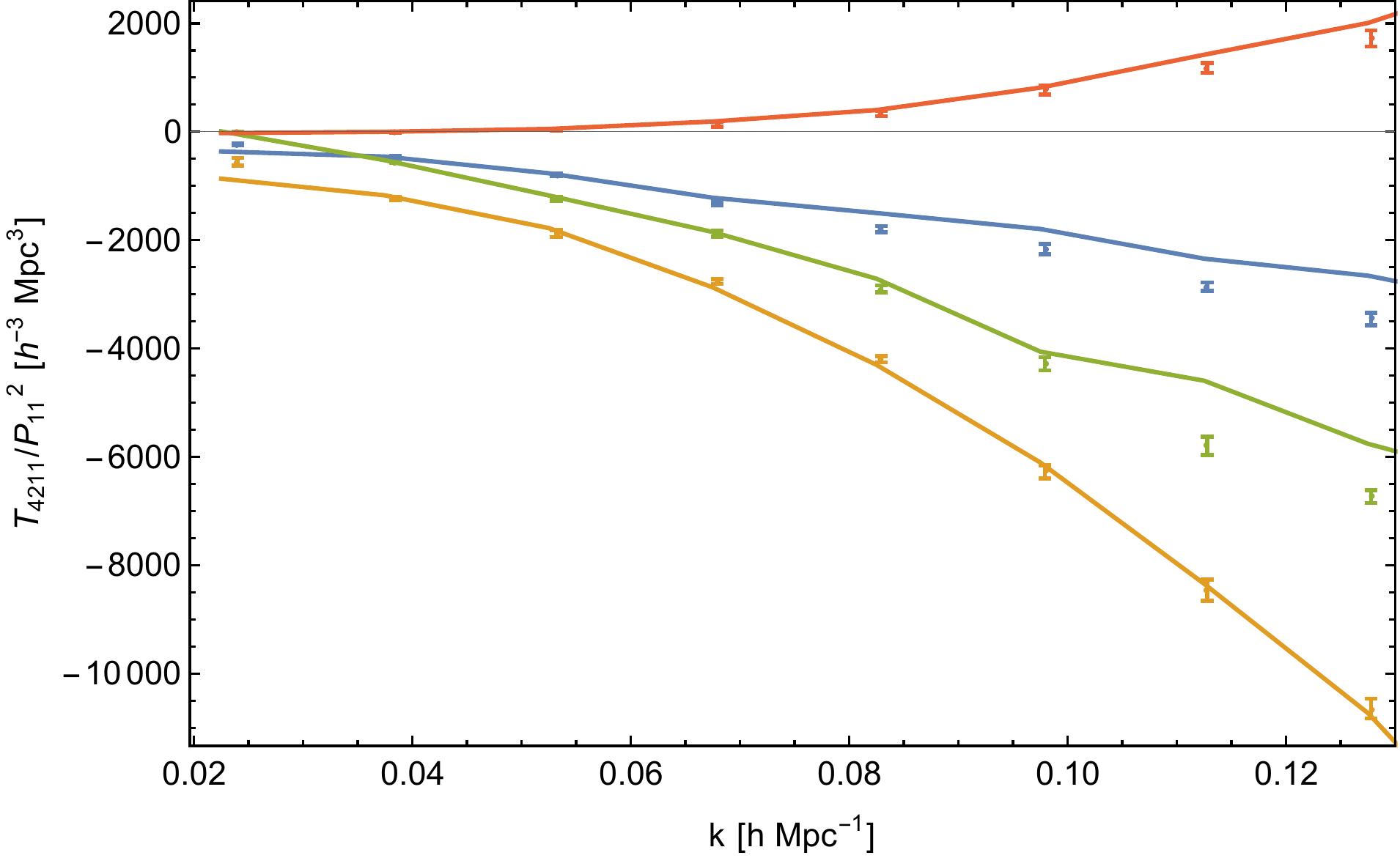}
	\includegraphics[width=0.49\textwidth]{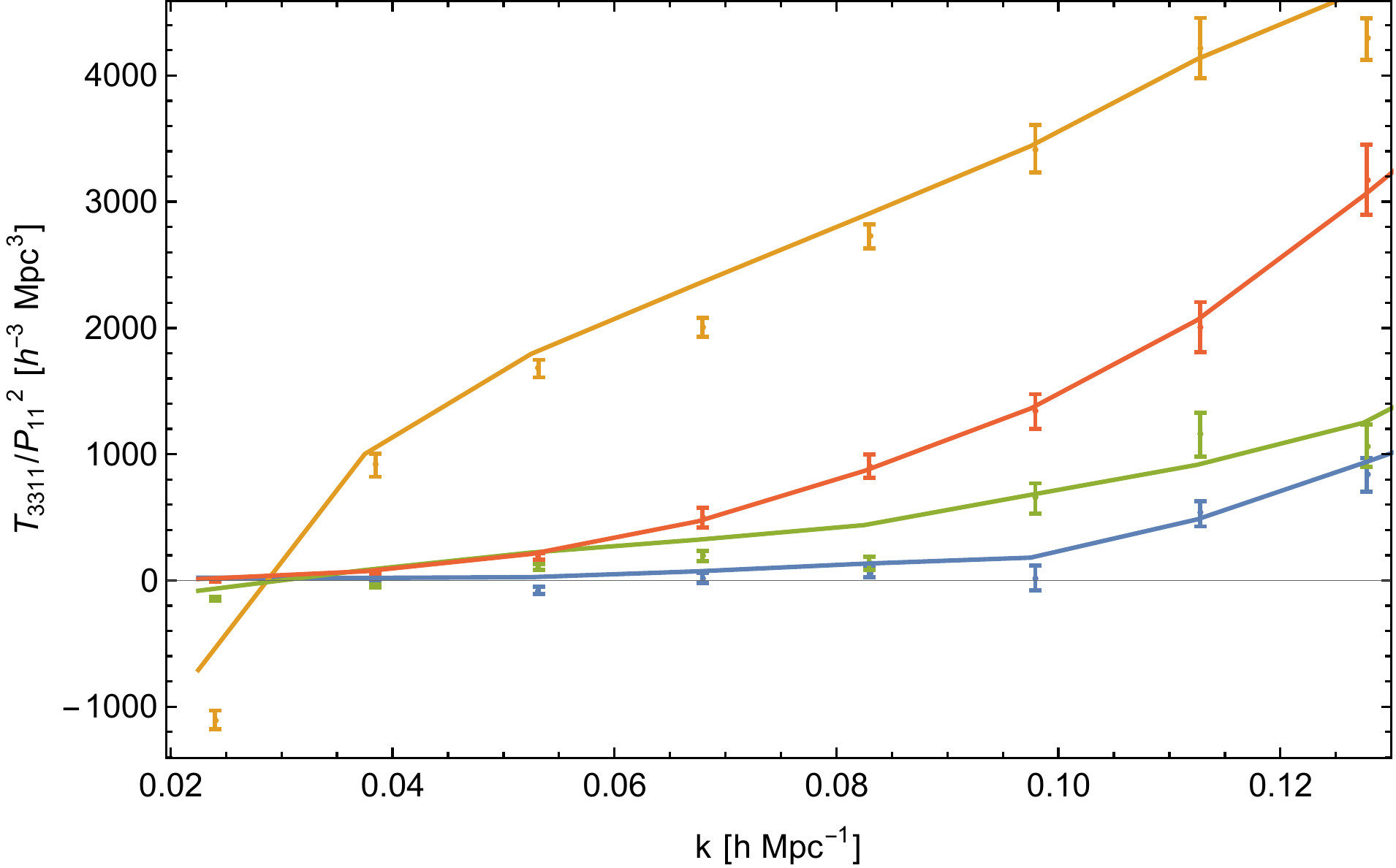}
	\caption{The contributions to the one-loop trispectrum after undergoing CVC through the removal of power spectra as measured in gridPT (points) against the calculations from standard perturbation theory (lines) with a cutoff of $\Lambda=0.3\ihMpc$.  The blue line is configuration PPM, the yellow line is PMM, the green line is MMM and the red line is PPP.}
	    \label{fig:tri}
\end{figure}

Fig.~\ref{fig:tri} shows each of the contributions to the one-loop trispectrum in the four studied configurations with a cutoff of $\Lambda=0.3\ihMpc$ as measured both from the simulations and from perturbation theory.  As can be seen, the grid perturbation theory agrees well with the analytic perturbation theory.  This constitutes a highly non-trivial validation of both our CUBA and gridPT calculations of the trispectra.

\subsection{Theoretical Errors}
Given that we are studying isolated counterterms in partially non-linear trispectra, we need only consider the tree level terms that are appropriate to the correlator under study.  Specifically, for the four non-linear terms we are studying, we have that
\begin{equation}
\begin{split}
\Delta T_{\mathrm{n}111,l}&=T_{3111}\left[\frac{k_\text{ext}}{k_\text{NL}}\right]^{(3+n_{\mathrm{NL}})l}~,~~~~~
\Delta T_{\mathrm{n}211,l}=T_{2211}\left[\frac{k_\text{ext}}{k_\text{NL}}\right]^{(3+n_{\mathrm{NL}})l}~,\\
\Delta T_{\mathrm{n}221,l}&=T_{1221}\left[\frac{k_\text{ext}}{k_\text{NL}}\right]^{(3+n_{\mathrm{NL}})l}~,~~~~~
\Delta T_{\mathrm{n}311,l}=T_{1311}\left[\frac{k_\text{ext}}{k_\text{NL}}\right]^{(3+n_{\mathrm{NL}})l}~,
\end{split}
\end{equation}
where the momentum in the denominator is $k_{\mathrm{ext}}=k_{1}$ because that is the momentum of the one-loop density field and we have chosen $n_{\mathrm{NL}}=-1.3$ to fit the magnitude of our one-loop theoretical errors to our measured one-loop contributions.

\begin{figure}[h!]
	\centering
	\includegraphics[width=0.49\textwidth]{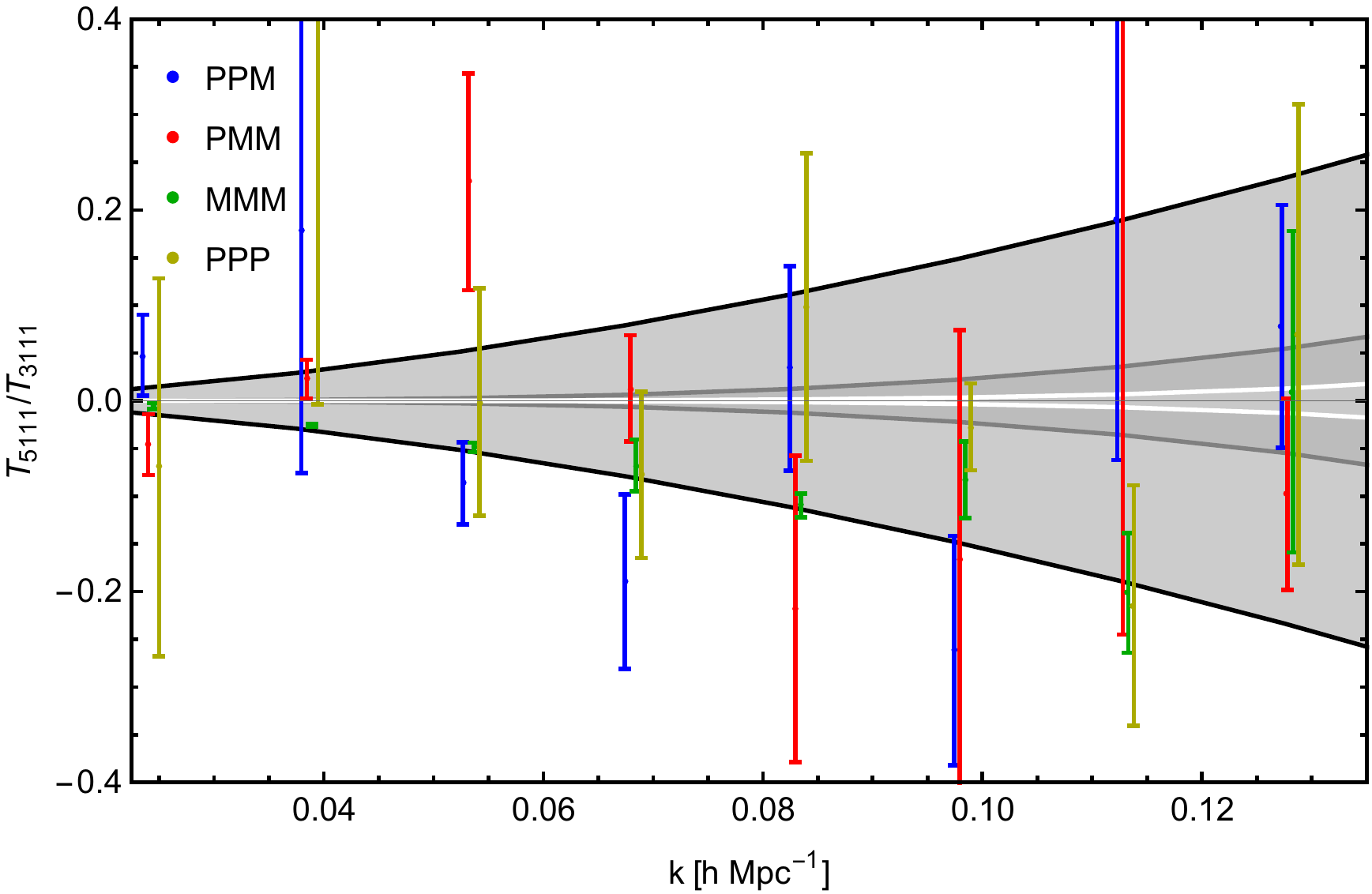}
	\includegraphics[width=0.49\textwidth]{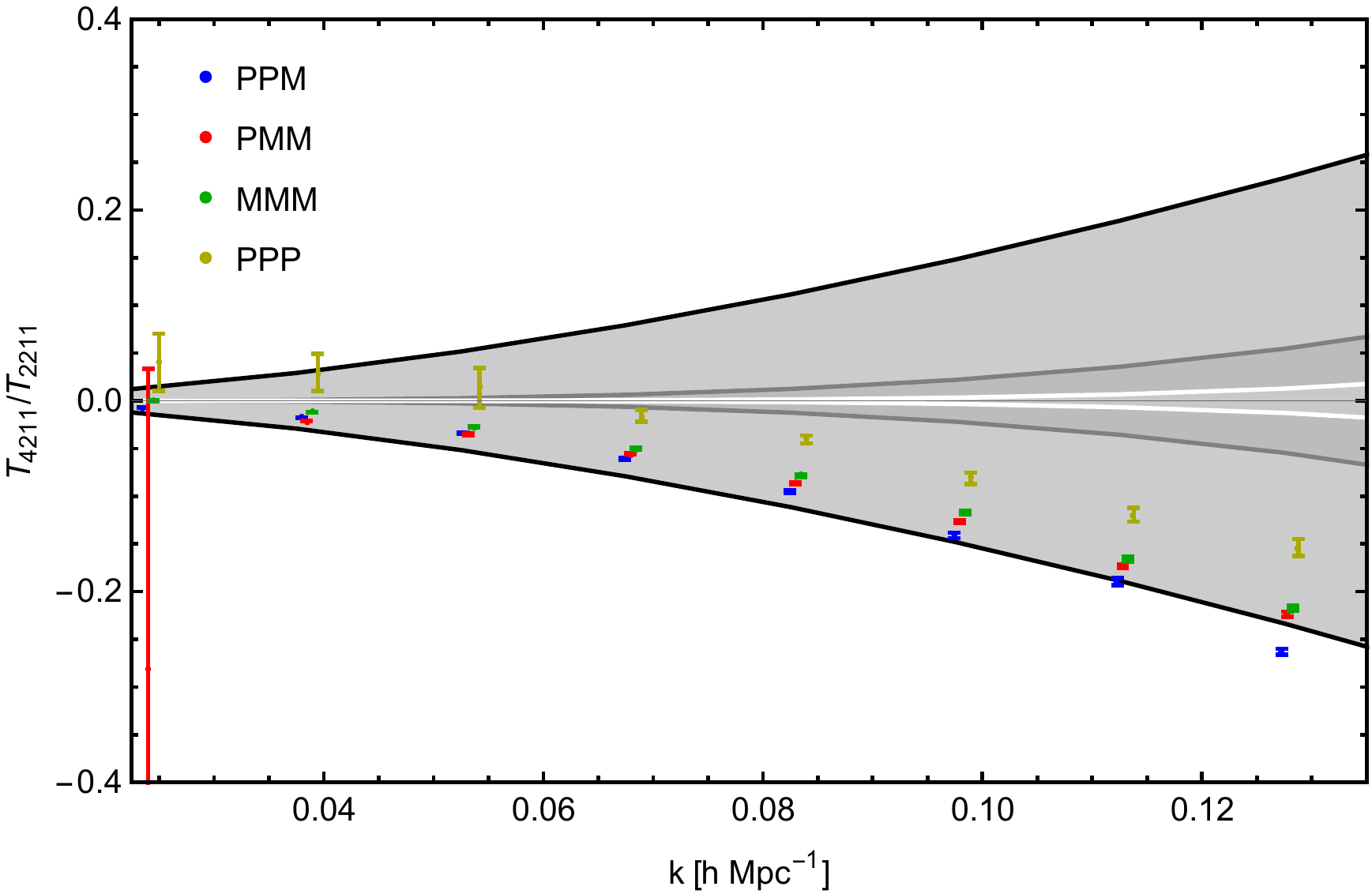}
	\includegraphics[width=0.49\textwidth]{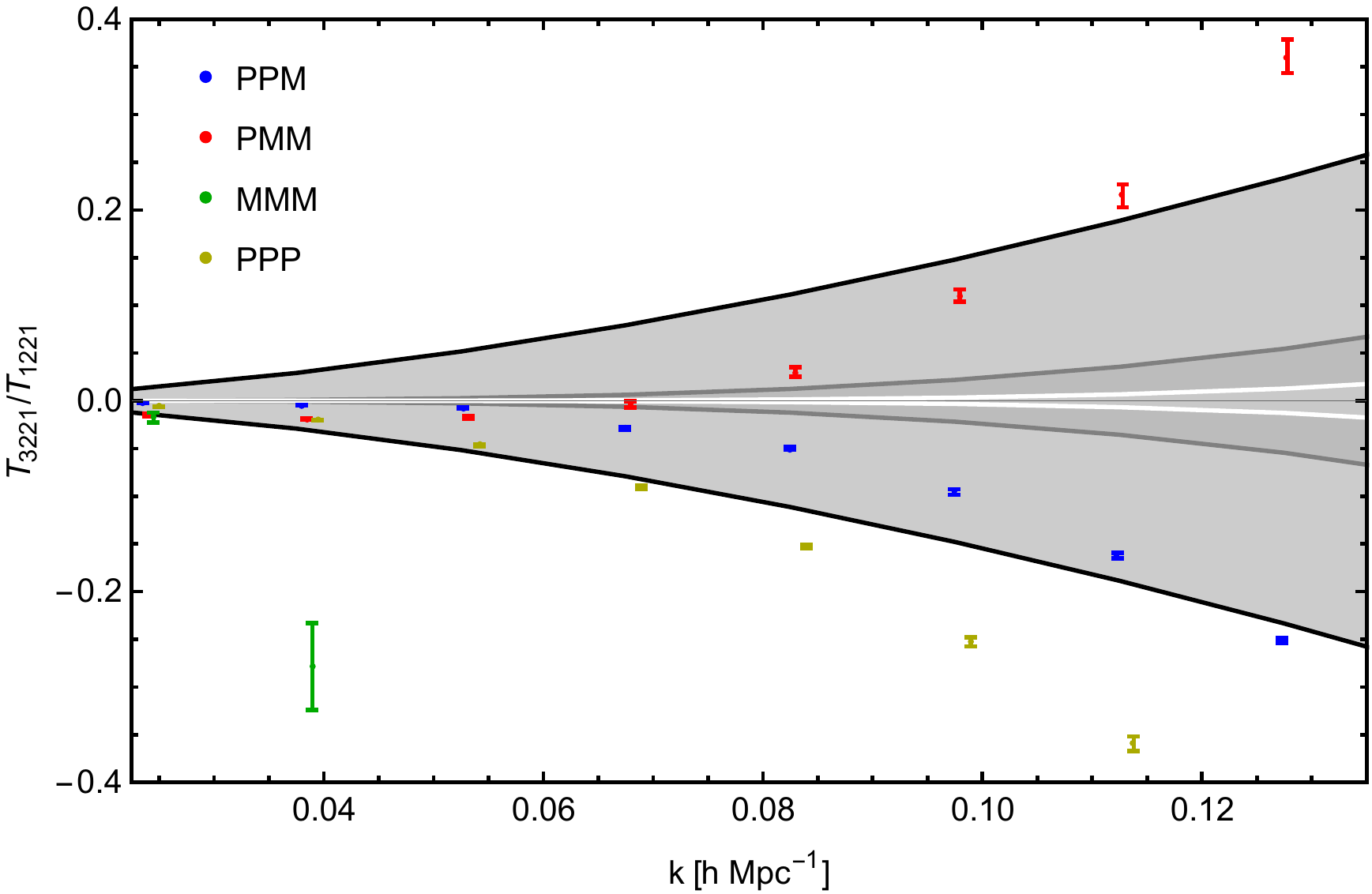}
	\includegraphics[width=0.49\textwidth]{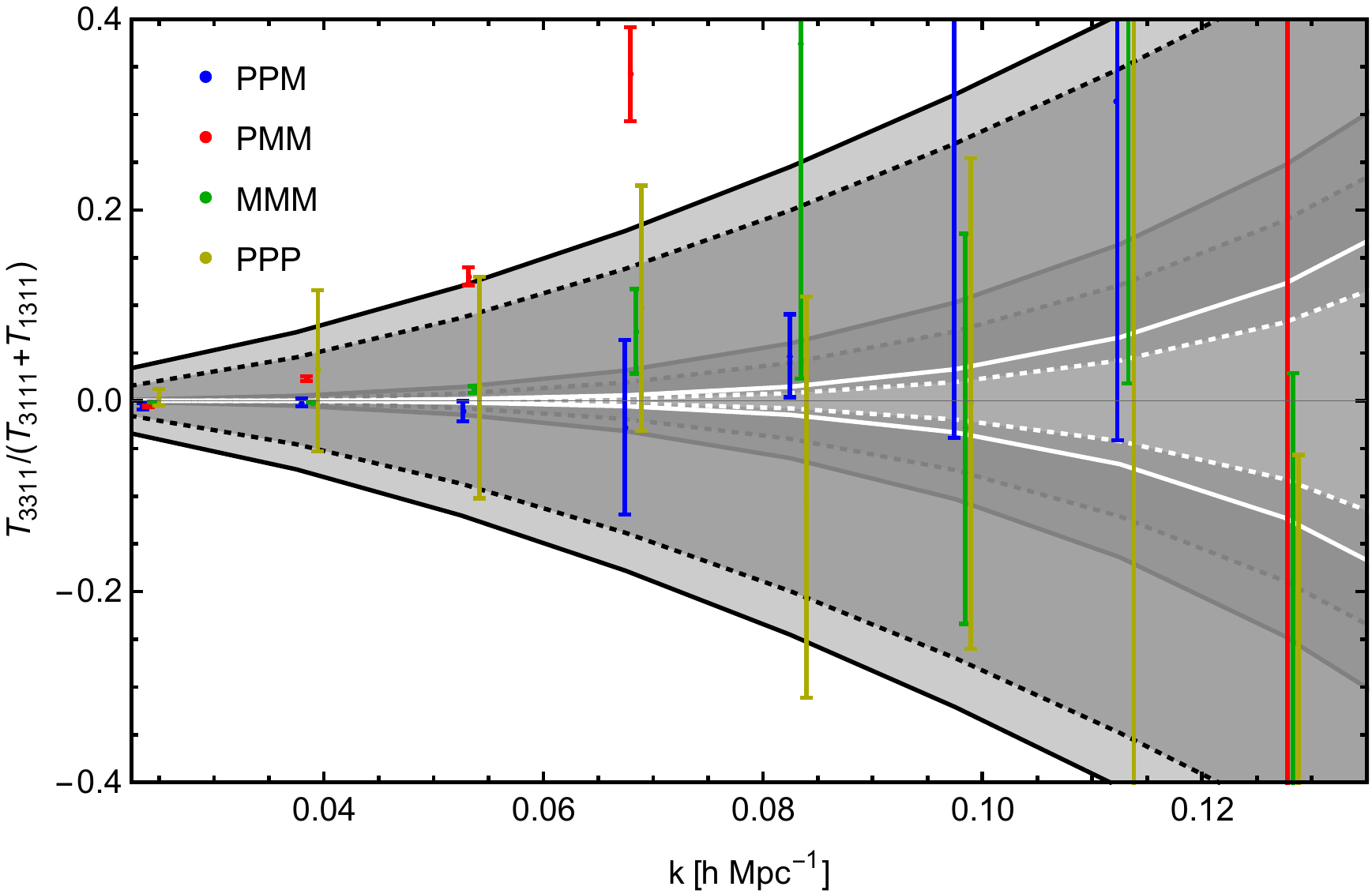}
	\caption{The theoretical errors at one (black), two (grey), and three (white) loops for the ratios one loop terms and their corresponding tree level terms in the four configurations considered.  In the case of $T_{3311}$, the loop momentum is an average of $k_{1}$ and $k_{2}$ such that the configurations PPM, PMM, and PPP have the same theoretical errors while those of MMM differ; those for MMM are shown as dotted lines while those for the other configurations are shown as solid.}
	\label{fig:err}
\end{figure}

As shown in Fig.~\ref{fig:err}, the measured values of the one-loop ratios fall within the expected errors for most configurations; the notable exceptions being the MMM configuration for $T_{3221}$ and the large scale values of the PPP configuration for the same trispectrum.  While undesirable, this is in keeping with the general difficulty we have faced with these configurations.  Also plotted are the expected magnitudes of the two and three loop contributions.  To avoid attempting to use a one-loop analysis in a region heavily affected by higher loop terms, we will generally limit our analysis to $k_{\mathrm{max}}=0.083\ihMpc$.

\subsection{The $\tilde{F}_{1}$ and $\tilde{F}_{2}$ Kernels}
\label{sec:expandedkernels}
In our recent paper \cite{Steele:2020tak} we defined a number of methods with which we constrained the parameters of $\tilde{F}_{1}$ and $\tilde{F}_{2}$ from the one-loop bispectrum.  Here we will take from that paper that $\gamma_{2}=2.332\hMpcsq$, a value obtained using what we referred to as method $B_{\mathrm{nnn}}$-1-S and which we refer to hereafter as $\gamma_{2,\mathrm{B}}$, leaving us with two parametrisations for the $\epsilon$ parameters of $\tilde{F}_{2}$ seen in Eq.~\ref{eq:symmf2tilde}, which we refer to as the UV and symmetry inspired parametrisations \footnote{In the UV inspired parametrisation we have $\epsilon_{1}=0$, $\epsilon_{2}=-0.565$, and $\epsilon_{3}=-1.699$, while in the symmetry inspired parametrisation we have $\epsilon_{1}=0.618$, $\epsilon_{2}=0.517$, and $\epsilon_{3}=2.978$.}, together with the approximation $\gamma_{1}=c_\text{s}^{2}$, giving a value of $\gamma_{1}=2.27\hMpcsq$.  We also take the UV approximations of the counterterms and of the components of $\tilde{F}_{2}$ calculated on the grid and fit them as counterterms, both with the values of $\gamma_{1}$ and $\gamma_{2}$ just mentioned and with independent fits to their respective residuals. 

With these parameters, we use the following methods for estimating the trispectrum counterterms containing $\tilde{F}_{1}$ and $\tilde{F}_{2}$:
\begin{itemize}
    \item We will use the chosen value of $\gamma_{2,\mathrm{B}}$ together with both corresponding parametrisations for the $\epsilon_{i}$ to calculate $T_{\tilde{2}211}$ perturbatively.
    \item We will use the chosen value of $c_{s}^{2}$ calibrated from the bispectrum  to calculate  $T_{\tilde{1}221}$ and $T_{\tilde{1}311}$ perturbatively.
    \item We will use the chosen value of $c_{s}^{2}$ as the amplitude of $\gamma_{1} T_{\tilde{1}221,\mathrm{grid}}$ and $\gamma_{1} T_{\tilde{1}311}$.
    \item We will use the chosen value of $\gamma_{2,\mathrm{B}}$ as the amplitude of $\gamma_{2} T_{\tilde{2}211,\mathrm{grid}}$.
    \item We will use a values of $\gamma_{2,\mathrm{B}}$ and $\epsilon_{1,2,3}$ as calibrated from the bispectrum as the amplitudes for the constituent functions of $\tilde{F}_{2}$ measured on the grid and sum these to estimate $T_{\tilde{2}211}$ with both symmetry and UV inspired parametrisations.  We refer to the parameters from the symmetry inspired parametrisation with the subscript S and those with the UV inspired parametrisation with the subscript U.
    \item We will fit for the amplitude of $\gamma_{1} T_{\tilde{1}221,\mathrm{grid}}$.
    \item We will fit for the amplitude of $\gamma_{1} T_{\tilde{1}311,\mathrm{grid}}$.
    \item We will fit for the amplitude of $\gamma_{2} T_{\tilde{2}211,\mathrm{grid}}$.
\end{itemize}

\begin{figure}[h!]
	\centering
	\includegraphics[width=0.49\textwidth]{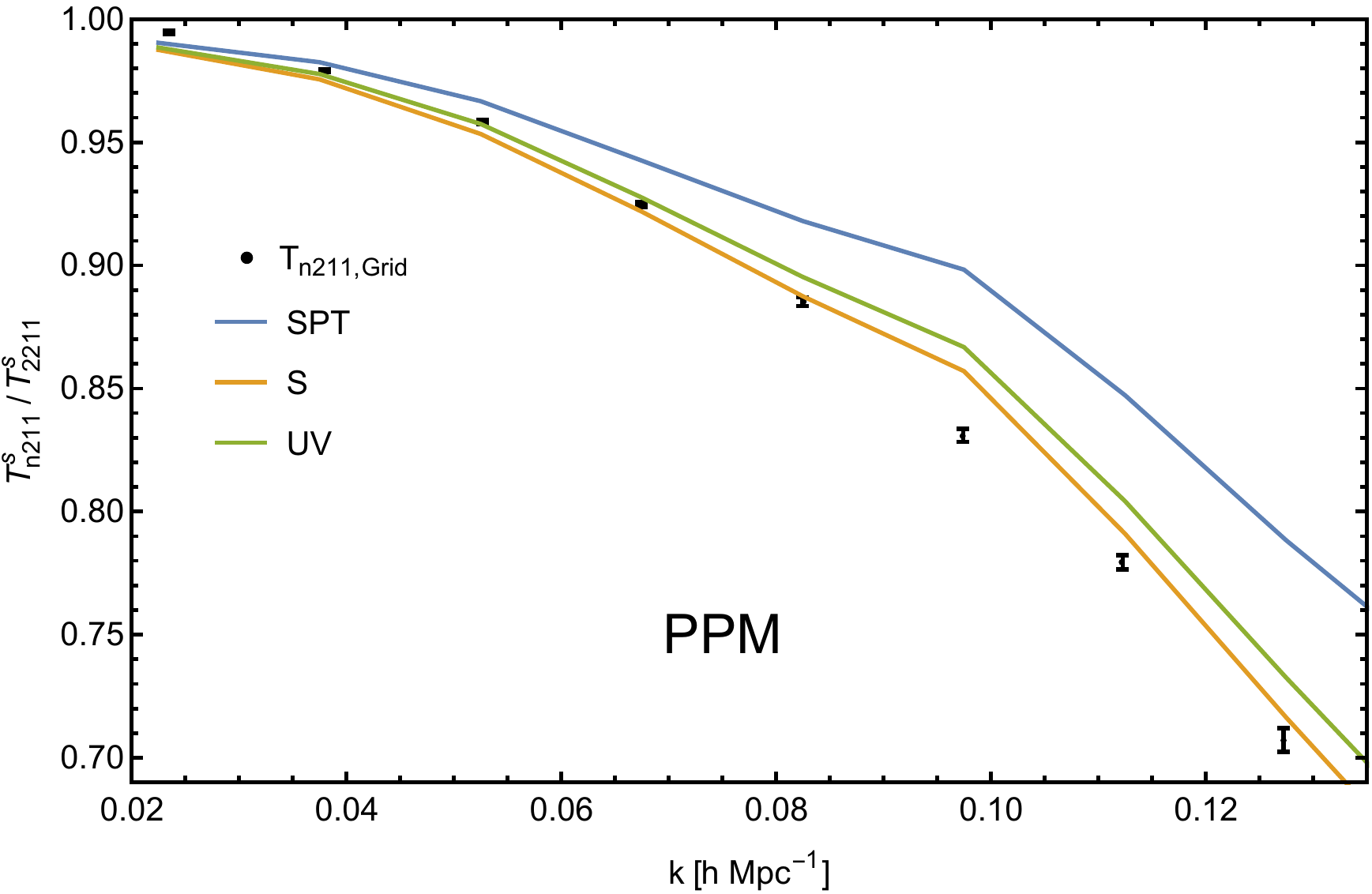}
	\includegraphics[width=0.49\textwidth]{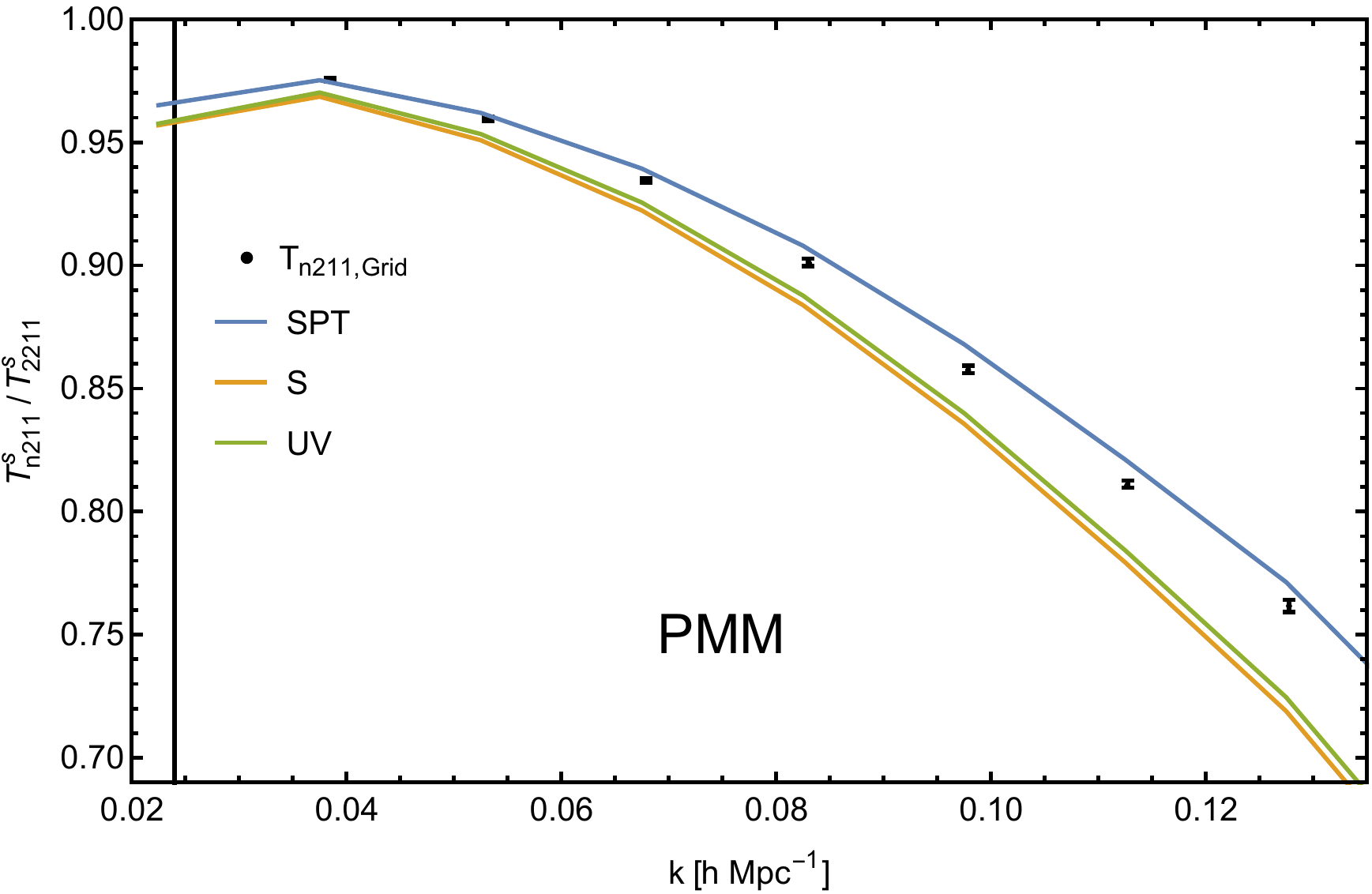}
	\includegraphics[width=0.49\textwidth]{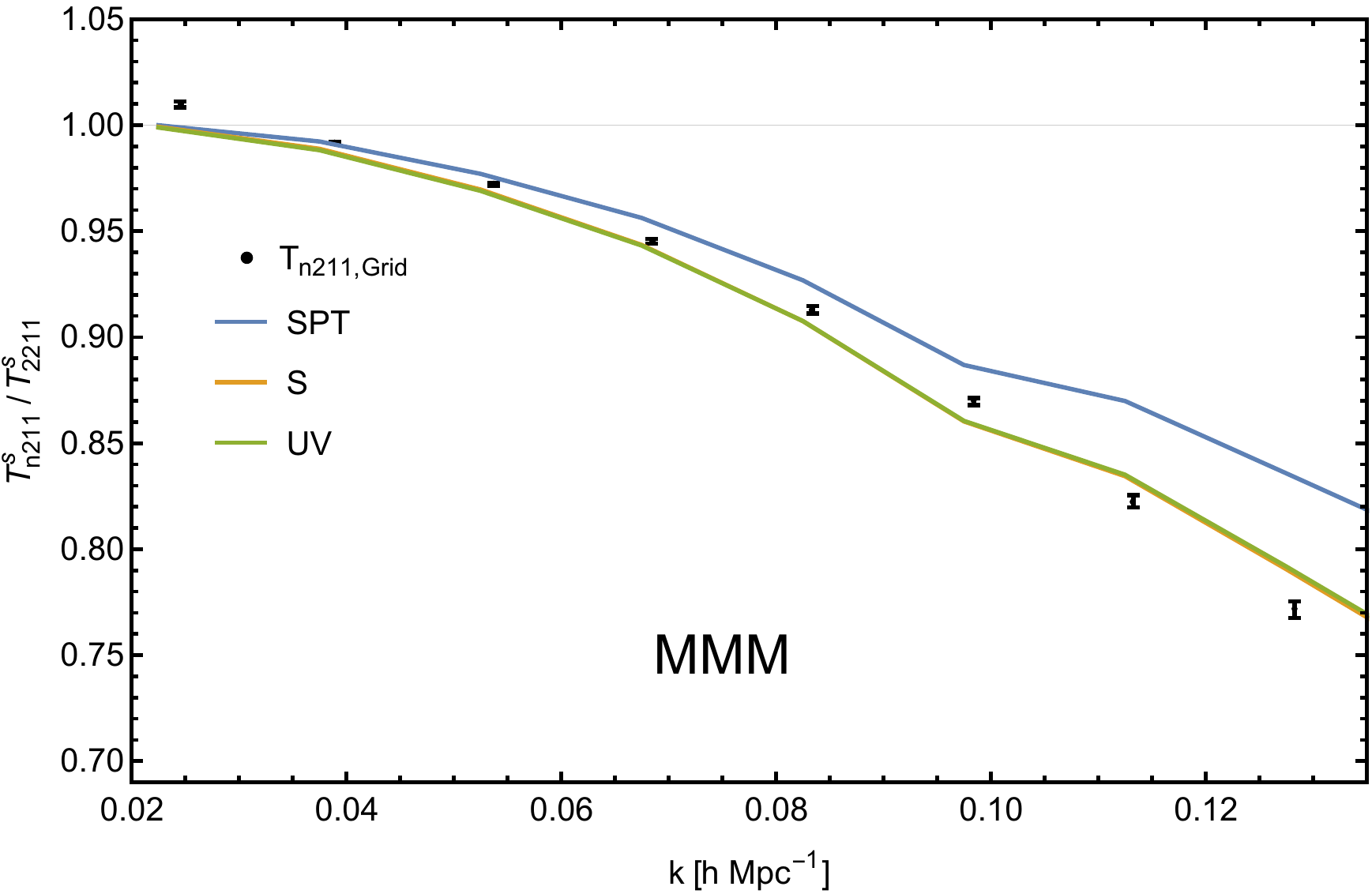}
	\includegraphics[width=0.49\textwidth]{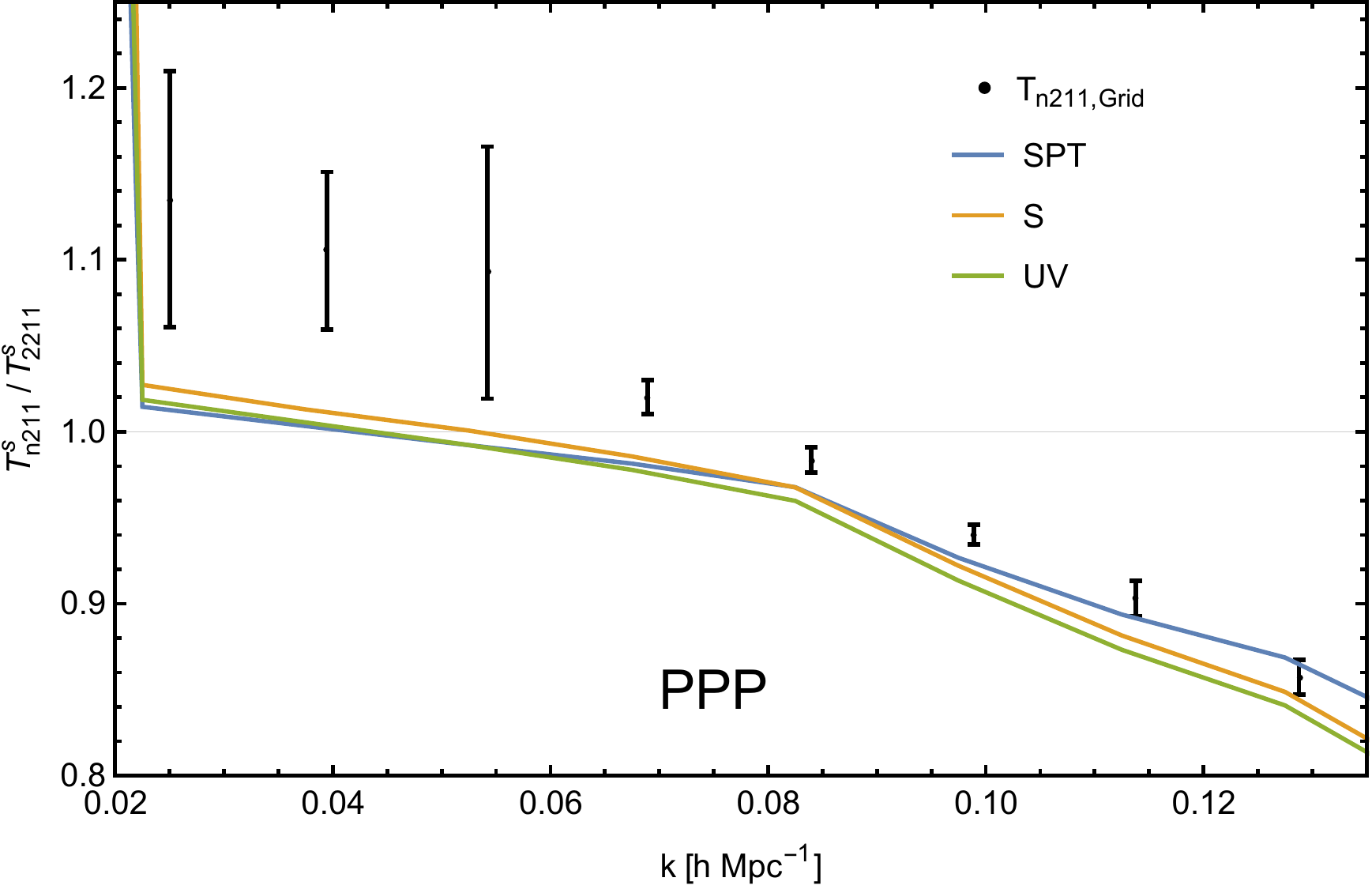}
	\caption{The calculated $T_{2211}+T_{4211}+T_{\tilde{2}211}$ with $\tilde{F}_{2}$ as calibrated from the bispectrum against the grid $T_{\mathrm{n}211}-T_{3211}-T_{1211}$ for the four configurations studied.  For PPM and MMM, the corrections are significant.  In the former case it is notable that the symmetry inspired fit works better than the UV inspired fit, a relation that was also seen in the study of the bispectra \cite{Steele:2020tak}, while for MMM the two parametrisations of the counterkernel seem to work equally well.  For PMM, the correction seems to be too large and is of debatable value, while for PPP the correction is extremely small. As this is a ratio over the tree-level terms, only studying perturbation theory up to tree level would have resulted in a constant line at one.  Configurations PPM, PMM, and MMM can only be described by this at the largest scales, deviating from one already by $k\sim0.03\ihMpc$, while the measurements in configuration PPP do not agree with tree level predictions at any scale.}
	\label{fig:T2t211b}
\end{figure}

The counterterms calculated in this way for $T_{\tilde{2}211}$ are plotted alongside the residuals in Fig.~\ref{fig:T2t211b} for the four considered configurations.  We can see that the UV approximations produce far more accurate results in analytic perturbation theory than the symmetry inspired fits, with the UV inspired models successfully approximating the residual on all scales for configurations PPM and PMM and on all but the largest scales for MMM. However, for PPP we found that we were unable to successfully regularise $T_{\mathrm{n}211}$ at any scale. These inaccuracies could be the result of simulation or analysis systematics which will be explored further in a future paper. We will also explore explicit trispectrum configurations without integrating over the diagonal legs in order to see if this averaging played a role in these errors.  In addition, we find that the two-loop errors are much larger than might have been expected, encompassing many of the counterterm calculations, indicating that two-loop terms would need to be taken into account for an accurate and precise calibration of these terms.

\begin{figure}[t]
	\centering
	\includegraphics[width=0.49\textwidth]{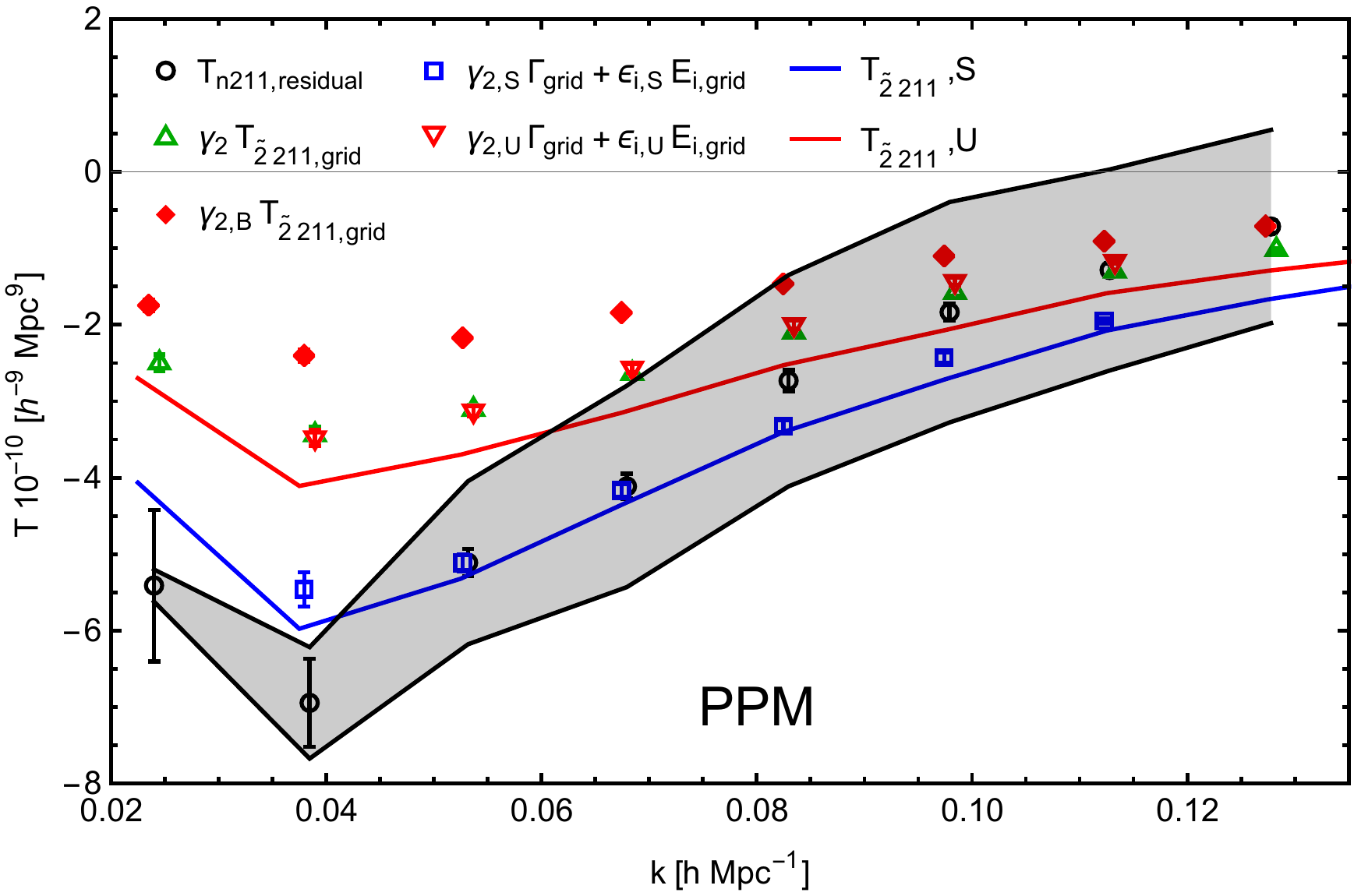}
	\includegraphics[width=0.49\textwidth]{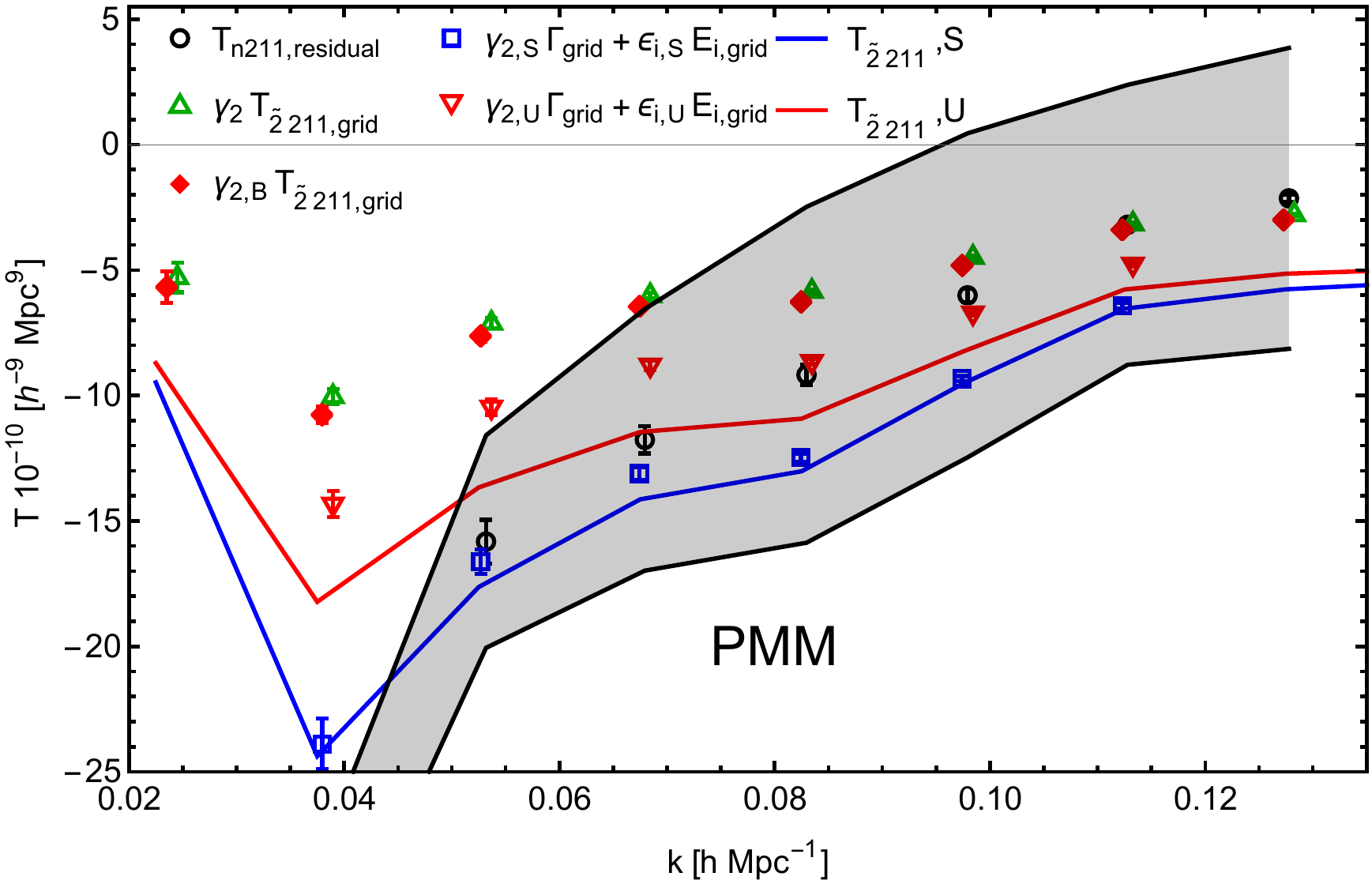}
	\includegraphics[width=0.49\textwidth]{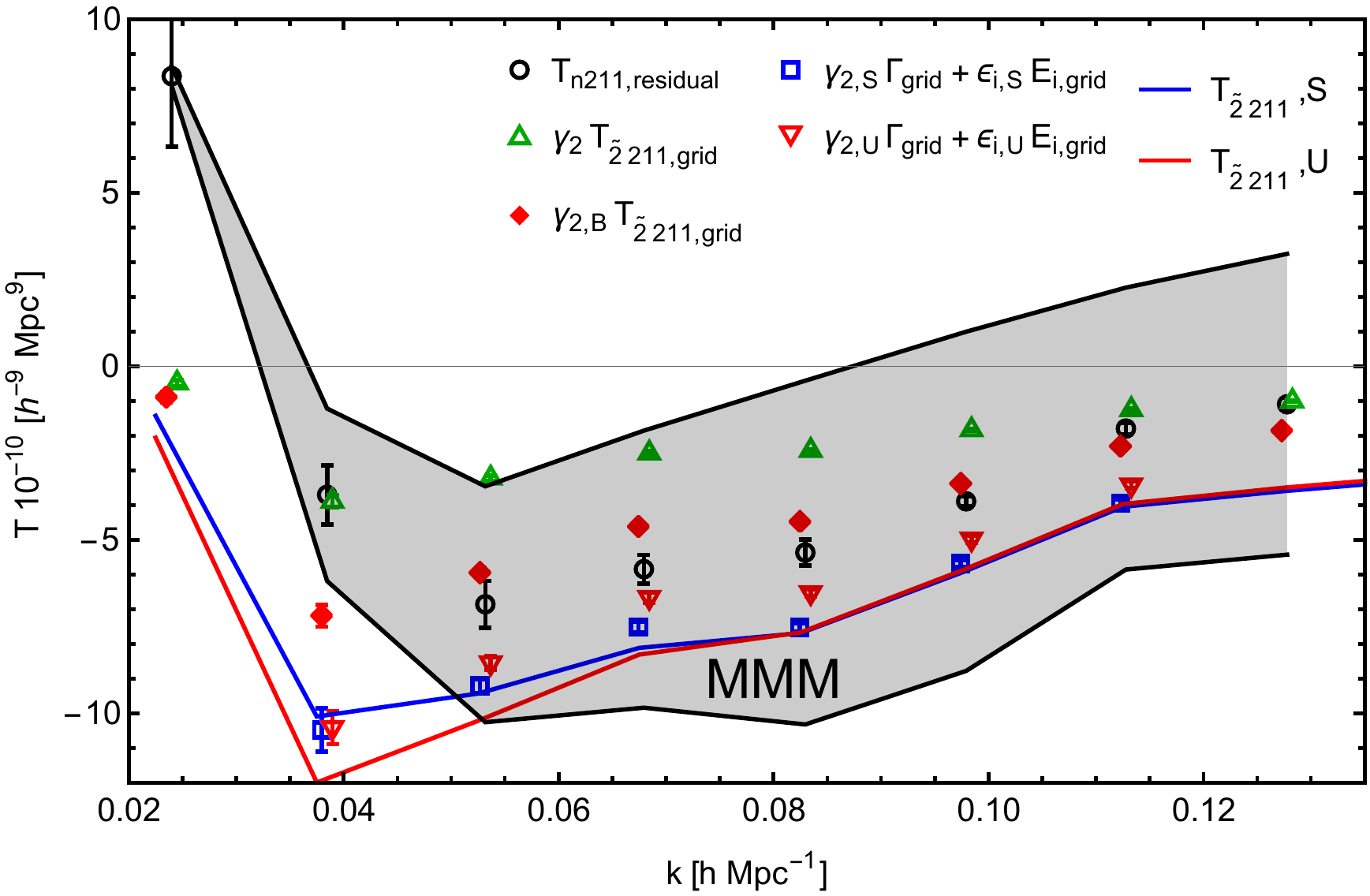}
	\includegraphics[width=0.49\textwidth]{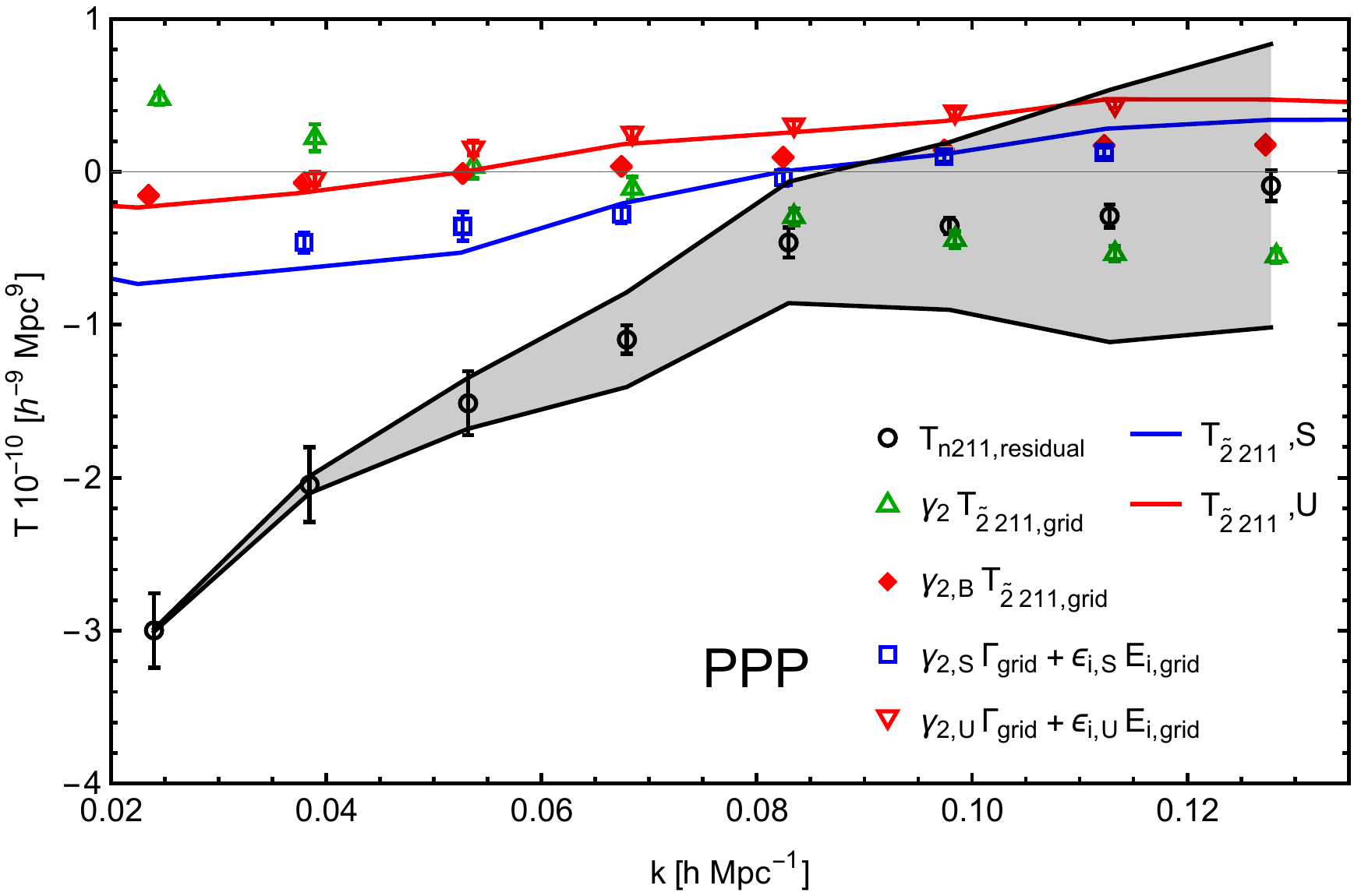}
	\caption{The calculated counterterms $T_{\tilde{2}211}$ with $\tilde{F}_{2}$ as calibrated from the bispectrum against the grid residual for configurations PPM (top left), PMM (top right), MMM (bottom left), PPP (bottom right).  The grey shaded region shows the theoretical error induced by two-loop corrections.  We apply a growth factor correction of $\Delta D_2\approx0.005$ inspired by our bispectrum measurements in \cite{Steele:2020tak} to account for time integration inaccuracies in the $N$-body results. }
\end{figure}

\begin{figure}[b]
	\centering
	\includegraphics[width=0.49\textwidth]{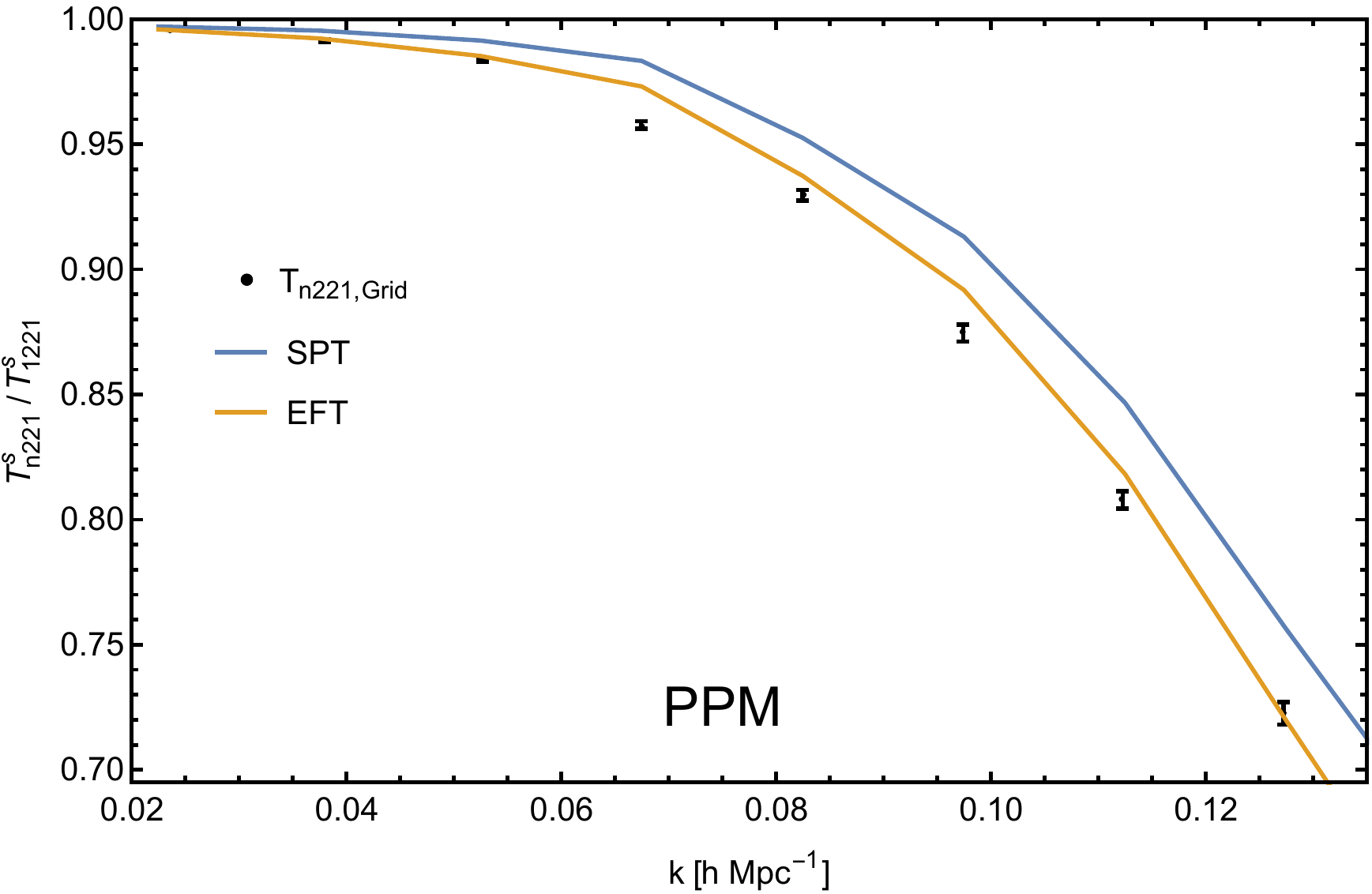}
	\includegraphics[width=0.49\textwidth]{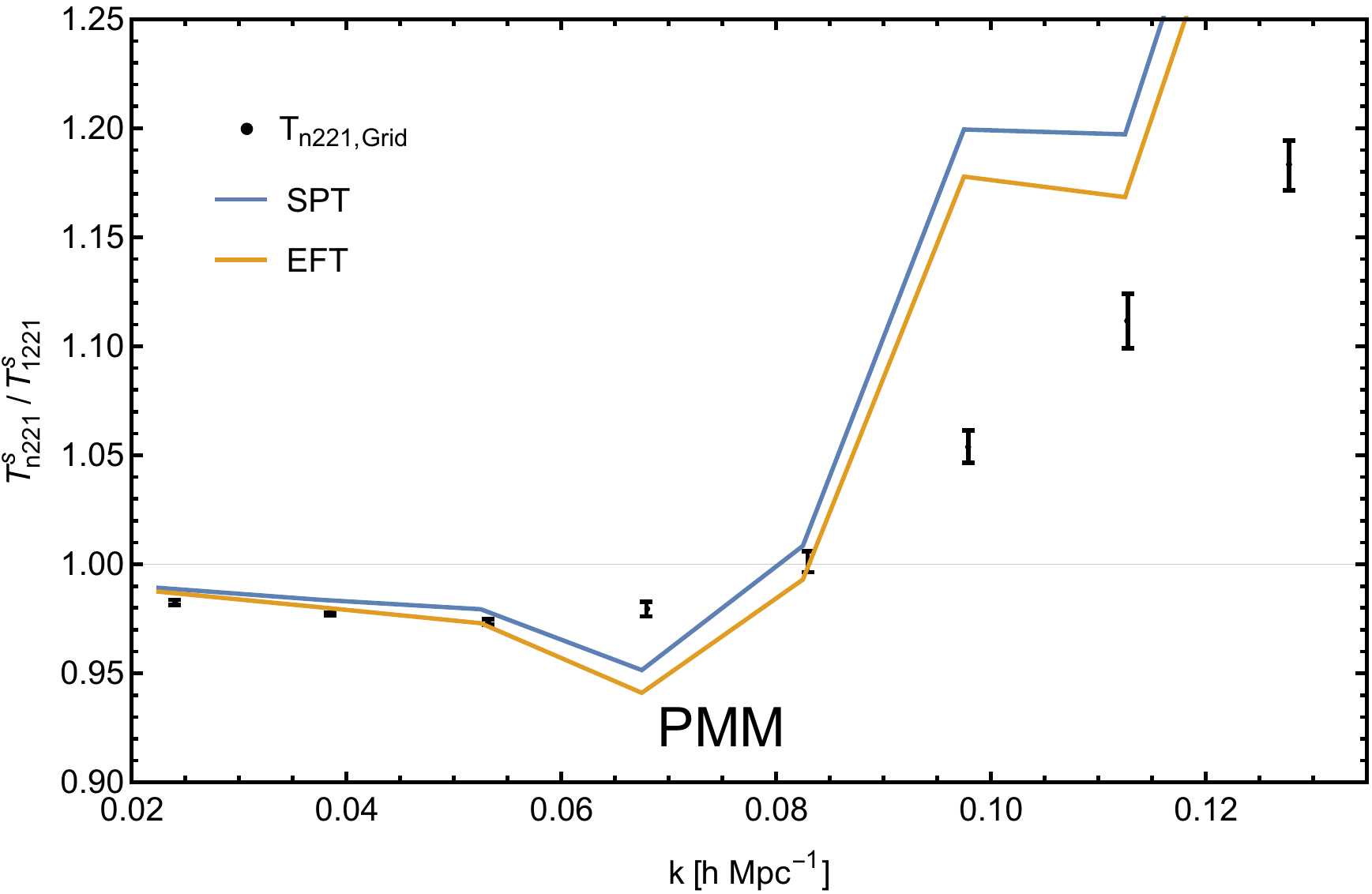}
	\includegraphics[width=0.49\textwidth]{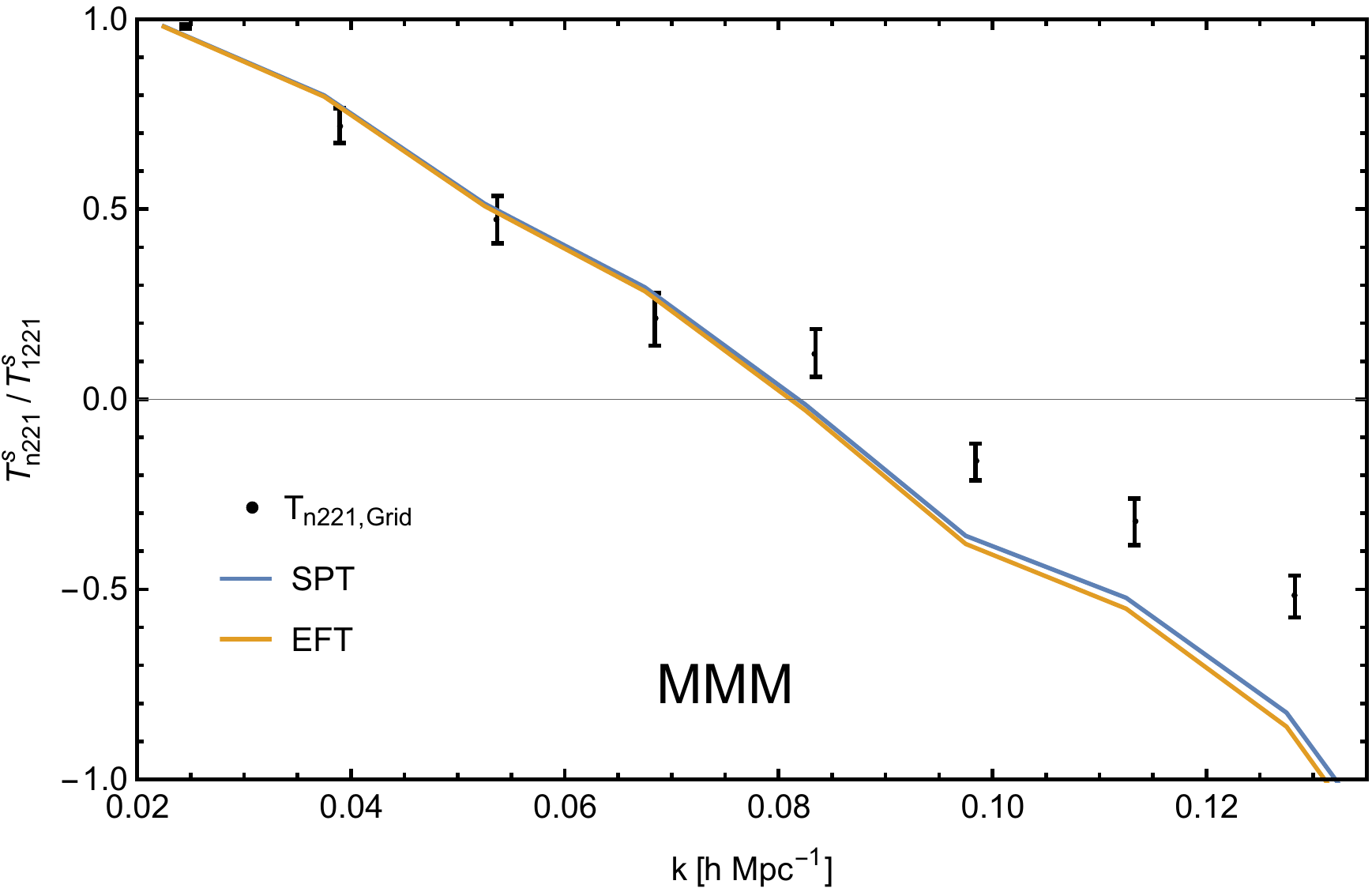}
	\includegraphics[width=0.49\textwidth]{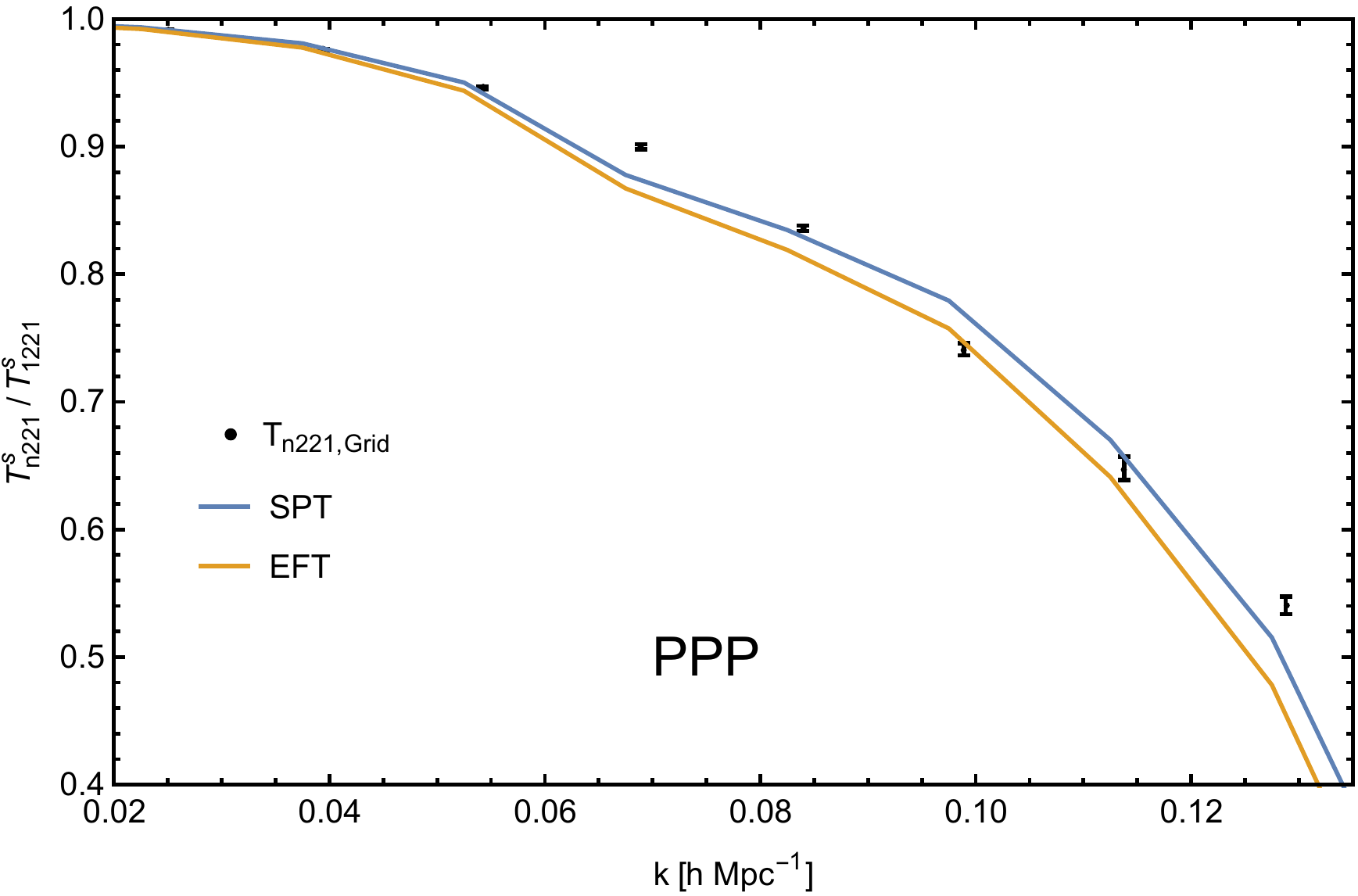}
	\caption{The calculated $T_{1221}+T_{3211}+T_{\tilde{1}221}$ with $\tilde{F}_{1}$ as calibrated from the bispectrum against the grid $T_{\mathrm{n}221}-T_{2221}$ for the four configurations studied.  The tree-level trispectrum corresponds to a horizontal line at one. The measurements clearly deviate from that starting from the largest scales for configurations PPM and PMM, with PPM and PPP deviating from tree level beginning at least at $k\sim0.03\ihMpc$.}
	\label{fig:T1t221b}
\end{figure}

The counterterms calculated in this way for $T_{\tilde{1}221}$ are plotted alongside the residuals in Fig.~\ref{fig:T1t221b} for the four considered configurations.  We see that the regularisation using  a number of methods works on very large scales but rapidly ceases to work beyond roughly $k\sim 0.04\ihMpc$, perhaps indicating that two loop terms make a much larger contribution to the trispectrum than they do to the bispectrum, where they do not have a large effect until roughly $k\sim 0.08\ihMpc$.  Most noticeably, in the case of configuration MMM the fit only works for the UV parametrisations with $\gamma_{1}=\cssq$ and even then only until $k\approx 0.04\ihMpc$, beyond which these models do not even have the same sign or shape as the residual.  The configuration dependence of our ability to regularise this trispectrum is noticeable; for configuration PPM we find that method 3 provides a very good regularisation up to about $k\sim 0.1\ihMpc$ and that the other two methods fail, while for configuration MMM we find that method 3 fails at all scales and that methods 1 and 2 provide good fits but only at very large scales.  Furthermore, the theoretical error envelope that predicts the magnitude of the two loop terms encompasses most of our results, indicating that without taking two loop terms into account, our results are likely to be inaccurate.

\begin{figure}[t]
	\centering
	\includegraphics[width=0.49\textwidth]{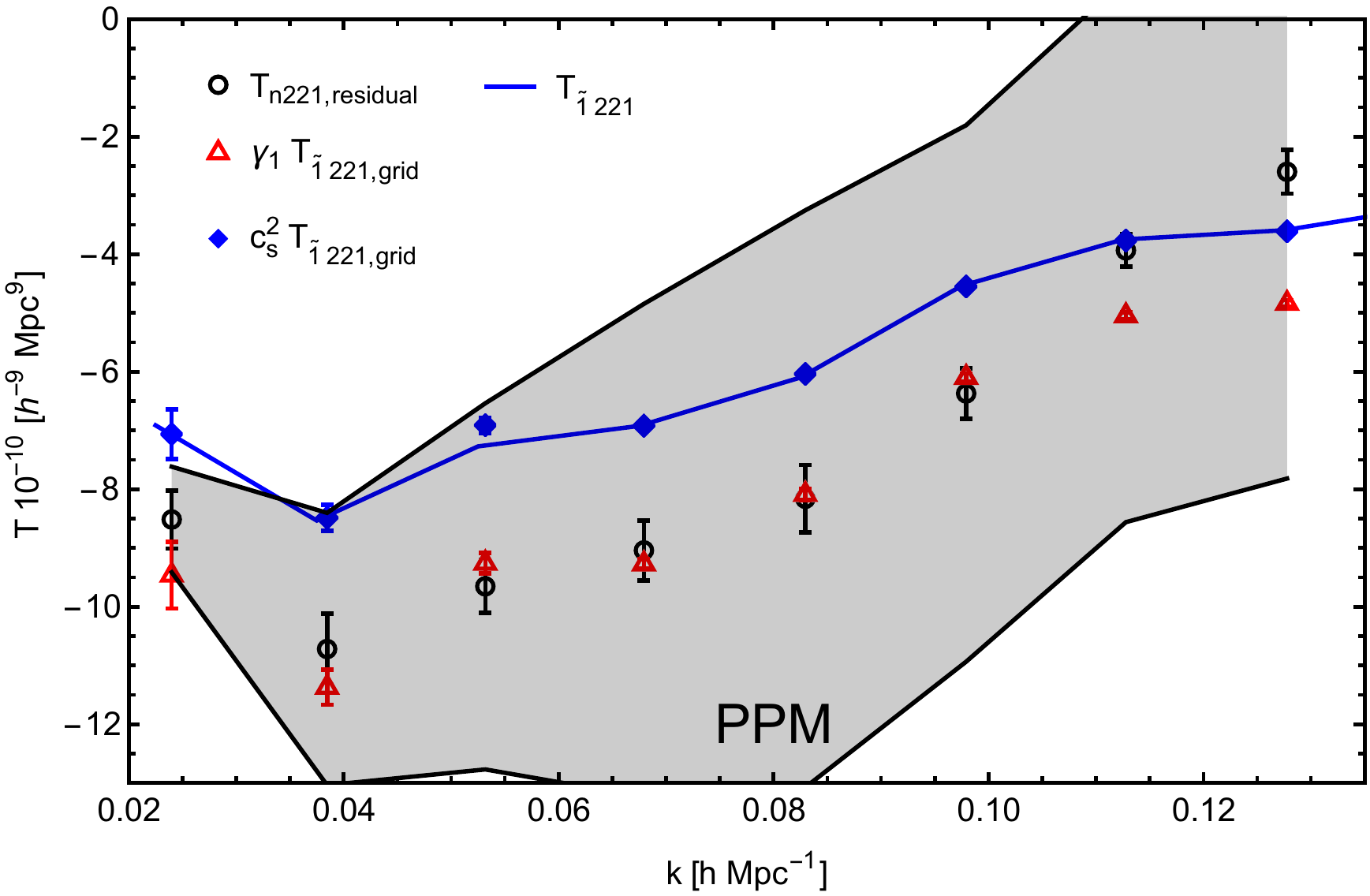}
	\includegraphics[width=0.49\textwidth]{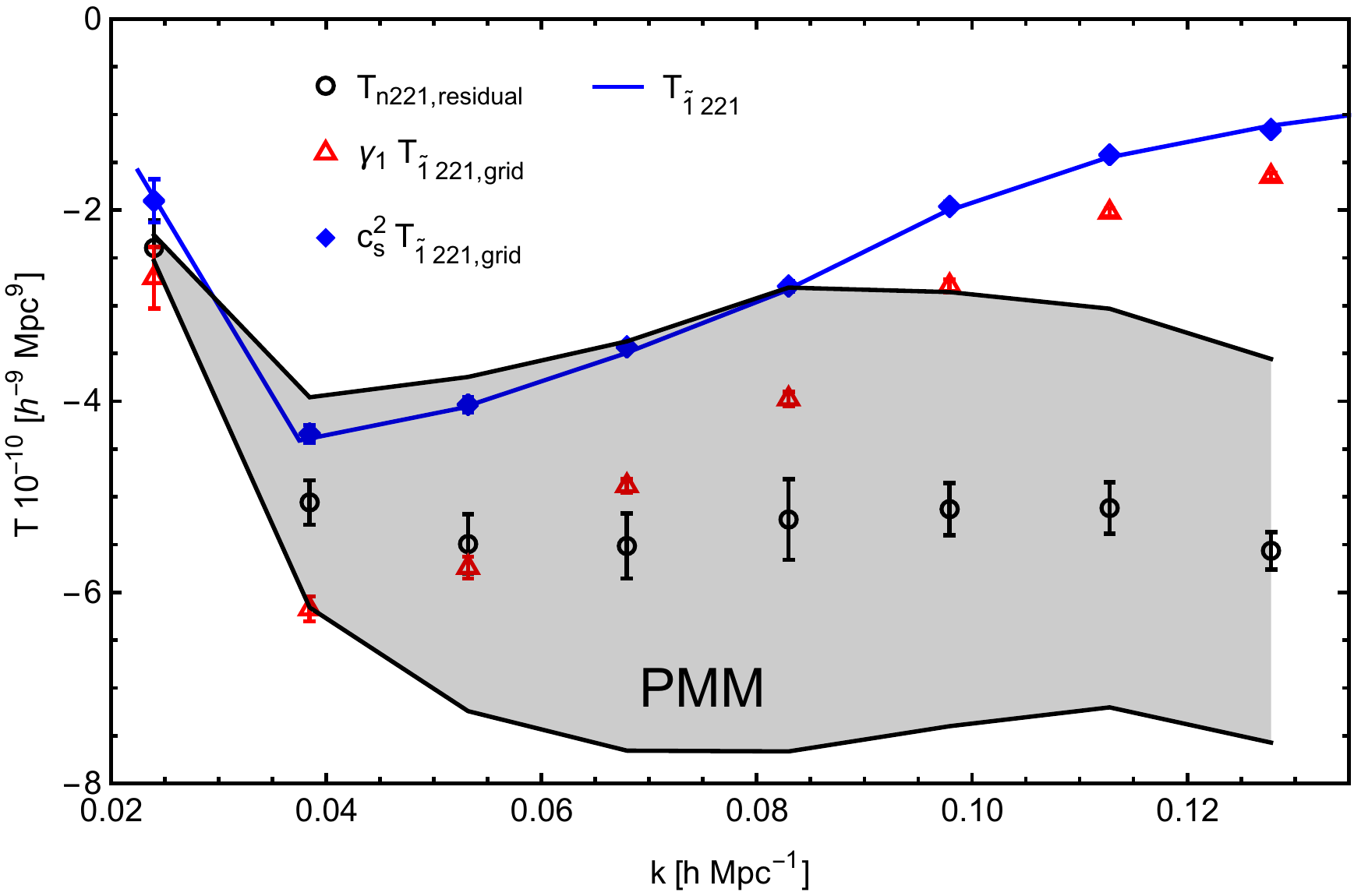}
	\includegraphics[width=0.49\textwidth]{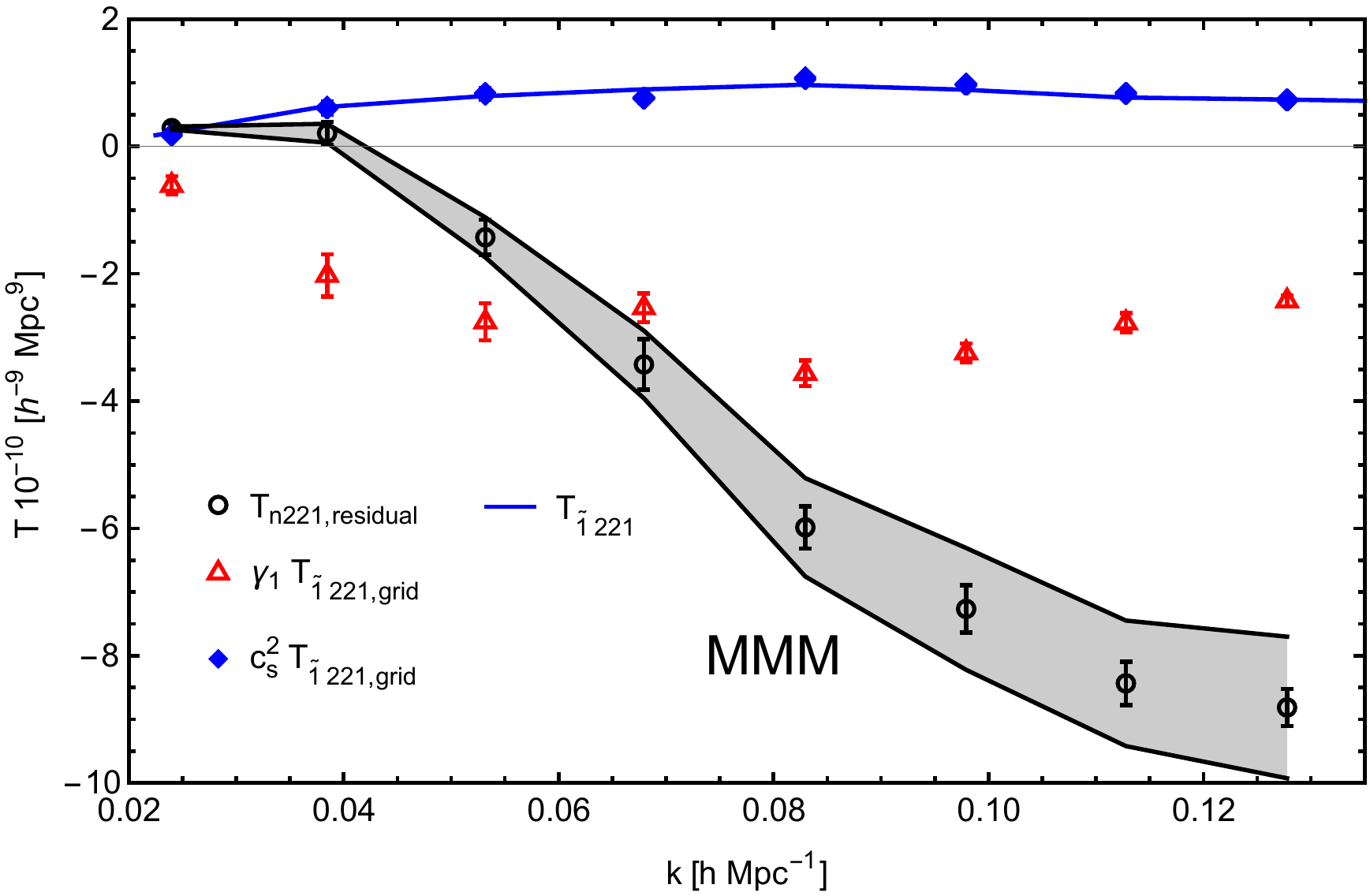}
	\includegraphics[width=0.49\textwidth]{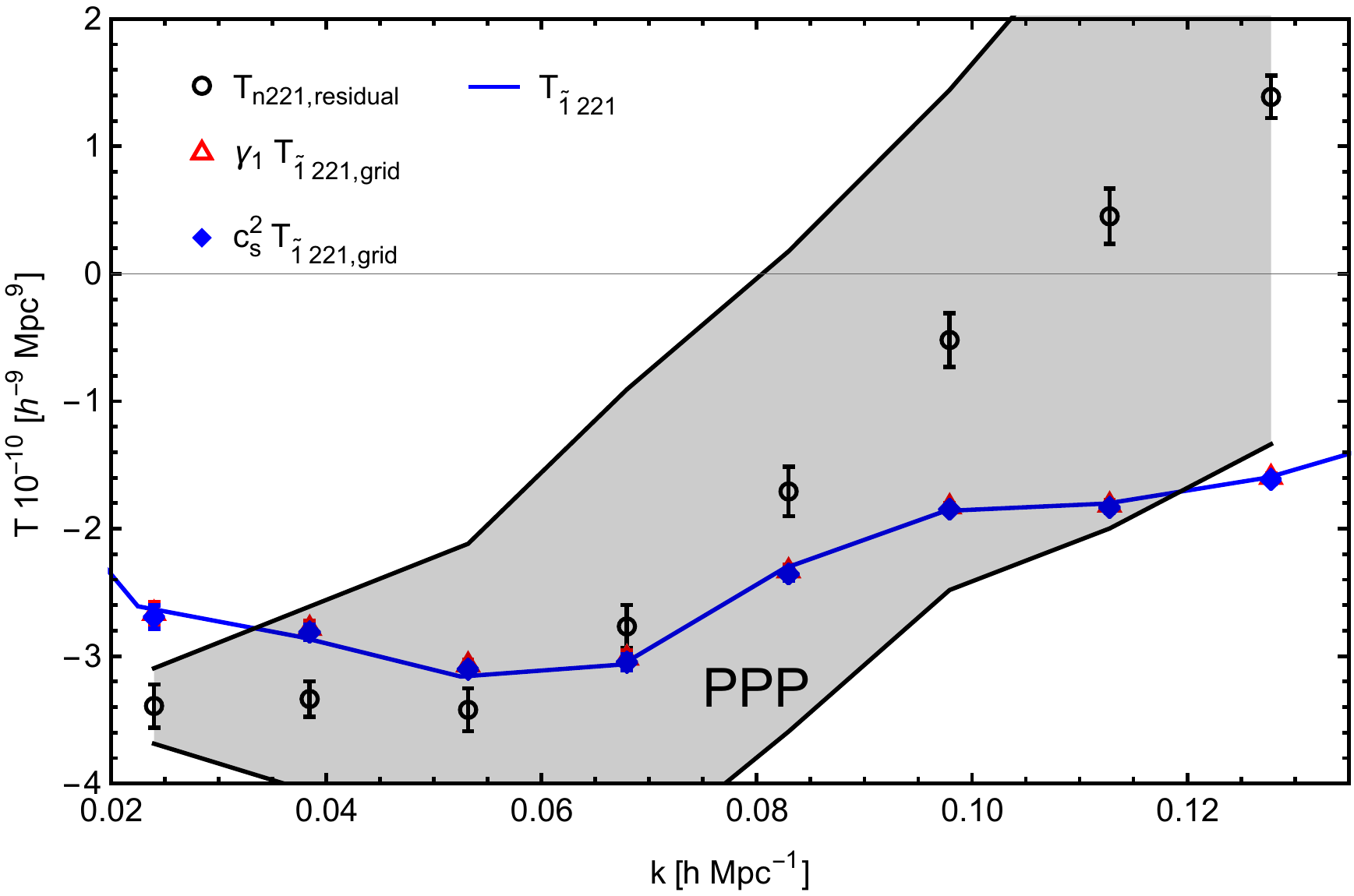}
	\caption{The calculated counterterms $T_{\tilde{1}221}$ with $\tilde{F}_{1}$ as calibrated from the bispectrum against the grid residual for configurations PPM (top left), PMM (top right), MMM (bottom left), PPP (bottom right).  The grey shaded region shows the theoretical errors for the two loop contributions.  We can see that the results work well for PPM, particularly with the amplitude fitted to the grid counterterm.  However, for the other configurations, particularly MMM, the counterterms fitted less well with the measured residuals.}
\end{figure}

\begin{figure}[b]
	\centering
	\includegraphics[width=0.49\textwidth]{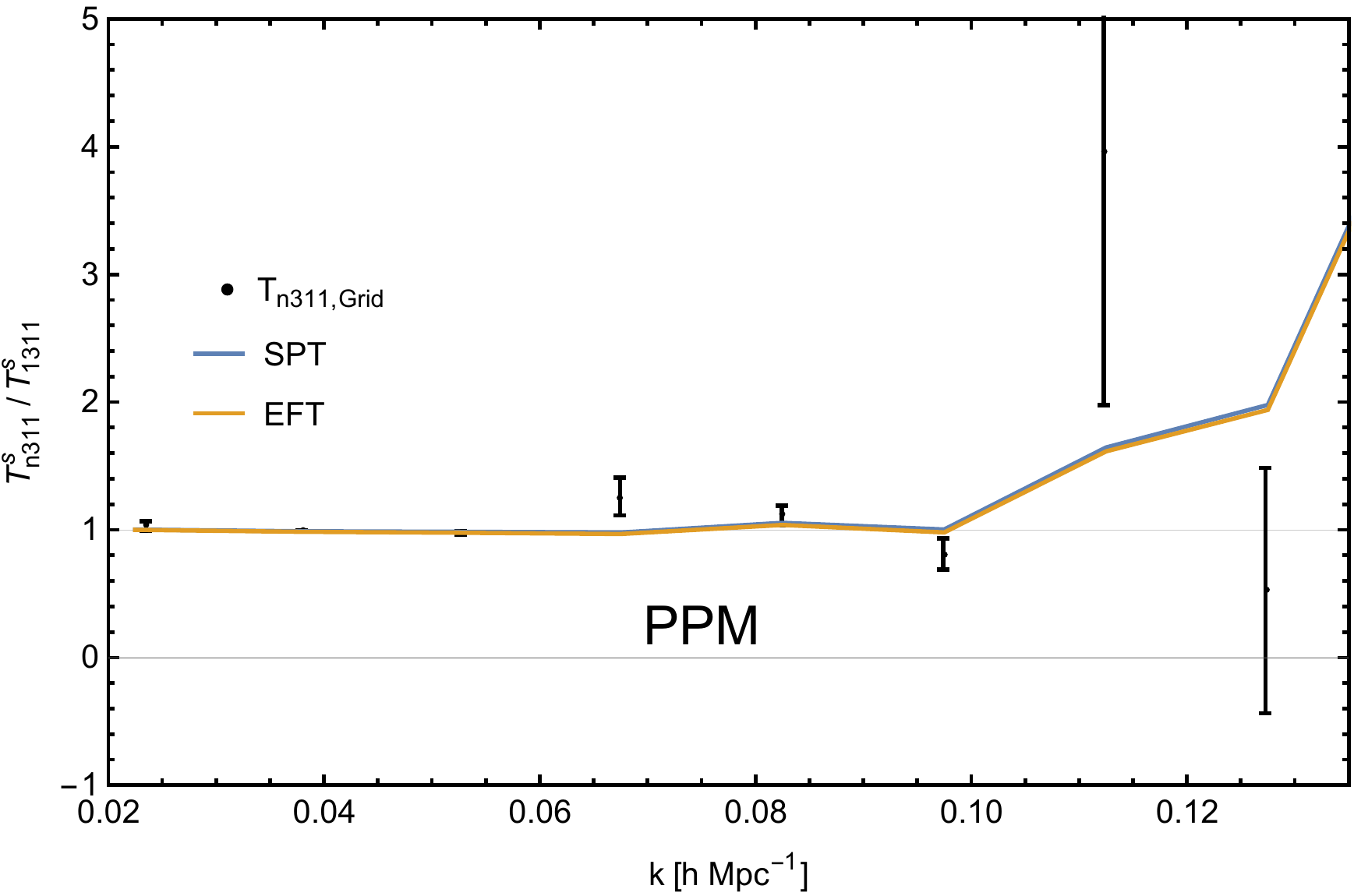}
	\includegraphics[width=0.49\textwidth]{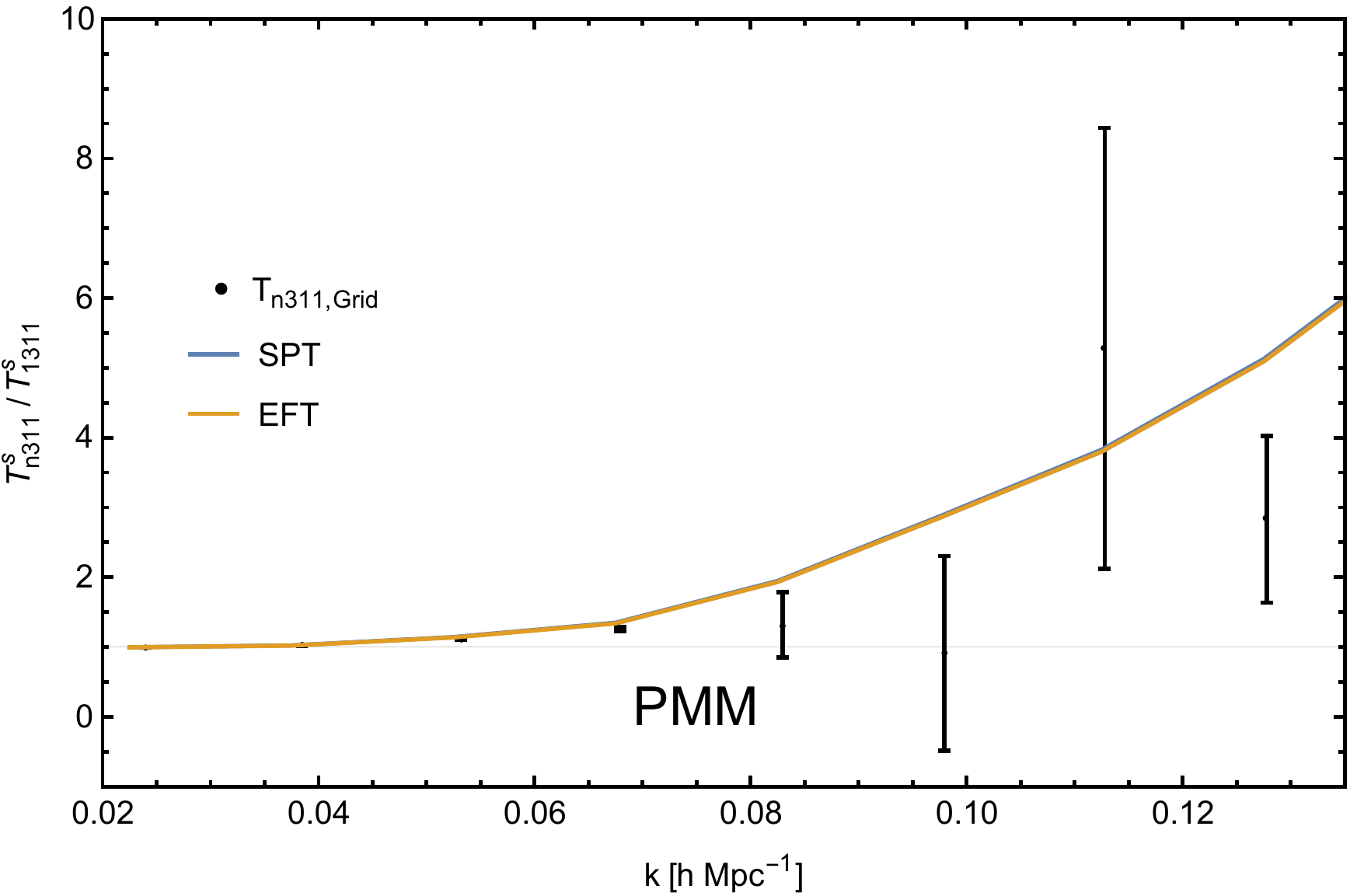}
	\includegraphics[width=0.49\textwidth]{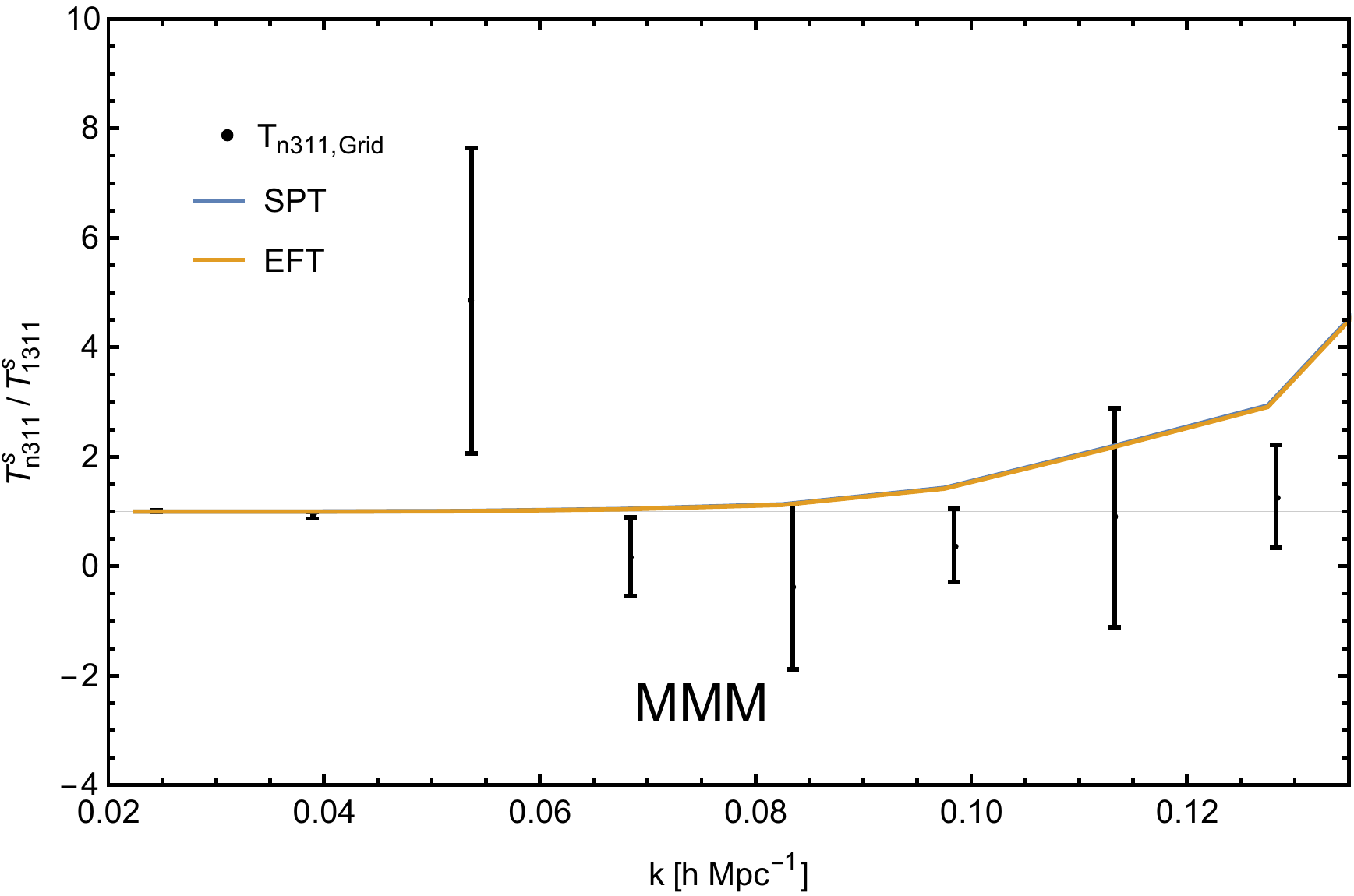}
	\includegraphics[width=0.49\textwidth]{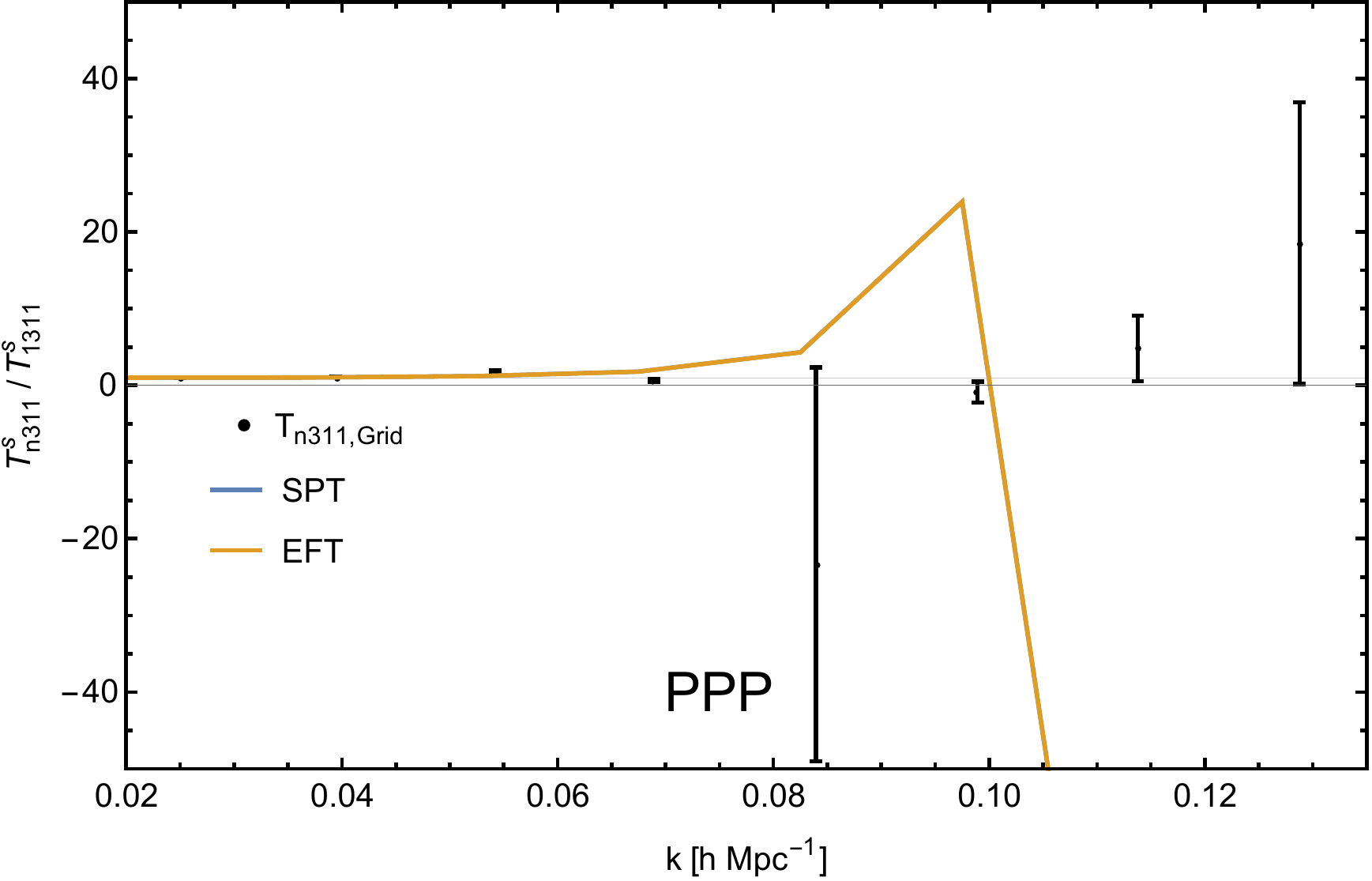}
	\caption{The calculated $T_{1311}+T_{3311}+T_{\tilde{1}311}$ with $\tilde{F}_{1}$ as calibrated from the bispectrum against the grid $T_{\mathrm{n}311}-T_{2311}$ for the four configurations studied.  We see that the counterterm estimated here was extremely small and made a subleading correction, such that the EFT results are almost indistinguishable from the SPT results.  Given the small sizes of the measured residuals for the four configurations studied, this could simply be because a larger correction was not required.  Note that this is a ratio over the residual, so if we had only studied perturbation theory up to tree level we would have predicted a constant line at one.  Unlike the other terms we have measured, $T_{\mathrm{n}311}$ does seem to be consistent with tree level up to about $k\sim 0.06\ihMpc$ for all studied configurations, which could explain the extremely small size of the EFT corrections.}
	\label{fig:T1t311b}
\end{figure}

The counterterms calculated in these ways for $T_{\tilde{1}311}$ are plotted alongside the residuals in Fig.~\ref{fig:T1t311b} for the four considered configurations.  We see that the various methods produce similar results to one another and all provide an adequate approximation to the one-loop residual of $T_{\mathrm{n}311}$ at least until $k\sim 0.06\ihMpc$.  However, once again we see that the two loop theoretical error is very large and encompasses many of our calculations.

\begin{figure}[t]
	\centering
	\includegraphics[width=0.49\textwidth]{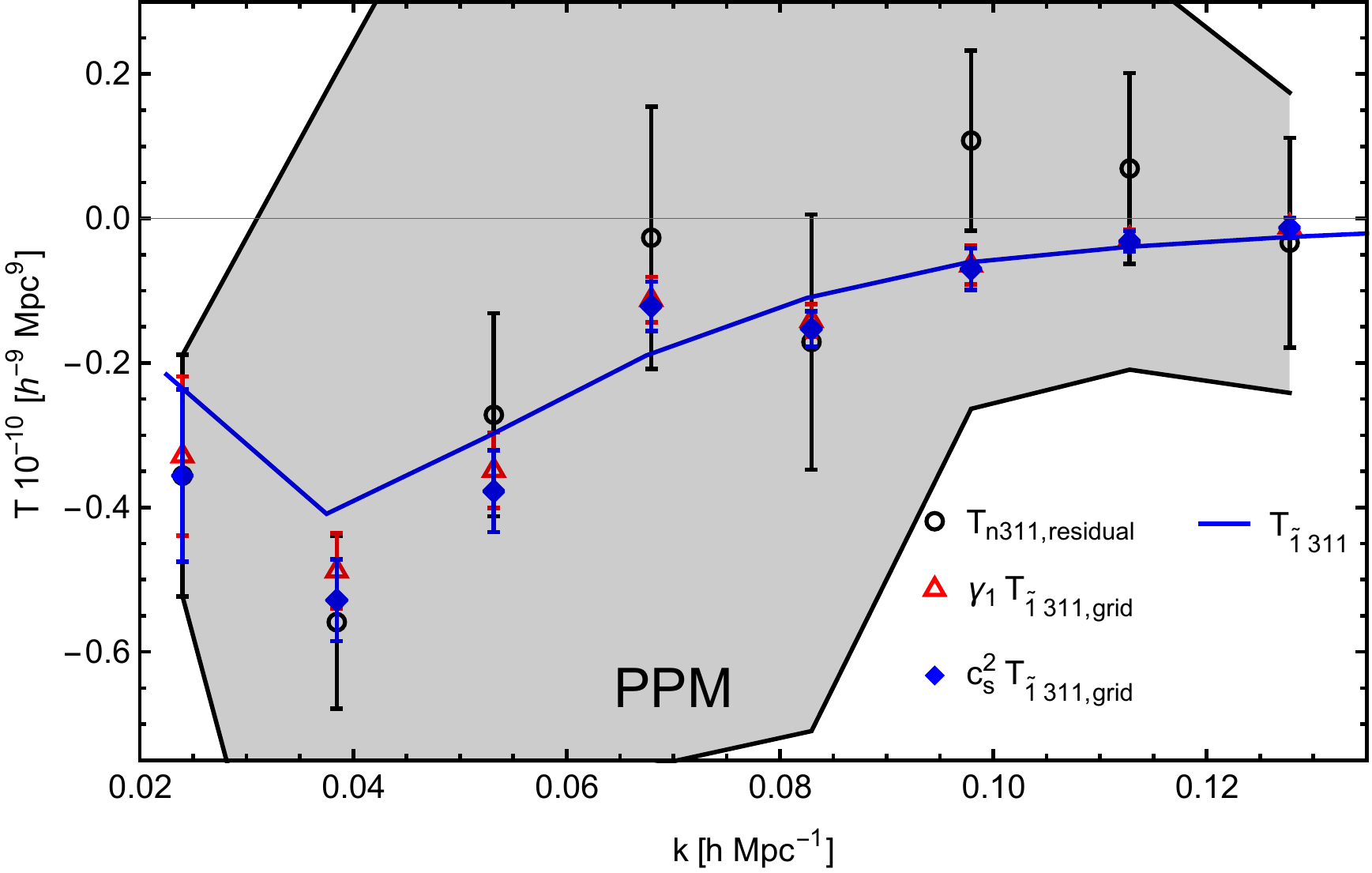}
	\includegraphics[width=0.49\textwidth]{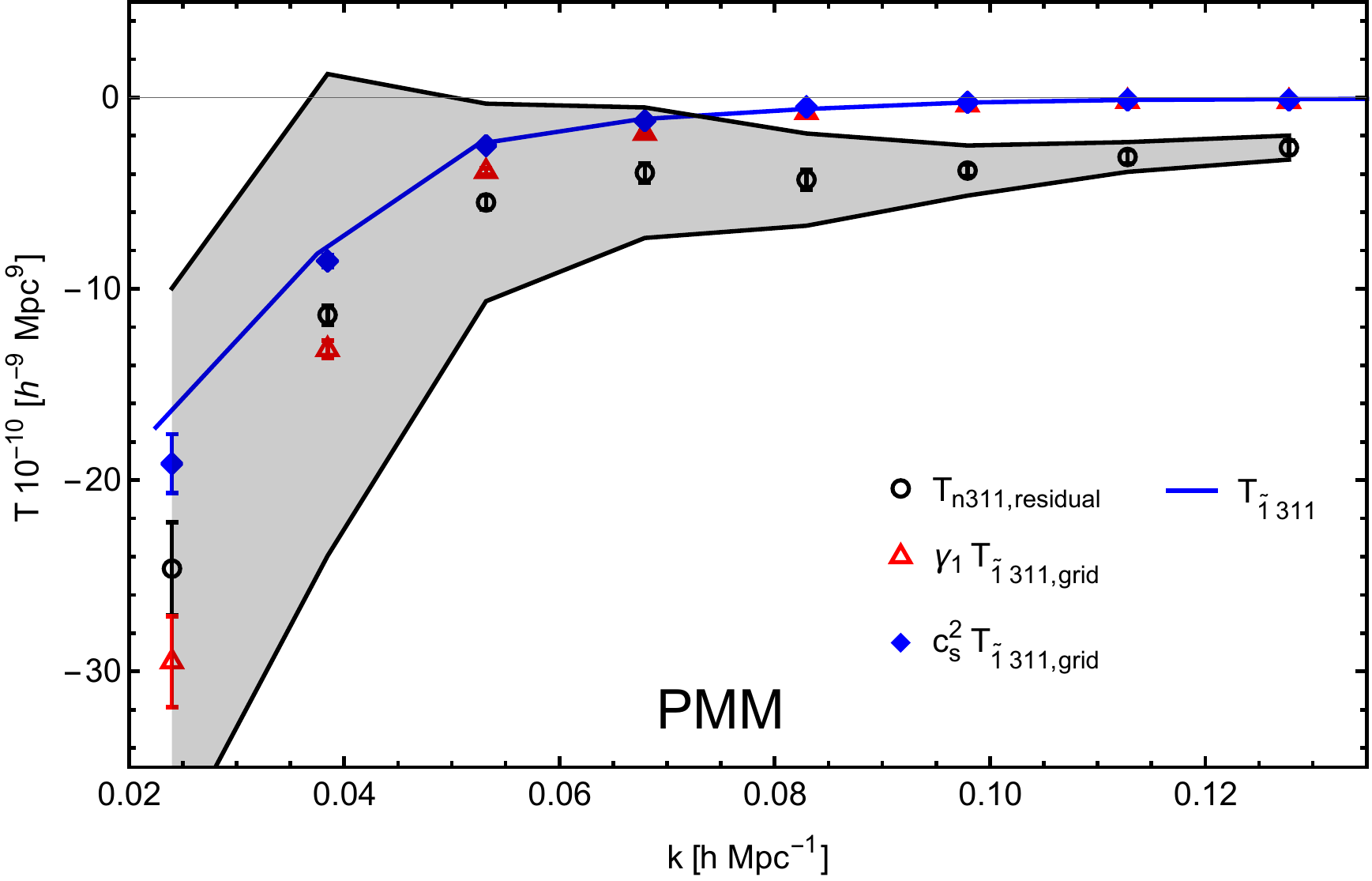}
	\includegraphics[width=0.49\textwidth]{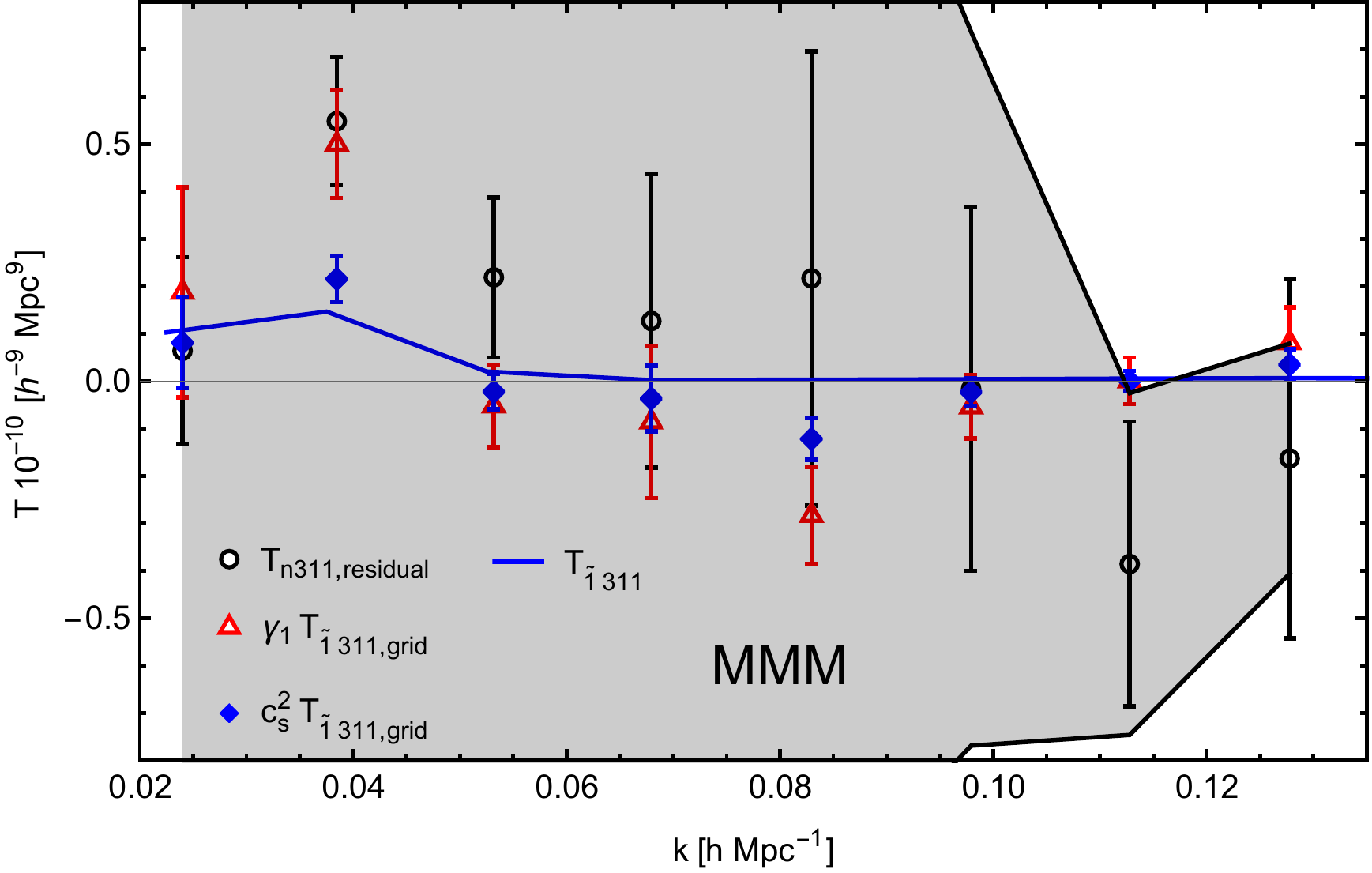}
	\includegraphics[width=0.49\textwidth]{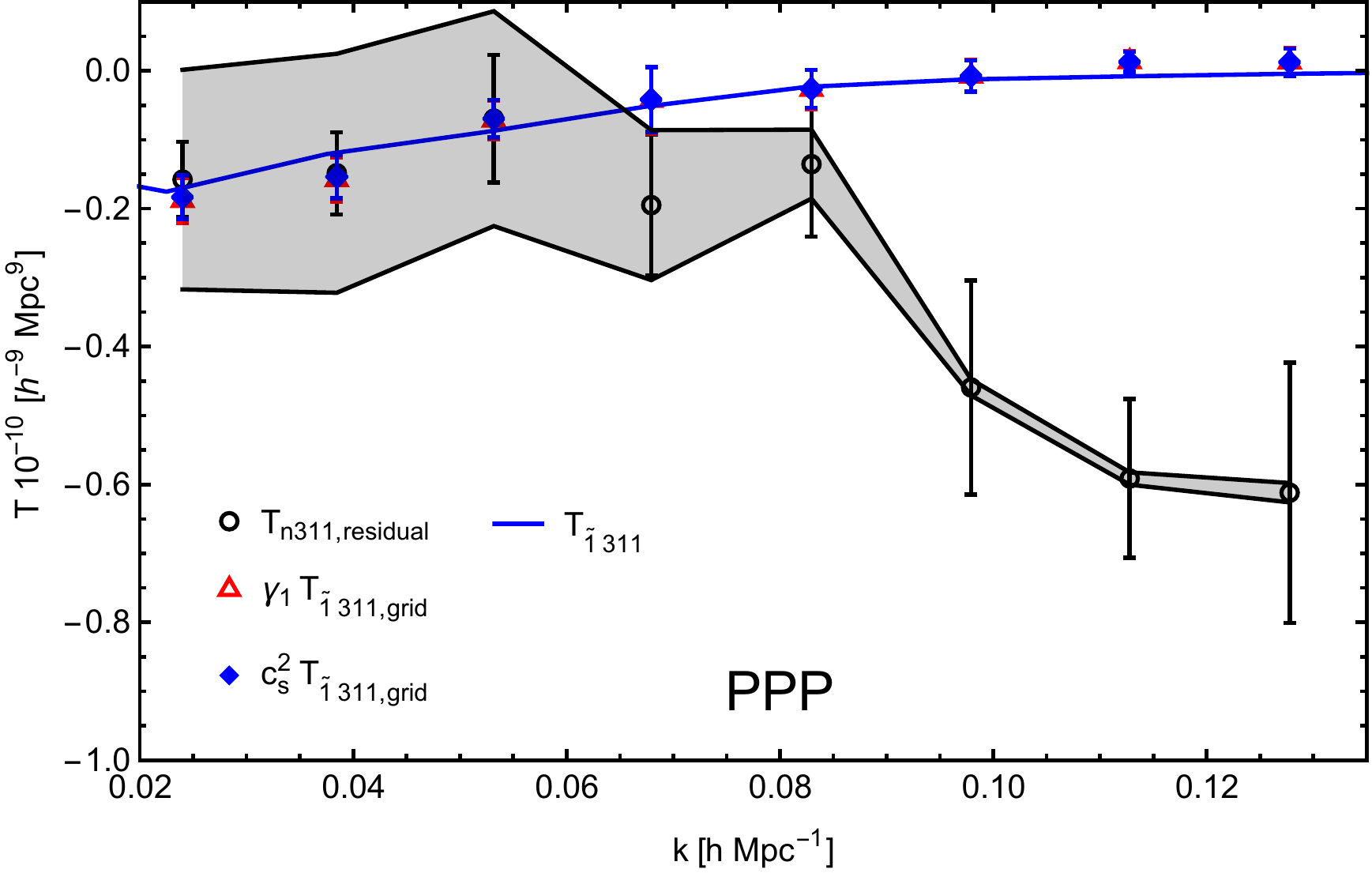}
	\caption{The theoretical counterterm $T_{\tilde{1}311}$ with $\tilde{F}_{1}$ as calibrated from the bispectrum against the grid residual for configurations PPM (top left), PMM (top right), MMM (bottom left), PPP (bottom right)  The grey shaded region shows the theoretical error for the two loop contributions.  For all configurations, we see that the grid counterterms, multiplied by some amplitude, fit well with the measured residuals at large scales.  For all configurations except MMM, we see the same for the analytically estimated $T_{\tilde{1}311}$.}
\end{figure}

 Overall, we see that this approach has a much higher rate of success in calculating the counterterms $T_{\tilde{2}211}$, $T_{\tilde{1}221}$, and $T_{\tilde{1}311}$ than the $\alpha$-parametrisation, with some methods working better than others, but still failed to completely account for the corresponding one-loop residuals up to the desired momenta for all configurations.  There are a number of possible explanations for this, including a larger than expected contribution from two-loop or higher order noise terms and problems with the simulation's time integration.  Nonetheless, we have found that our methods are able to regularise all of the isolated trispectra considered at at least some momenta for most configurations sampled.  However, the large theoretical errors for the two-loop terms indicate that our one-loop counterterm calculations are likely to be rendered ineffective without two loop terms being accounted for; this is left for a future project.

\subsection{The $\tilde{F}_{3}$ Kernel}
\label{sec:alphabeta}
The counterkernel $\tilde{F}_{3}$ was not constrained in \cite{Steele:2020tak} and is the new interaction studied in this paper.  We propose two simple one-parameter models for the counterterm $T_{\tilde{3}111}$ by defining what we call the $\alpha$ and $\beta$ parametrisations, as will be discussed in this section.

\begin{figure}[h!]
	\centering
	\includegraphics[width=0.49\textwidth]{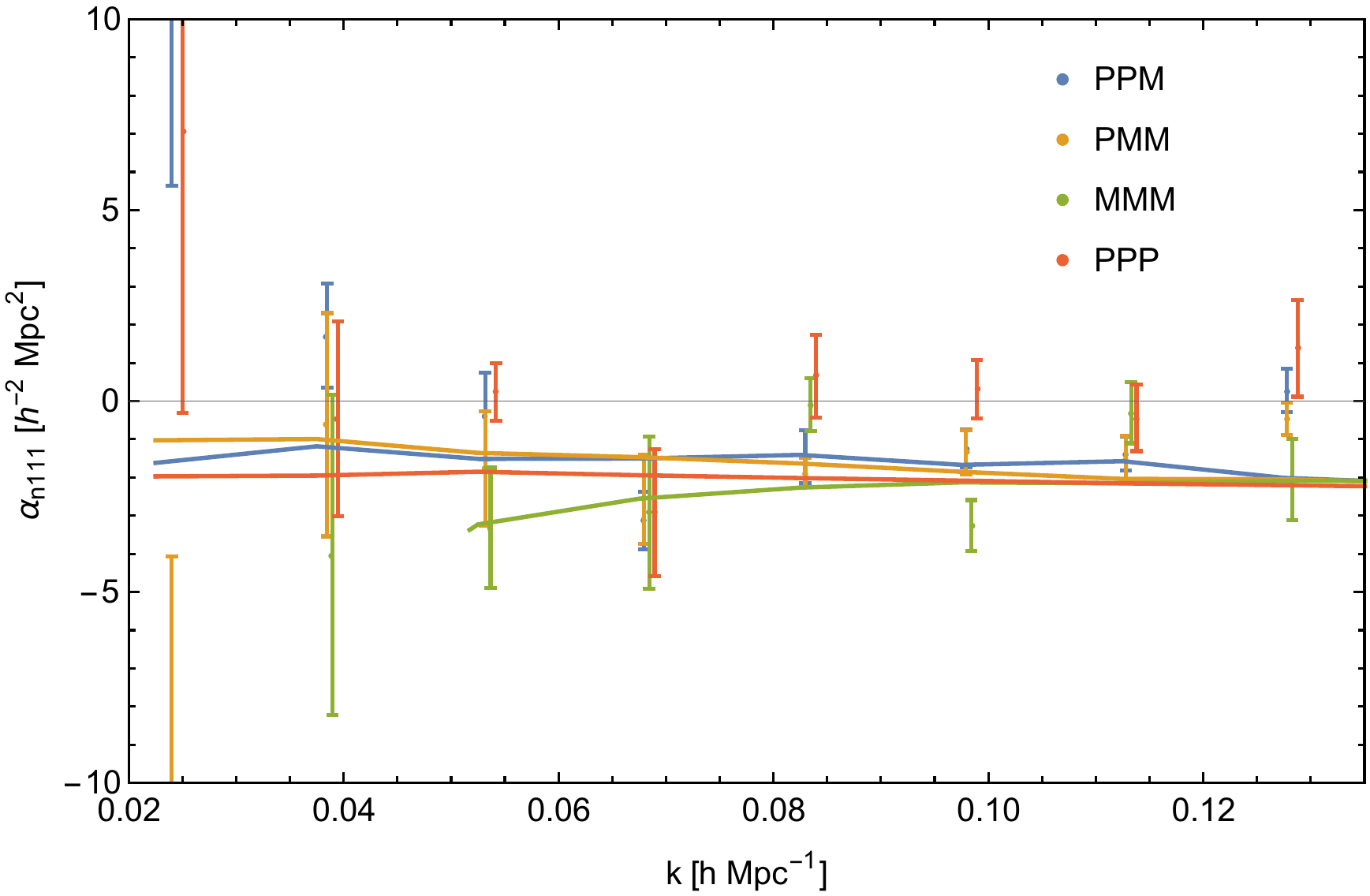}
	\includegraphics[width=0.49\textwidth]{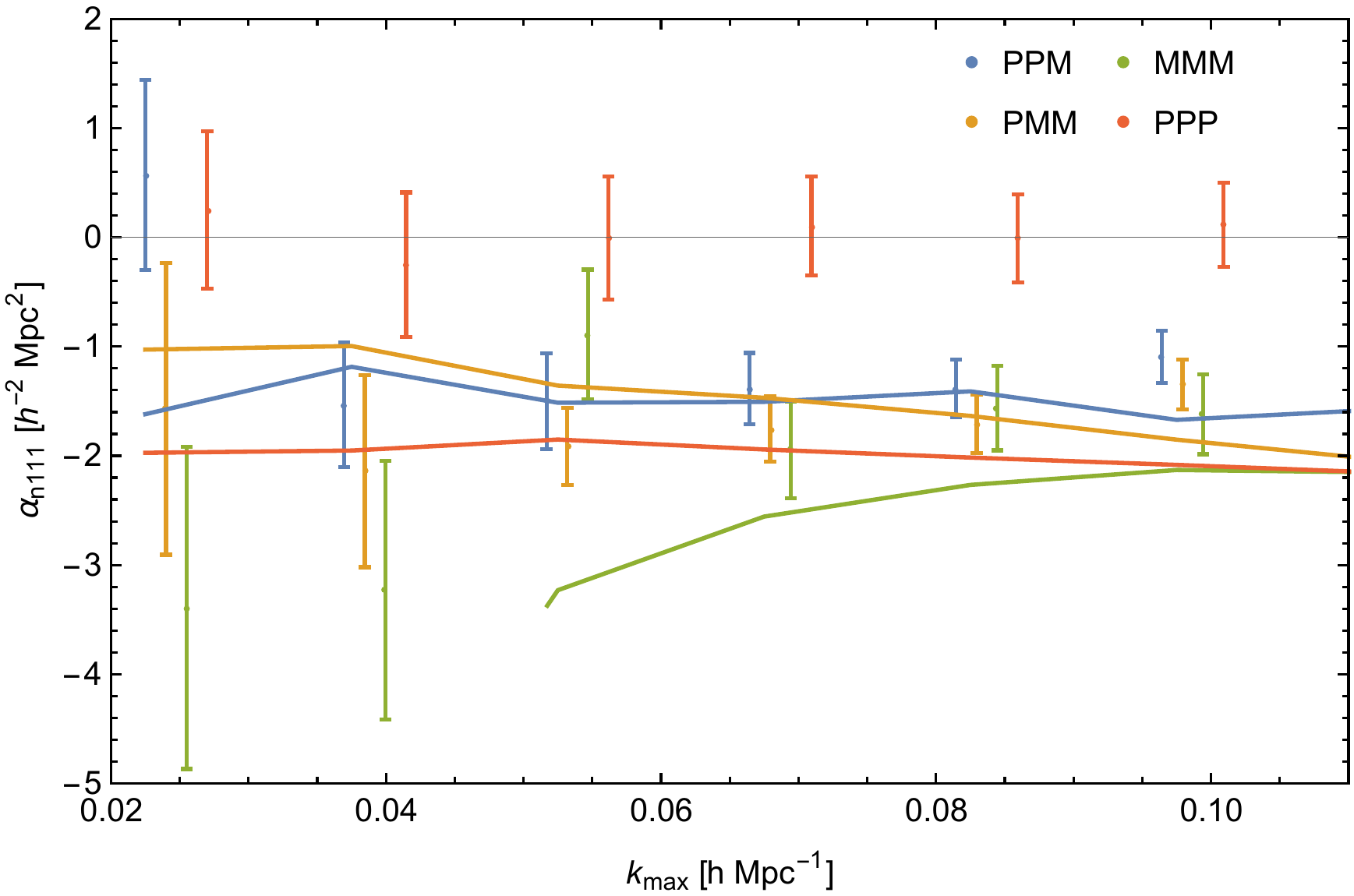}
	\caption{\emph{Left hand panel:} $\alpha_{\mathrm{n}111}$ calculated at each point in isolation.  \emph{Right hand panel:} The $\chi^{2}$ minimisation up to $k_{\mathrm{max}}$ of the same.  We can see that the counterterm parameter takes values of the expected order of magnitude for all configurations except PPP, in which it is consistent with zero.}
	\label{fig:alpha}
\end{figure}

Our first and simplest one parameter ansatz is to define
\begin{align}
T_{\mathrm{n}111}(\bkone,\bktwo,\bkthree,\bkfour)=&T_{3111}(\bkone,\bktwo,\bkthree,\bkfour)+T_{5111}(\bkone,\bktwo,\bkthree,\bkfour,\Lambda)+\alpha_{\mathrm{n}111}(\Lambda)k_{1}^{2}P_{11}(k_{2})P_{11}(k_{3})P_{11}(k_{4})~.
\end{align} 
This simply acknowledges that the counterterm is expected to scale as $k^{2}P_{11}^{3}$ and leaves a free parameter, $\alpha_{\mathrm{n}111}$, to fit that curve to the measured residual.  By rearranging this equation and removing unphysical terms, we can estimate the $\alpha$ parameter as
\begin{align}
\alpha_{\mathrm{n}111}(\Lambda)&=\frac{T_{\mathrm{n}111}(\bkone,\bktwo,\bkthree,\bkfour)-\sum_{i=1}^{5}T_{i111}(\bkone,\bktwo,\bkthree,\bkfour)}{k_{1}^{2}P_{11}(k_{2})P_{11}(k_{3})P_{11}(k_{4})}~.
\end{align}
In Fig.~\ref{fig:alpha} we plot the calculated values of $\alpha_{\mathrm{n}111}$ for all four studied configurations, both as a function of $k$ and as a $\chi^{2}$ minimisation up to a $k_{\mathrm{max}}$.

Secondly, we define the $\beta$ parameters.  By defining
\begin{equation}
T_{5111\mathrm{s},i-j}=T_{5111}(\Lambda_i)-T_{5111}(\Lambda_j)~,
\end{equation}
for $T_{5111}$ with cutoffs of $\Lambda_{i}$ and $\Lambda_{j}$, we have found a way of isolating and encapsulating the cutoff dependence of $T_{5111}$.  We can then define the parameter $\beta_{i-j}$, which we call the amplitude of $T_{5111\mathrm{s},i-j}$, such that we can minimise
\begin{equation}
    \chi^{2}_{\beta_{i-j}}=\sum_{k=k_{\mathrm{min}}}^{k_{\mathrm{max}}}\frac{\left[T_{\mathrm{n}111}(\bk_{1},\bk_{2},\bk_{3},\bk_{4})-T_{\mathrm{SPT}}(\bk_{1},\bk_{2},\bk_{3},\bk_{4})-\beta_{i-j}(k_{\mathrm{max}})T_{5111\mathrm{s},i-j}(\bk_{1},\bk_{2},\bk_{3},\bk_{4})\right]^{2}}{\Delta T_{\mathrm{n}111}(\bk_{1},\bk_{2},\bk_{3},\bk_{4})^{2}}
\end{equation}
for the amplitude and take $\beta_{i-j}(0.083\ihMpc)T_{5111\mathrm{s},i-j}(\bk_{1},\bk_{2},\bk_{3},\bk_{4})$ as a counterterm estimator for the one-loop trispectrum propagator.  Rather than simply fitting a curve with defined scaling to the residual as we did with the $\alpha$ parameter, the $\beta$ parameters explicitly account for the cutoff dependence of the one-loop terms; by capturing the difference between the measured terms with different cutoffs, they provide us with a curve that scales exactly as a counterterm should scale in order to capture any given term's cutoff dependence.

\begin{figure}[h!]
	\centering
	\includegraphics[width=0.49\textwidth]{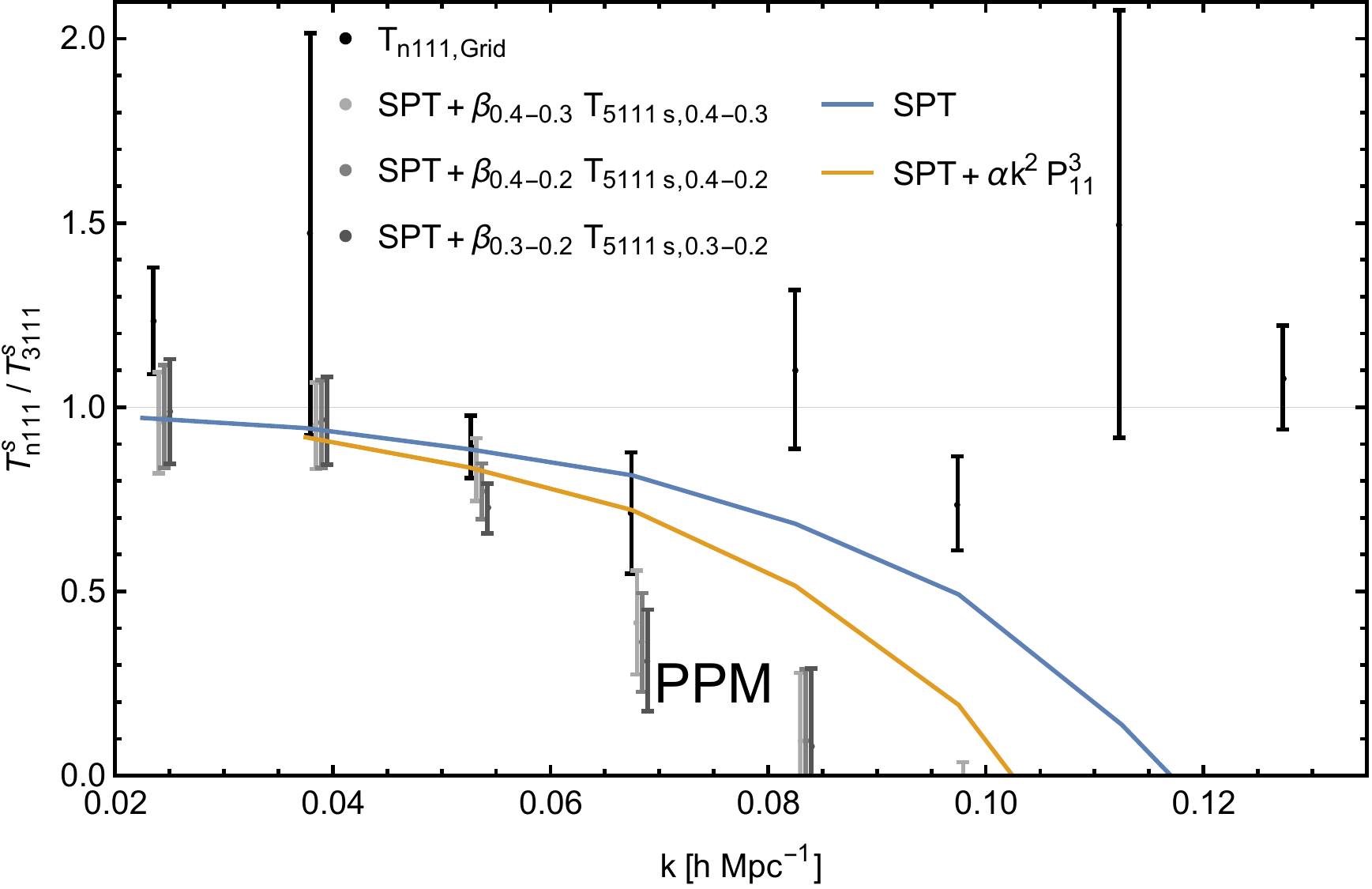}
	\includegraphics[width=0.49\textwidth]{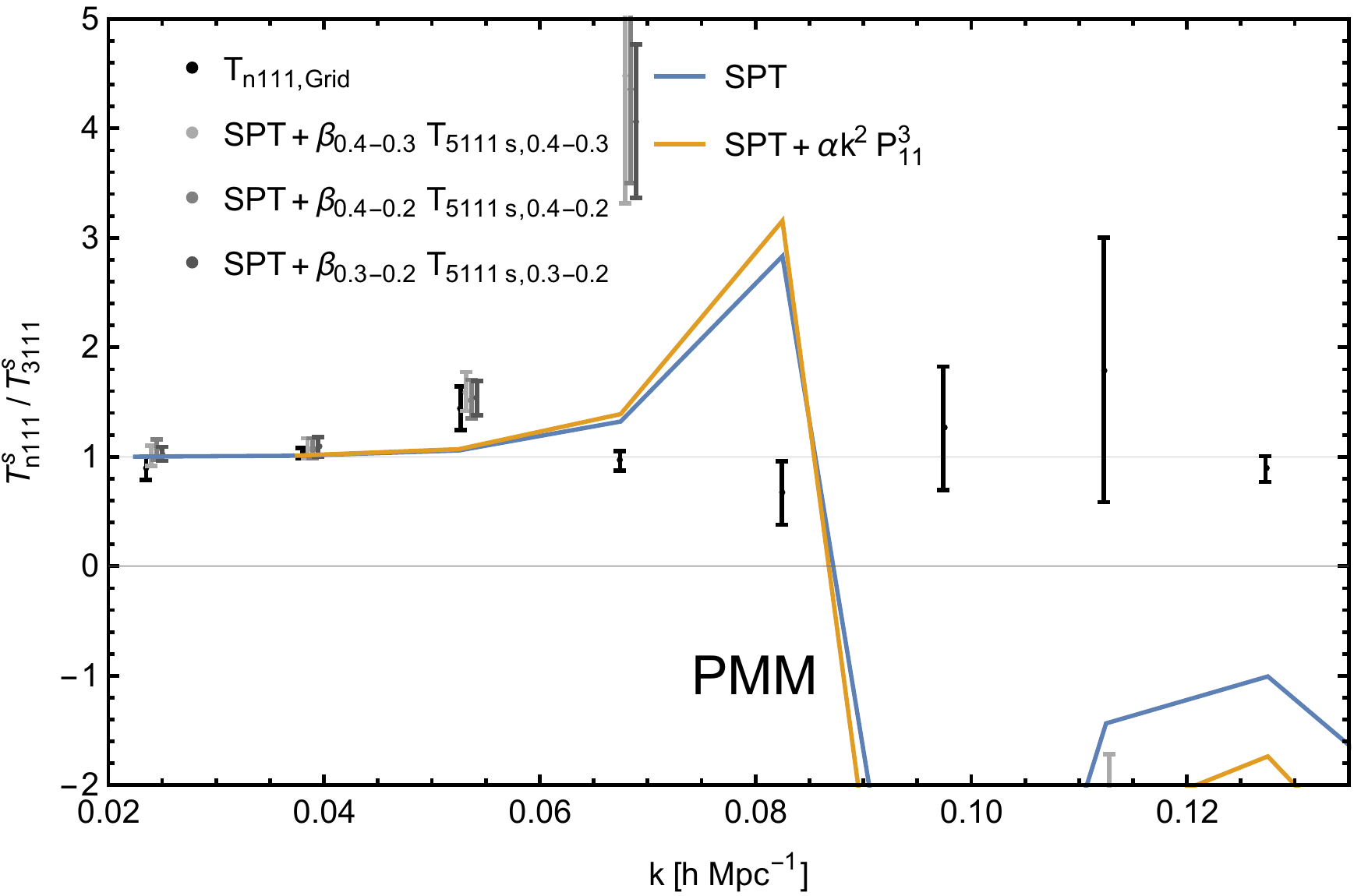}
	\includegraphics[width=0.49\textwidth]{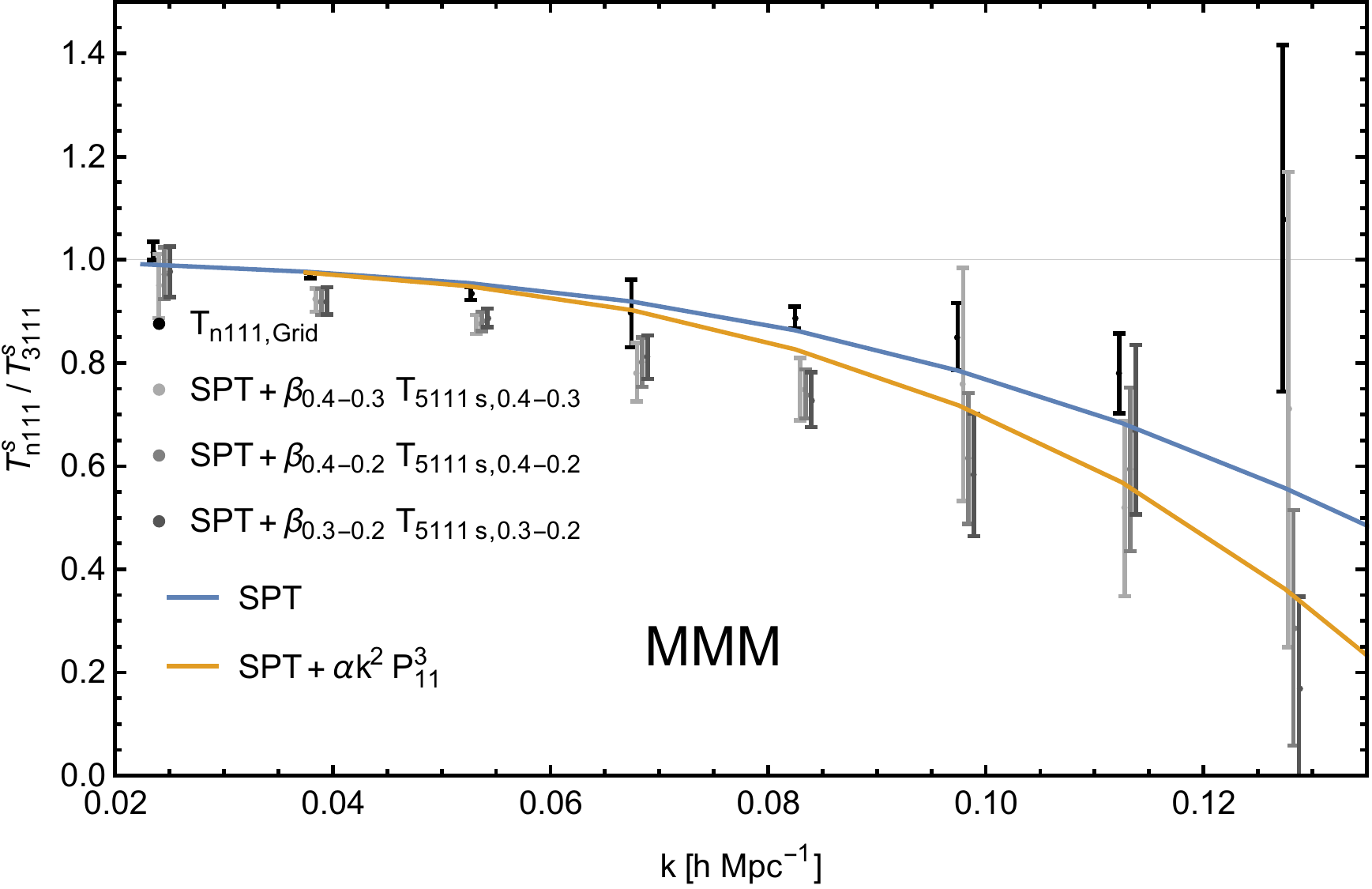}
	\includegraphics[width=0.49\textwidth]{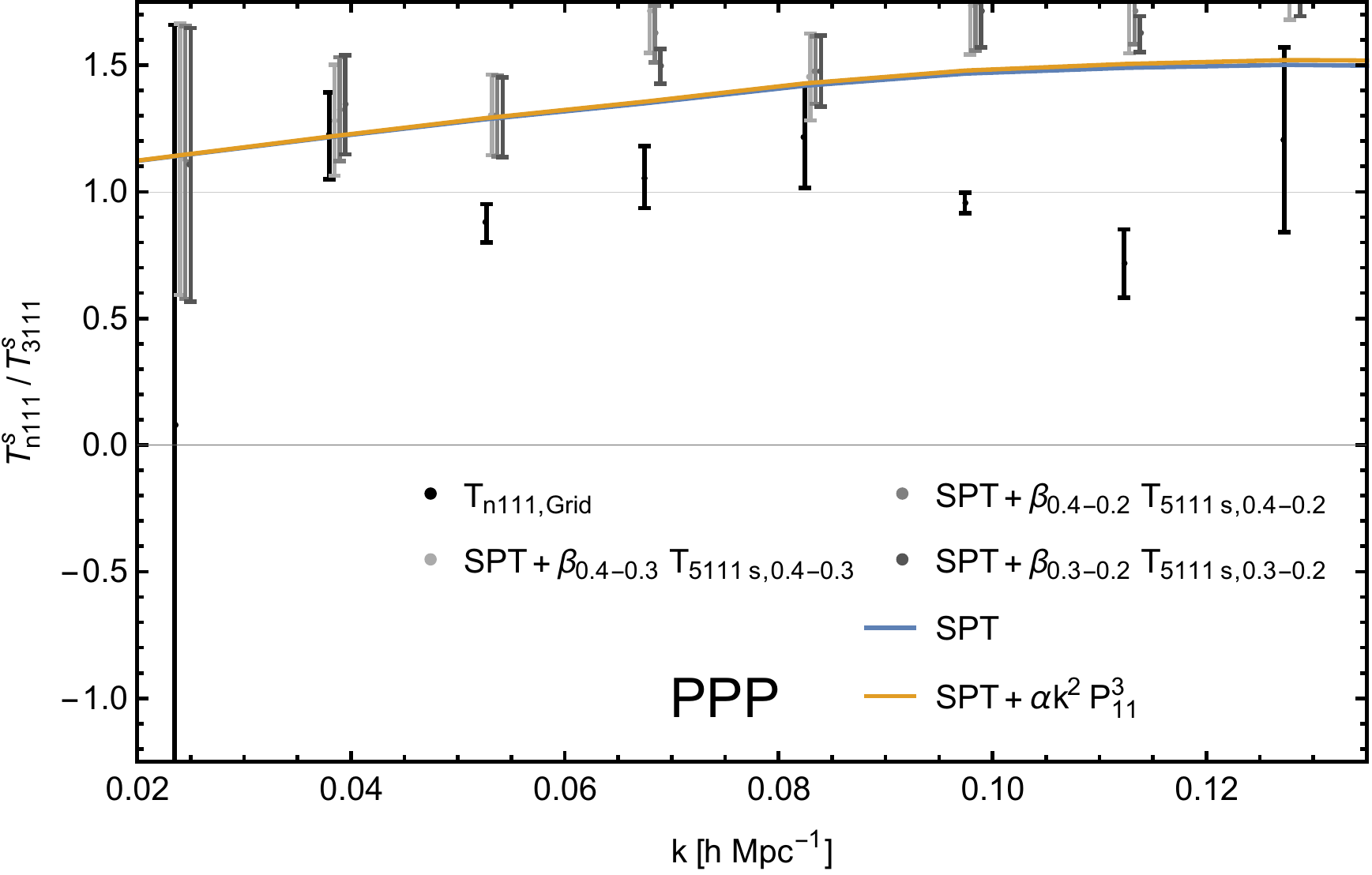}
	\caption{The calculated $T_{3111}+T_{5111}+T_{\tilde{3}111}$ with $\tilde{F}_{3}$ as calibrated from the bispectrum against the grid $T_{\mathrm{n}111}-T_{4111}-T_{2111}-T_{1111}$ for the four configurations studied.  In all cases the measured trispectra had $\Lambda=0.3\ihMpc$ except those used for calculating the $\beta$ parameters which had the cutoffs represented in those parameters' subscripts.  The parameters were taken as those with the values $k_{\mathrm{max}}=0.08 \ihMpc$.  Note that this is a ratio over the tree-level terms such that, had we only studied perturbation theory up to tree level, we would have a constant line at one.  It is noticeable that this would not have given up an accurate estimator for $T_{\mathrm{n}111}$ in configurations PPM and MMM, but would have sufficed up to even $k\sim 0.1\ihMpc$ for PPM and PPP, perhaps explaining the smaller sizes of the EFT corrections in those configurations.}
	\label{T3t111b}
\end{figure}

In Fig.~\ref{T3t111b} we plot the calculated $T_{3111}+T_{5111}+T_{\tilde{3}111}$ against the measured $T_{\mathrm{n}111}-T_{4111}-T_{2111}-T_{1111}$ using our calculated $T_{\tilde{3}111}$ with both the $\alpha$ and $\beta$ parametrisations for the four configurations studied.  We find that both parametrisations work reasonably well at the scales of interest, with the $\beta$ parameters outperforming the $\alpha$ parameters in configurations PMM and PPP and the $\alpha$ parameters performing better for the configurations PPM and MMM.  In both cases, at least one parametrisation was capable of providing a good fit to the measured residuals of $T_{\mathrm{n}111}$ up to $k\sim 0.07\ihMpc$, where we would expect to see two loop effects strongly coming into effect.

\section{Discussion}
This paper presents the first clear detection of the trispectrum counterterms for the EFTofLSS through a direct measurement.  We have developed an estimator which allows us to sample various trispectrum configurations in a computationally efficient manner. This permits calibration of the trispectrum counterterms through a study of the trispectrum itself, rather than through the covariance of the measured power spectrum.  Furthermore, we have shown that realization based cosmic variance cancellation techniques allow for significantly more precise results than could otherwise have been obtained for the given sample size.

We have implemented two ways for estimating trispectra on the grid, one direct summation method and one Fourier method.  We have shown that these methods produce consistent results but that the Fourier space routines are significantly less computationally intensive.  These estimators might be useful for extracting cubic bias parameters or cubic primordial non-Gaussianity from simulations and surveys.

The method of gridPT allows us to measure perturbative contributions to the trispectrum from the same seeds which initialise the simulations. We compared the grid based perturbative results for the trispectrum with analytical calculations and found excellent agreement. Using gridPT we were able to estimate the connected trispectrum with excellent signal-to-noise from a modest simulation volume. The gridPT approach could thus be used to test and calibrate models for the covariance matrix at a managable computational cost, avoiding the need for thousands of $N$-body simulations.

Using the gridPT methodology, we have shown that correlators with an odd number of fields, i.e. terms with a vanishing mean and a non-zero variance, heavily affect the measurements in finite volumes, making it hard to calibrate counterterms  without subtracting them from the measurements. Using cosmic variance cancellation we have been able to attain significant detections of the counterterm amplitudes with only fourteen realisations, making our analysis faster than previous studies of the four point function in the EFTofLSS that required thousands of realisations to compensate for the variance of the covariance.

We have also shown that when studying the trispectrum it is necessary to use the correct $\Lambda$CDM growth factors for the density perturbations $\delta_{2}$ and $\delta_{3}$ in the tree-level terms, as the growth factor corrections are of the same order of magnitude as the one-loop corrections.

We have found that a simple single parameter approximation for the counterkernel $\tilde{F}_{3}$ was sufficient to regularise the one-loop residual of $T_{\mathrm{n}111}$. This counterterm for $T_{5111}$ is a new interaction uniquely probed by the trispectrum but not by the power spectrum and bispectrum. Calibrations of $\tilde{F}_{1}$ and $\tilde{F}_{2}$ taken from the one-loop bispectrum were able to successfully regularise the one-loop residuals of $T_{\mathrm{n}211}$, $T_{\mathrm{n}221}$, and $T_{\mathrm{n}311}$.  Even after finding methods that could regularise each term, we found that our ability to calculate the counterterms was heavily configuration dependent, with some configurations being much harder to regularise than others, and that our results often only worked on the largest physical scales.  We have proposed a number of possible explanations for this, including the notion that higher order terms, both multi-loop and noise terms, could be having a larger impact on the one-loop trispectra than they have on the one-loop bi- and power spectra, or that there may be interference from numerical errors in the simulations.  Nonetheless, our results offer the first successful regularisation of the one-loop trispectrum in the EFTofLSS and have been shown to be successful for most configurations up to a non-trivial maximum wavenumber $k_\text{max}$.  

While this study was being finalized, \cite{Gualdi:2020eag} presented a detailed measurement of the trispectrum from a suite of 5000 simulations and assessed the trispectrum's potential to improve constraints on primordial non-Gaussianity. Within their error bars the tree-level calculation suffices to describe the trispectrum. Our significantly smaller error bars however, allow us to show that loop corrections significantly improve the modelling of the trispectrum.

We leave it to a future paper to study the full configuration space of the trispectrum and to constrain the parameters of the stress tensor explicitly.  We also leave it to a future paper to study the full trispectrum with all of its counterterms being constrained simultaneously in a manner analogous to our $B_{\mathrm{nnn}}$ methods in \cite{Steele:2020tak}.  In particular, we will be investigating whether or not using these new methods will allow us to overcome the problems encountered in this paper.

\begin{acknowledgements}
We would like to thank M. Garny, T. Nishimichi, E. Pajer, F. Schmidt, M. Simonovic, A. Taruya and M. Zaldarriaga for helpful discussions and K. Kornet for excellent computing support. This research made use of the COSMOS supercomputer at the Department of Applied Mathematics and Theoretical Physics, Cambridge. TS acknowledges support through the Science and Technology Facilities Council Doctoral Training Centre in Data Intensive Science. TB is supported by the Stephen Hawking Advanced Fellowship at the Center for Theoretical Cosmology.
\end{acknowledgements}

\clearpage

\appendix

\section{UV Tests}
We define a number of tests which allow us to confirm that the calibration of our $\chi^{2}$ calculations is correct; without them being properly calibrated we would have to assume that all of the results we gain from them may be inaccurate.  

\begin{figure}[h!]
	\centering
	\includegraphics[width=0.49\textwidth]{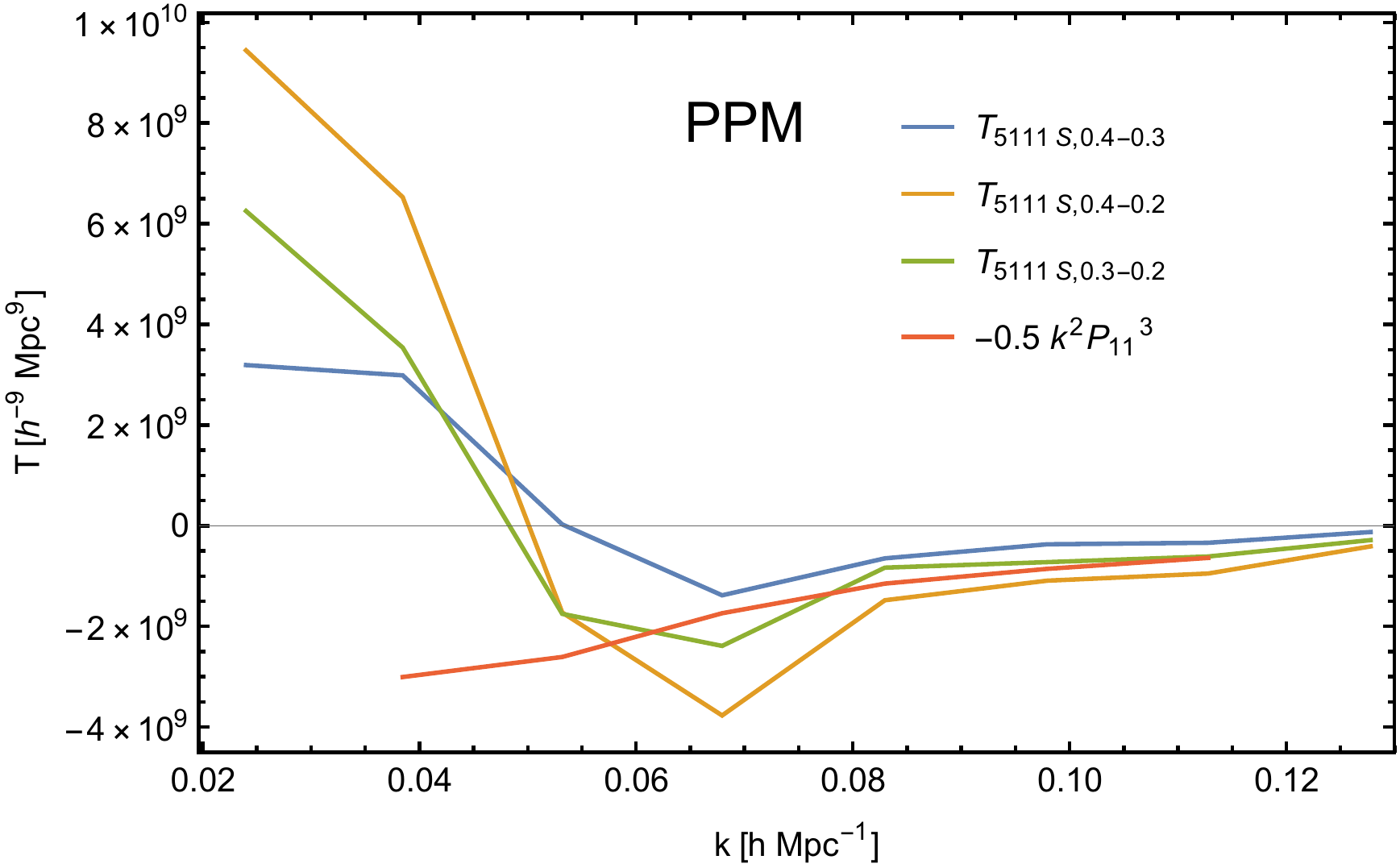}
	\includegraphics[width=0.49\textwidth]{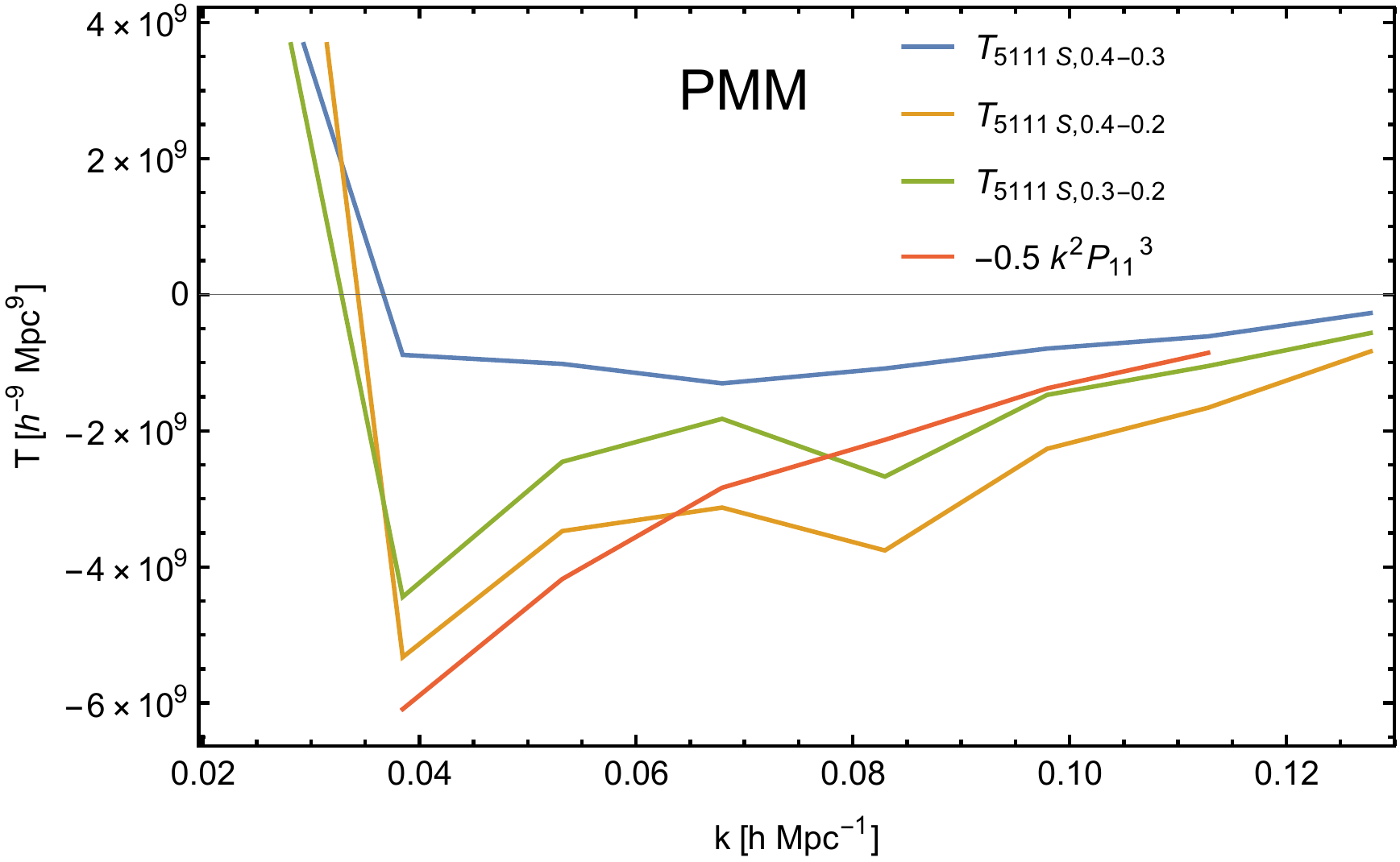}
	\includegraphics[width=0.49\textwidth]{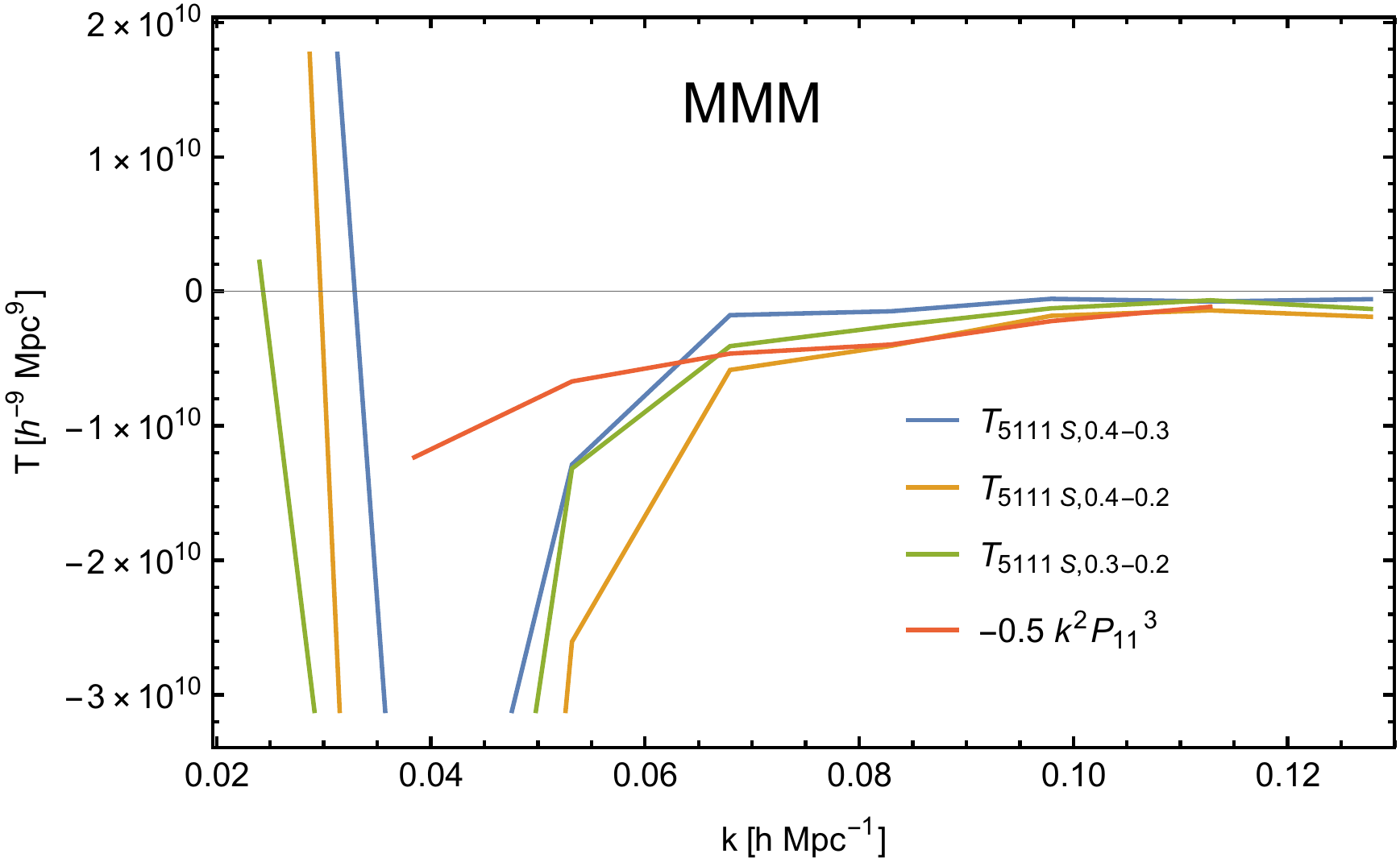}
	\includegraphics[width=0.49\textwidth]{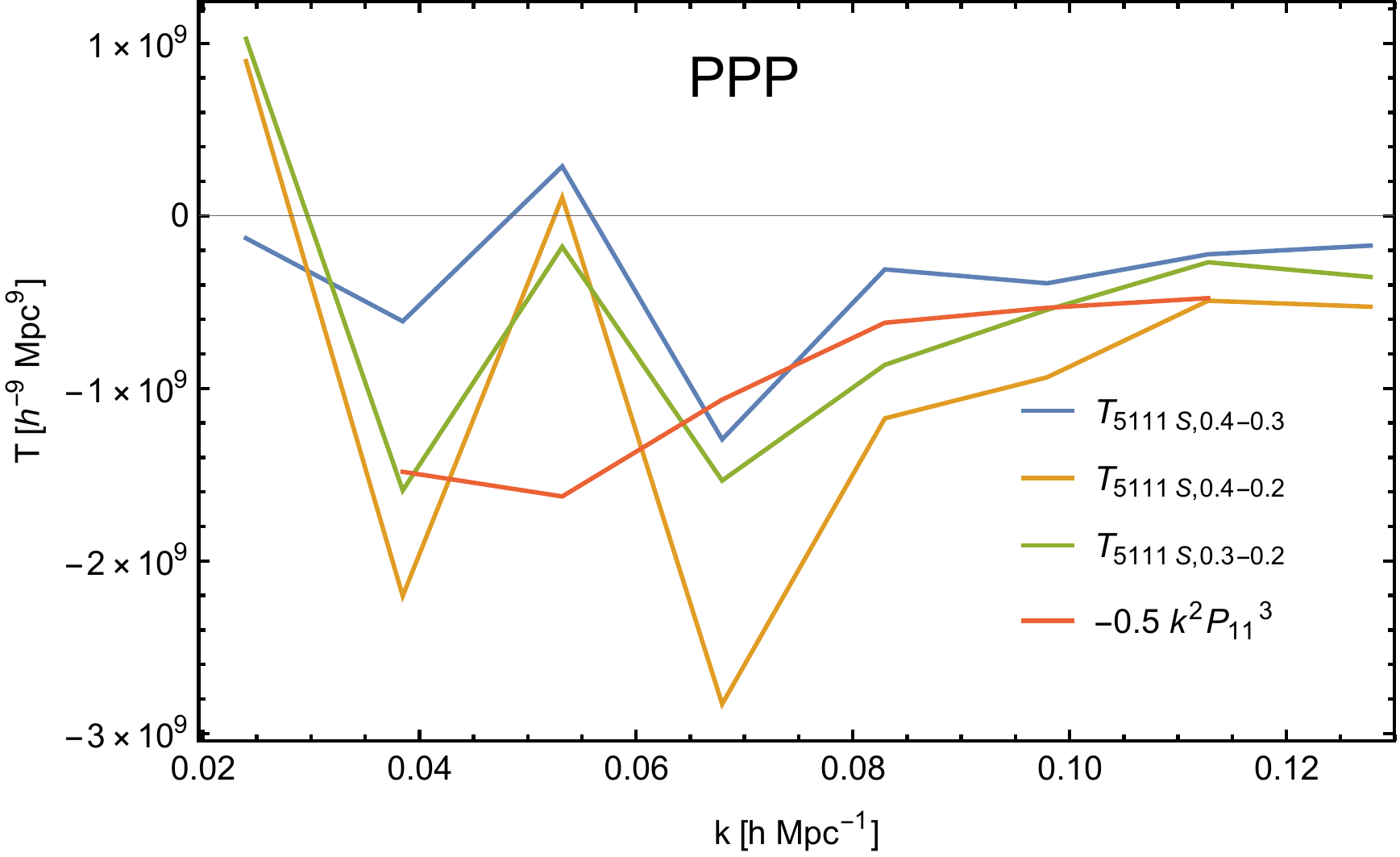}
	\caption{The measured $T_{5111\mathrm{s},i-j}$ for three combinations of cutoffs against $-0.5 k^{2}P_{11}^{3}$, showing the similar scaling.  This constitutes a validation of the calibration of our measurements and calculations, showing that the terms scale as they would be expected to.}
	\label{fig:T5111s}
\end{figure}

If our routines are properly calibrated, we should expect the scale dependent $T_{5111,\mathrm{S},i-j}$ to scale as $k^{2}P_{11}^{3}$.  As shown in Fig.~\ref{fig:T5111s}, we find that for the three possible combinations of trispectra with $\Lambda\epsilon\{0.2,0.3,0.4\}\ihMpc$, we have that
\begin{equation}
T_{5111}(\Lambda_2)-T_{5111}(\Lambda_1)\approx -0.5k^2 P^3~.
\end{equation}

\section{Generalised Trispectrum Estimators}
\label{App:FFT}
While Eq.~\eqref{eq:FFT1} integrates over the four external legs and allows the diagonals to vary, which is what we do for the four configurations we are going to focus on, we also developed equivalent algorithms for other forms of configuration.  Eq.~\eqref{eq:FFT2} allows us to measure a trispectrum configuration which has one diagonal leg fixed and is only allowing the other to vary, giving only one unspecified degree of freedom,
\begin{equation}
\begin{split}
\hat{T}&=\frac{1}{\mathcal{N}}\int \frac{\derd^{3}\bk_{1}}{(2\pi)^{3}} \int \frac{\derd^{3}\bk_{2}}{(2\pi)^{3}} \int \frac{\derd^{3}\bk_{3}}{(2\pi)^{3}} \int \frac{\derd^{3}\bk_{4}}{(2\pi)^{3}} \int \frac{\derd^{3}\bk_{5}}{(2\pi)^{3}}~(2\pi)^{6}\delta^{(\mathrm{D})}(\bk_{1}+\bk_{2}-\bk_{5})\delta^{(\mathrm{D})}(\bk_{3}+\bk_{4}+\bk_{5}) \delta(\bk_{1})\delta(\bk_{2})\delta(\bk_{3})\delta(\bk_{4})\\
&=\frac{1}{\mathcal{N}} \int \derd^{3}\bx \int \derd^{3}\bx' \int \frac{d^{3}\bk_{1}}{(2\pi)^{3}} \int \frac{\derd^{3}\bk_{2}}{(2\pi)^{3}} \int \frac{\derd^{3}\bk_{3}}{(2\pi)^{3}} \int \frac{\derd^{3}\bk_{4}}{(2\pi)^{3}}\int \frac{\derd^{3}\bk_{5}}{(2\pi)^{3}} ~ e^{i\bx(\bk_{1}+\bk_{2})}e^{-i\bk_{5}\bx}e^{i\bx'(\bk_{3}+\bk_{4})}e^{i\bk_{5}\bx'}\delta(\bk_{1})\delta(\bk_{2})\delta(\bk_{3})\delta(\bk_{4})\\
&\equiv\frac{1}{\mathcal{N}}\int \derd^{3}\bx  \int \derd^{3}\bx' \int \frac{\derd^{3}\bk_{5}}{(2\pi)^{3}} ~f_{1}(\bx)f_{2}(\bx)f_{3}(\bx')f_{4}(\bx')e^{-i\bk_{5}\bx}e^{i\bk_{5}\bx'}\\
&=\frac{1}{\mathcal{N}}\int \frac{\derd^{3}\bk_{5}}{(2\pi)^{3}} \int \derd^{3}\bx  ~f_{1}(\bx)f_{2}(\bx)e^{-i\bk_{5}\bx}\int \derd^{3}\bx' f_{3}(\bx')f_{4}(\bx')e^{i\bk_{5}\bx'}\\
&\equiv\frac{1}{\mathcal{N}}\int \frac{\derd^{3}\bk_{5}}{(2\pi)^{3}} ~f_{12}(\bk_{5})f_{34}(\bk_{5})~,
\end{split}
\label{eq:FFT2}
\end{equation}
where $\bk_{5}$ can trivially be replaced by $\bk_{6}$ with the appropriate changes in the Dirac functions, while Eq.~\eqref{eq:FFT3} allows us to measure a trispectrum in which all six legs are specified, limiting us to a single configuration \cite{PhysRevD.71.063001}:
\begin{equation}
\begin{split}
\hat{T}&=\frac{1}{\mathcal{N}}\int \frac{\derd^{3}\bk_{1}}{(2\pi)^{3}} \int \frac{\derd^{3}\bk_{2}}{(2\pi)^{3}} \int \frac{\derd^{3}\bk_{3}}{(2\pi)^{3}} \int \frac{\derd^{3}\bk_{4}}{(2\pi)^{3}} \int \frac{\derd^{3}\bk_{5}}{(2\pi)^{3}}\int \frac{\derd^{3}\bk_{6}}{(2\pi)^{3}} ~(2\pi)^{12}\delta^{(\mathrm{D})}(\bk_{1}+\bk_{2}-\bk_{5})\delta^{(\mathrm{D})}(\bk_{3}+\bk_{4}+\bk_{5}) \\&~~~~~\times\delta^{(\mathrm{D})}(\bk_{1}+\bk_{4}+\bk_{6})\delta(\bk_{1})\delta(\bk_{2})\delta(\bk_{3})\delta(\bk_{4})\\
&=\frac{1}{\mathcal{N}} \int \derd^{3}\bx \int \derd^{3}\bx' \int \derd^{3}\bx'' \int \frac{\derd^{3}\bk_{1}}{(2\pi)^{3}} \int \frac{\derd^{3}\bk_{2}}{(2\pi)^{3}} \int \frac{\derd^{3}\bk_{3}}{(2\pi)^{3}} \int \frac{\derd^{3}\bk_{4}}{(2\pi)^{3}} \int \frac{\derd^{3}\bk_{5}}{(2\pi)^{3}}\int \frac{\derd^{3}\bk_{6}}{(2\pi)^{3}} ~ e^{i\bx\cdot(\bk_{1}+\bk_{2})}e^{-i\bk_{5}\cdot\bx}\\&~~~~~\times e^{i\bx'\cdot(\bk_{3}+\bk_{4})} e^{i\bk_{5}\cdot\bx'}e^{i\bx''\cdot(\bk_{1}+\bk_{4})}e^{i\bk_{6}\cdot\bx''} \delta(\bk_{1})\delta(\bk_{2})\delta(\bk_{3})\delta(\bk_{4})\\
&=\frac{1}{\mathcal{N}} \int \frac{\derd^{3}\bk_{6}}{(2\pi)^{3}}\int \derd^{3}\bx''e^{\ii \vec k_6\cdot \vec x''} 
 \int \frac{\derd^{3}\bk_{5}}{(2\pi)^{3}} \int \derd^{3}\bx e^{-\ii \vec k_5\cdot \vec x}f_1(\vec x+\vec x'')f_2(\vec x)
 \int \derd^{3}\bx'
e^{\ii \vec k_5\cdot \vec x'}
f_3(\vec x')f_4(\vec x'+\vec x'')\\
&=\frac{1}{\mathcal{N}} \int \frac{\derd^{3}\bk_{6}}{(2\pi)^{3}}\int \derd^{3}\bx''e^{\ii \vec k_6\cdot \vec x''} 
\int \frac{\derd^{3}\bk_{5}}{(2\pi)^{3}}
f_{12}(\vec k_5,\vec x'')
f_{34}(\vec k_5,\vec x'')\\
&=\frac{1}{\mathcal{N}} \int \frac{\derd^{3}\bk_{6}}{(2\pi)^{3}}\int \derd^{3}\bx''e^{\ii \vec k_6\cdot \vec x''} 
f_{1234}(\vec x'')\\
&=\frac{1}{\mathcal{N}} \int \frac{\derd^{3}\bk_{6}}{(2\pi)^{3}}
f_{1234}(\vec k_6)\\
&=\frac{1}{\mathcal{N}} f_{1234}(k_6)~.
\end{split}
\label{eq:FFT3}
\end{equation}
The formalism that allows to fix both diagonals is numerically more demanding than the algorithms marginalizing over one or both diagonals. Fixing both diagonals requires Fourier transforms of $f_1(\vec x+\vec x'')f_2(\vec x)$ and $f_3(\vec x')f_4(\vec x'+\vec x'')$ for all $\vec x''$ in the third line.  This is similar to the method suggested in Eq. (13) of \cite{Tomlinson:2019bjx}, which begins with a different set of Dirac functions but accomplishes a similar effect.

\bibliography{bibliography}

\end{document}